\documentclass[aps,prl,reprint]{revtex4-2}
\usepackage{amsmath,amssymb,graphicx,numprint,times}
\usepackage[usenames,dvipsnames,svgnames,table]{xcolor}
\definecolor{prlblue}{rgb}{0.18,0.18,0.573}
\usepackage[colorlinks=true, urlcolor=prlblue, linkcolor=prlblue, citecolor=prlblue, pdftex]{hyperref}

\npthousandsep{\,}
\npdecimalsign{.}

\def\<{\langle}
\def\>{\rangle}
\newcommand{\prlsection}[1]{{\it #1.}---}

\newcommand{\PRL}{Phys.~Rev.~Lett.}

\newcommand{\PRB}{Phys.~Rev. B}

\newcommand{\PRE}{Phys.~Rev. E}
\newcommand{\PRX}{Phys.~Rev. X}

\newcommand{\JPCM}{J.~Phys.: Condens.~Matter}
\newcommand{\JPAOLD}{J.~Phys.~A: Math.~Gen.}

\newcommand{\physrep}{Phys.~Rep.}

\newcommand{\NPB}{Nucl.~Phys.~B}

\newcommand{\RMP}{Rev.~Mod.~Phys}

\newcommand{\beq}{\begin{equation}}
\newcommand{\eeq}{\end{equation}}
\newcommand{\bea}{\begin{eqnarray}}
\newcommand{\eea}{\end{eqnarray}}

\begin{document}

\title{Boundary Criticality of the 3D O($N$) Model: From Normal to Extraordinary}
\author{\firstname{Francesco} \surname{Parisen Toldin}}
\email{francesco.parisentoldin@physik.uni-wuerzburg.de}
\affiliation{\mbox{Institut f\"ur Theoretische Physik und Astrophysik, Universit\"at W\"urzburg, Am Hubland, D-97074 W\"urzburg, Germany}}
\author{\firstname{Max~A.} \surname{Metlitski}}
\email{mmetlits@mit.edu}
\affiliation{Department of Physics, Massachusetts Institute of Technology, Cambridge, Massachusetts 02139, USA}
\begin{abstract}
It was recently realized that the three-dimensional O($N$) model possesses  an extraordinary boundary universality class for a finite range of $N \ge 2$. For a given $N$, the existence and universal properties of this class are predicted to be controlled by certain  amplitudes of the normal universality class, where one applies an explicit symmetry breaking field to the boundary. In this Letter, we  study the normal universality class for $N = 2, 3$ using Monte Carlo simulations on an improved lattice model and extract these universal amplitudes. Our results are in good agreement with direct Monte Carlo studies of the extraordinary universality class serving as a nontrivial quantitative check of the connection between the normal and extraordinary classes.
\end{abstract}

\maketitle

\prlsection{Introduction}
When a physical systems is
in the vicinity of a continuous phase transition, various observables develop power-law singularities which have a character of universality: they are determined by the gross features of the system, such as dimensionality and symmetry, and not by the details of local interactions.
Renormalization-group (RG) theory allows us to understand the emergence of universality as the result of the existence of fixed points in a suitably defined flow of Hamiltonians.
Accordingly, systems exhibiting identical critical behavior define a given universality class (UC) \cite{Cardy-book}.
The presence of a boundary gives rise to rich phenomena, which have attracted a large amount of experimental \cite{Dosch-book} and theoretical \cite{Binder-83,Diehl-86,Pleimling-review} studies.
General RG arguments show that a given bulk UC class, describing the critical behavior far away from the boundary, potentially admits different surface UCs \cite{Cardy-book}.
Further, surface critical exponents  and other universal data generally differ from those of the bulk \cite{Binder-83,Diehl-86}.
Surface UCs also determine the critical Casimir force \cite{FG-78,Krech-94,Krech-99,BDT-00,Gambassi-09,GD-11,MD-18}.
While boundary criticality is a mature subject, it has recently received renewed attention driven in part by advances in conformal field theory \cite{McAvity:1995zd, Liendo2013, Gliozzi2015, Billo:2016cpy, Liendo:2016ymz,Lauria:2017wav, Mazac:2018biw, Kaviraj:2018tfd, Dey:2020lwp, Behan:2020nsf, Gimenez-Grau:2020jvf} and developments in topological phases of quantum matter. Many topological phases including quantum Hall states, topological insulators, and certain quantum spin liquids  possess protected boundary states. While it was initially thought that this protection relies on the presence of a bulk energy gap, examples where the boundary state survives in some form even as the bulk gap closes were later discovered \cite{TarungaplessSPT, MaissamCFL, MaissamGenon, MaissamMajorana,ScaffidiGapless, ParkerGapless, RyangaplessSPT, RubengaplessSPT, RubenIntGapless}. The study of such ``gapless topological states'' and their boundaries lies in the domain of boundary critical phenomena. As gapless topological states were investigated in the context of quantum magnets \cite{SS-12,ZW-17,DZG-18,WPTW-18,WW-19,JXWX-20,ZDZG-20,WW-20,DZGZ-21}, it was realized that even for the simplest model of {\it classical} magnets---the O($N$) model---basic questions about the boundary phase diagram remain open \cite{Metlitski-20,PT-20,HDL-21}.

The much investigated classical O($N$)  model \cite{PV-02} provides a prototypical example of boundary criticality. In three dimensions, for
$N=1$, $2$, the bulk-surface phase diagram hosts a surface transition line, where the bulk is disordered and the surface critical behavior belongs to that of 2D O($N$) UC. This line terminates at the bulk transition line dividing it into ordinary and extraordinary surface UCs; the termination point is the so-called special UC \cite{Binder-83,Diehl-86}.
Surprisingly, the surface phase diagram for $N > 2$ is still not fully settled. For $d=3$ and $N>2$ there is no surface transition for a disordered bulk \cite{PV-02}, thus the topology of the phase diagram does not necessarily dictate  the existence of the extraordinary UC or the special multicritical point \cite{Krech-00,DBN-05}. Yet, a recent field-theoretical analysis in Ref.~\cite{Metlitski-20} has pointed out that if one treats $N$ as a continuous parameter, the extraordinary UC survives for a range  $2 < N < N_c$, where $N_c$ is a currently  unknown constant. Further, the extraordinary UC in the region  $2 \leq N < N_c$ exhibits a surface order parameter correlation function that falls off as
\begin{equation} \langle \vec{\phi}({\bf x}) \cdot \vec{\phi}(0)\rangle \sim \frac{1}{(\log {\bf x})^q}, \label{eq:qlog}\end{equation}
thus, it was labeled the ``extraordinary-log'' UC in Ref.~\cite{Metlitski-20}; this should be contrasted to the extraordinary transition for $N=1$ or in $d >3$ where the above correlation function approaches a constant at large separation.   In fact, for $N  = 3$ a recent numerical simulation \cite{PT-20} finds firm evidence of a special transition with exponents differing from those of the ordinary UC and a phase consistent with the extraordinary-log UC, implying $N_c > 3$ \footnote{Earlier a hint of the special transition for $N = 3$ was seen in Ref.~\cite{DBN-05}}. For $N=2$, the ``logarithmic'' character of the extraordinary phase was also verified numerically~\cite{HDL-21}.

Reference~\cite{Metlitski-20} showed that for a given $N$ the existence of the extraordinary-log phase and its properties [such as the exponent $q$ in Eq.~(\ref{eq:qlog})]  are determined by certain universal amplitudes of the normal boundary UC. The latter is realized when an explicit symmetry breaking field is applied to the boundary \cite{Binder-83,Diehl-86,BM-77,BC-87}.

Motivated by these recent developments, in this Letter we study the normal surface UC of the three dimensional  O($N$) model, for $N=2$ and $N=3$,
by means of Monte Carlo (MC) simulations of an {\it improved} lattice model \cite{PV-02}, where the leading bulk irrelevant scaling field is suppressed.
Through a finite-size scaling analysis of MC data we determine certain universal amplitudes of the normal UC.
Such amplitudes are {\it per se} of interest, as they provide a quantitative description of the normal UC;
for $N=1$ they have been studied in Ref.~\cite{PTD-10}.
Furthermore, exploiting the analysis of Ref.~\cite{Metlitski-20}, our results confirm the existence of the extraordinary-log UC for $N=3$, and allow us to compute the universal exponent $q$ in Eq.~(\ref{eq:qlog}) for $N=2$ and $N=3$.
Our results are in good agreement with the value of $q$ found in direct studies of the extraordinary phase in Refs.~\cite{PT-20, HDL-21}.

\prlsection{Model}
We study the classical lattice $\phi^4$ model by means of MC simulations. It is defined on a three-dimensional $L_\parallel\times L_\parallel\times L$ lattice, with periodic boundary conditions (BCs) along the lateral directions with size $L_\parallel$, and open BCs along the remaining direction.
The reduced Hamiltonian ${\cal H}$, such that the Gibbs weight is $\exp(-\cal H)$, is
\begin{equation}
  \begin{split}
    &{\cal H} = -\beta\sum_{\< i\ j\>}\vec{\phi}_i\cdot\vec{\phi}_j
    -\beta_{s}\sum_{\< i\ j\>_{s}}\vec{\phi}_i\cdot\vec{\phi}_j -\vec{h}_{s}\cdot\sum_{i\in s} \vec{\phi}_i\\
    &
    +\sum_i[\vec{\phi}_i^{\,2}+\lambda(\vec{\phi}_i^{\,2}-1)^2],
  \end{split}
  \label{model}
\end{equation}
where $\vec{\phi}_x$ is an $N-$components real field on the lattice site $x$ and the first sum extends over the nearest-neighbor pairs where at least one site belongs to the inner bulk. The second and third sums extend over the lattice sites on the surface.
The last term in Eq.~(\ref{model}) is summed over all lattice sites. In Eq.~(\ref{model}) the coupling constant $\beta$ determines the critical behavior of the bulk, while $\beta_{s}$ 
controls the surface coupling. Finally, we have introduced a symmetry-breaking boundary field $\vec{h}_{s}= h_s \vec{e}_N$ along the $N$th direction.  

For $\lambda\rightarrow\infty$, the Hamiltonian (\ref{model}) reduces to the hard spin O($N$) model. In the $(\beta, \lambda)$ plane, the bulk exhibits a second-order transition line in the O($N$) UC \cite{CHPRV-02,PV-02,CHPV-06}.
For $N=2$ the model is {\it improved} for $\lambda=2.15(5)$ \cite{CHPV-06}, i.e., the leading bulk irrelevant scaling field with dimension $y_i=-0.789(4)$ \cite{Hasenbusch-19} is suppressed. At $\lambda=2.15$ the model is critical for $\beta=\numprint{0.50874988}(6)$ \cite{PT-21}.
For $N=3$ the model is improved for $\lambda=5.17(11)$ and the suppressed leading irrelevant scaling field has dimension $y_i=-0.759(2)$ \cite{Hasenbusch-20}. At $\lambda=5.2$, the model is critical at $\beta=\numprint{0.68798521}(8)$ \cite{Hasenbusch-20}.
Improved models are instrumental to obtain accurate results in critical phenomena \cite{PV-02}, in particular in boundary critical phenomena
\cite{Hasenbusch-09b,Hasenbusch-10c,PTD-10,Hasenbusch-11,Hasenbusch-11b,Hasenbusch-12,PTTD-13,PT-13,PTTD-14,PTAW-17,PT-20}, because the broken translational invariance generically gives rise to additional scaling corrections, which cumulate to those arising from bulk irrelevant operators.
The latter are suppressed for improved lattice models, hence enabling a more accurate analysis.

In the MC simulations presented here we set $\beta$ and $\lambda$ to the central value of the bulk critical point in the improved models, and $\beta_s=\beta$. 
Note that the boundary parameters $\beta_s$, $h_s$ are chosen to be identical on the two surfaces: this realizes the normal UC on both surfaces, and allows us to compute improved estimators of surface observables by averaging them over the two surfaces.
The geometry is fixed by $L=L_\parallel$.
MC simulations are performed by combining Metropolis, overrelaxation, and Wolff single-cluster updates \cite{Wolff-89}; details of the algorithm are reported in Ref.~\cite{PT-20},
and the implementation of the Wolff algorithm in the presence of a symmetry-breaking surface field is discussed in Ref.~\cite{PTD-10}.
The inclusion of a boundary field breaks the O($N$) symmetry to O($N-1$). Accordingly, we distinguish the components of $\vec{\phi}$ defining $\vec{\phi} \equiv (\vec{\varphi}, \sigma)$, where $\sigma$ is the component parallel to the surface field, and $\vec{\varphi}$ is a $(N-1)$-component vector orthogonal to it.
As discussed below, we measure the magnetization profile $\<\sigma\>$ and various surface-surface and surface-bulk two-point functions.

Besides the model realizing the normal UC, we also perform some MC simulations of the $\phi^4$ model with periodic BCs, with the aim of determining the bulk field normalization. In this case, the Hamiltonian is as in Eq.~(\ref{model}), without the surface terms.

\prlsection{Normal universality class}
In this section, we discuss the normal surface UC of the O($N$) model in $d = 3$. Unless otherwise stated, all  operators in this section [e.g. $\vec{\phi} = (\vec{\varphi}, \sigma)$] denote continuum fields; when referring to fields of the lattice model (\ref{model}), we use the subscript ``$\rm{lat}$''. To leading order, the bulk field $\vec{\phi}_{\mathrm{lat}} \propto \vec{\phi}$. 
The boundary operator spectrum  contains two ``protected'' operators whose existence is mandated by bulk conservation laws and whose scaling dimensions are known exactly \cite{BC-87,BM-77, Cardy-90}: 
\begin{itemize}
\item The ``tilt'' operator $\mathrm{t}^i$ of dimension $\hat{\Delta}_{\mathrm{t}} = d-1 = 2$, which is an O(N-1) vector ($i = 1 \ldots N-1$). This operator is induced on the boundary when the symmetry breaking field $\vec{h}_s$ is tilted---thus the nomenclature \cite{Padayasi}.
\item The displacement operator $\mathrm{D}$ of dimension $\hat{\Delta}_{\mathrm{D}} = d = 3$, which is an O(N-1) scalar.  Perturbing the boundary with this operator is equivalent to moving the location of the boundary, justifying the name ``displacement''.
\end{itemize}
These are believed to be the two lightest boundary operators. In particular, on the lattice the boundary field $\varphi^i_{\rm{lat}} \propto {\rm t}^i$.
The boundary operator product expansion (OPE) holds for $z\rightarrow 0$:
\begin{eqnarray} \sigma({\bf x}, z) &=& \frac{a_\sigma}{(2 z)^{\Delta_{\phi}}}  + b_{\mathrm{D}} (2 z)^{3 - \Delta_{\phi}} {\rm D}({\bf x}) + \ldots,\nonumber \\
\varphi^i({\bf x}, z) &=& b_{\mathrm{t}} (2 z)^{2 - \Delta_\phi} {\rm t}^i({\bf x}) + \ldots, \label{eq:BOPE} \end{eqnarray}
where $\Delta_\phi$ is the bulk scaling dimension of $\vec{\phi}$. The coefficients $a_\sigma$, $b_{\mathrm{t}}$, $b_{\mathrm{D}}$ are universal, assuming that the bulk and boundary operators are normalized.  $a_\sigma$ and $b_{\mathrm{t}}$ will be the main target of this Letter---as was shown in Ref.~\cite{Metlitski-20}, their ratio controls the existence and universal properties of the extraordinary-log phase (in the absence of a boundary magnetic field). Defining 
\begin{equation} \alpha \equiv \frac{\pi}{2}\left(\frac{a_\sigma}{4\pi b_{\mathrm{t}}}\right)^2 - \frac{N-2}{2\pi},\label{eq:alpha}\end{equation}
the extraordinary-log phase exists when $\alpha > 0$. Further, the exponent $q$ in Eq.~(\ref{eq:qlog}) is given by 
\begin{equation} q = \frac{N-1}{2 \pi \alpha}. \label{eq:qalpha}\end{equation}

We extract $a_\sigma$ and $b_{\mathrm{t}}$  from the following correlators, which in a semi-infinite geometry take the form
\begin{equation} \langle \sigma(z) \rangle = \frac{a_\sigma}{(2z)^{\Delta_\phi}}, \quad \langle {\rm t}^i(0) \varphi^j({\bf x}, z)\rangle  = \delta_{ij} b_{\rm t} \frac{(2 z)^{2 - \Delta_{\phi}}}{({\bf x}^2 + z^2)^2}. \label{tvarphi} \end{equation}
The bulk field $\phi^a$, $a = 1\ldots N$, is normalized so that in an infinite geometry $\langle \phi^a(x) \phi^b(0)\rangle = \delta^{ab} x^{-2 \Delta_{\phi}}$, while ${\rm t}^i$ is normalized so that in a semi-infinite geometry $\langle \mathrm{t}^i({\bf x} ) \mathrm{t}^j(0)\rangle = \delta^{ij} {\bf x}^{-4}$. Thus, in a lattice model, to fix the normalizations above and to find $a_\sigma$, $b_{\rm t}$ we will need to measure four different correlators. On the lattice, we expect both finite size scaling corrections and corrections to scaling \cite{SM}.

\prlsection{Results}
\begin{figure}
  \centering
  \includegraphics[width=\linewidth]{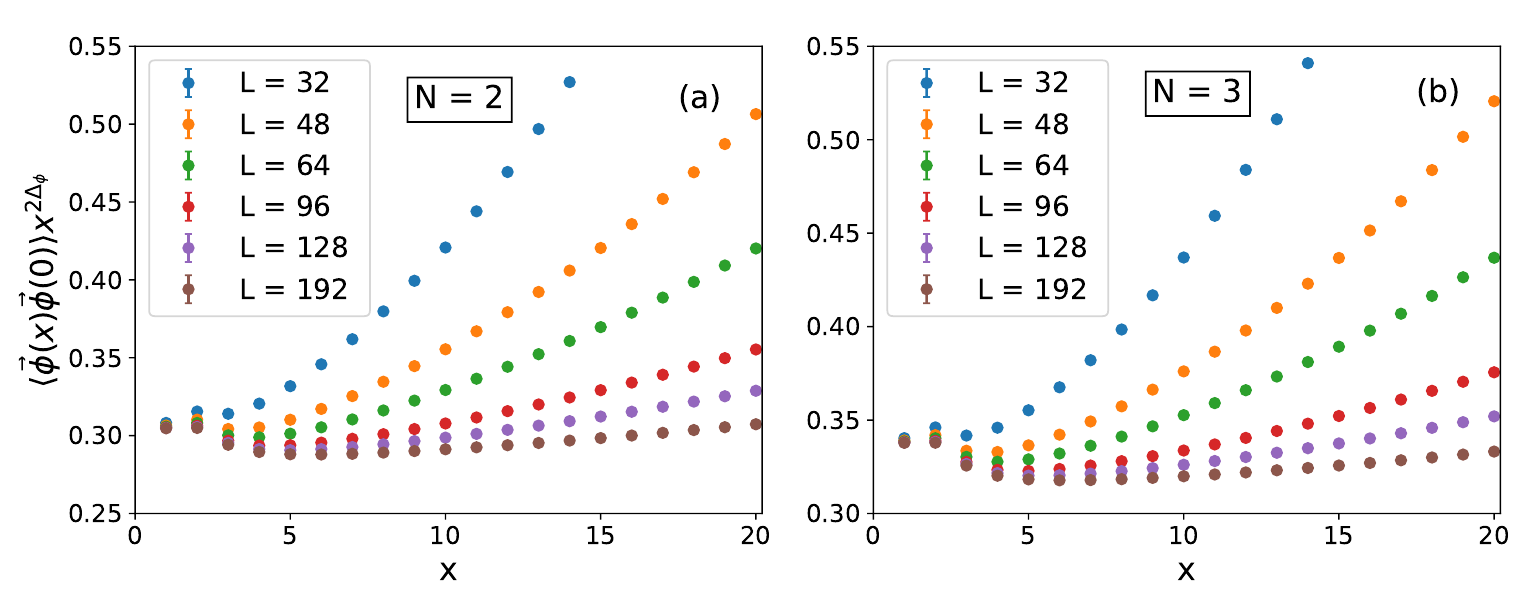}
  \caption{Bulk two-point function for (a) $N=2$ and (b) $N=3$, rescaled to the large-distance decay exponent $2\Delta_\phi$. Error bars are smaller than symbol size.}
  \label{plots_bulk}
\end{figure}
We study first the $XY$ UC.
In order to determine the normalization of the bulk field,
we have simulated the $\phi^4$ model for $N=2$, with periodic BCs and at the critical point, for lattice sizes $L=32-192$.
In Fig.~\ref{plots_bulk}(a) we show the two-point function $\<\vec{\phi}(x)\cdot \vec{\phi}(0)\>$, rescaled to the expected decay $x^{-2\Delta_\phi}$.
Here and below we use $\Delta_\phi=0.519088(22)$, $\Delta_\epsilon=1.51136(22)$ \cite{CLLPSDSV-19}.
We fit the MC data to
\begin{equation}
  \langle\vec{\phi}(x)\cdot\vec{\phi}(0)\rangle = {\cal N}_{\rm bulk} x^{-2\Delta_\phi}\left[1 + B_\epsilon\left(\frac{x}{L}\right)^{\Delta_\epsilon} + C x^{-2}\right],
  \label{phi_bulk_ansatz}
\end{equation}
    where the leading finite-size correction $\propto L^{-\Delta_\epsilon}$
is due to the energy operator in the OPE of $\phi^a\times\phi^a$, while
      the correction $\propto x^{-2}$ comes from the next-to-leading irrelevant operator in the action and from descendant operators in the expansion of the lattice field $\vec{\phi}_i$ in terms of continuum fields \cite{SM}.
Equation (\ref{phi_bulk_ansatz}) holds
for $(x/L)\ll 1$ and $x \gtrsim x_0$, with $x_0$ a nonuniversal length governing the two-point function at short distance.
The analysis of various fits \cite{Young_notes,Michael-94,Seibert-94} to Eq.~(\ref{phi_bulk_ansatz})
allows us to infer \cite{SM}:
\begin{equation}
    {\cal N}_{\rm bulk} = 0.28152(15),\qquad
    B_\epsilon = 2.758(10).
  \label{Nbulk_xy}
\end{equation}

\begin{figure}
  \centering
  \includegraphics[width=\linewidth]{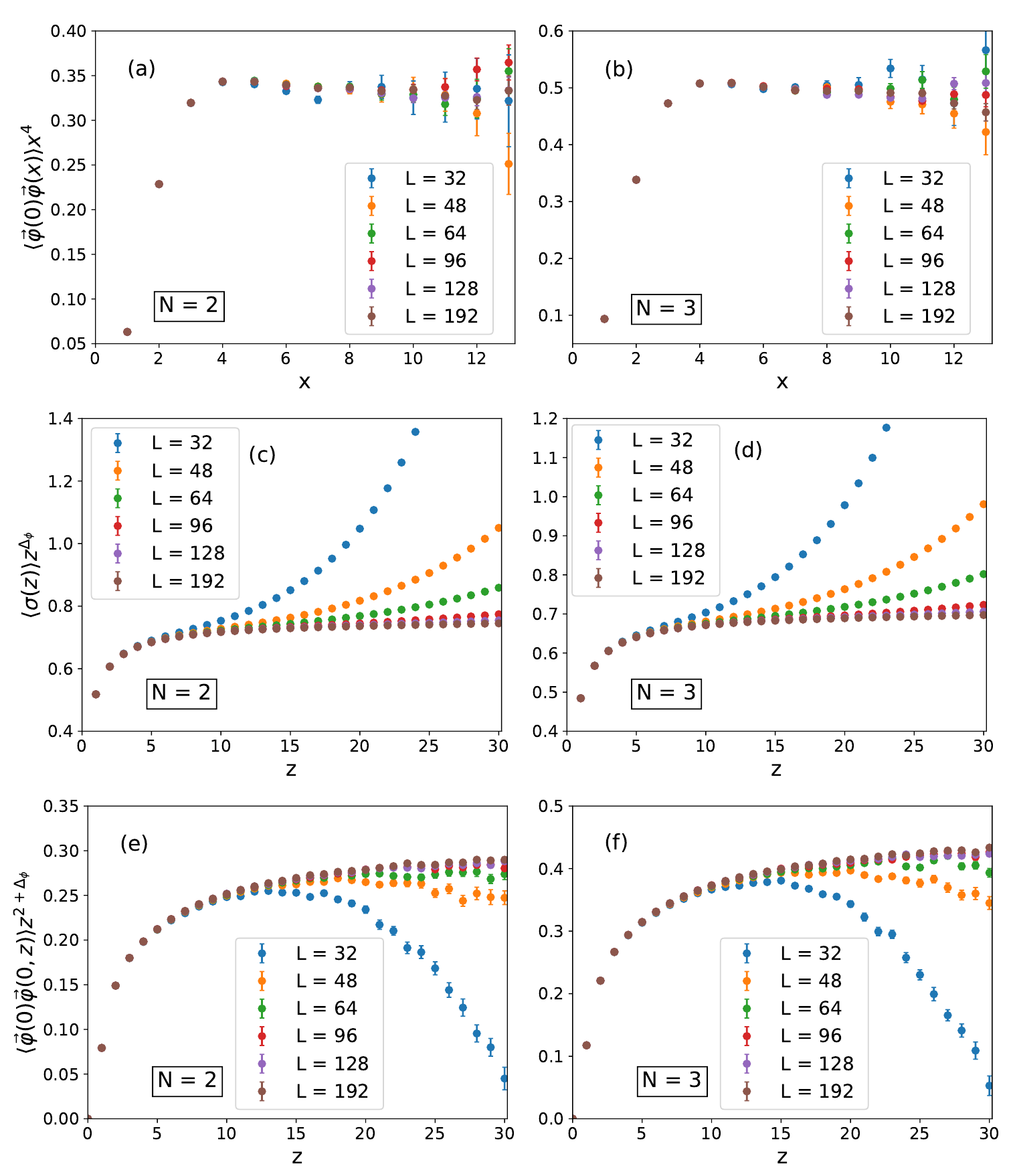}
  \caption{Plots of surface observables. (a) and (b): Two-point functions of the surface field component $\varphi$, for $N=2$ and $N=3$, rescaled to the large-distance decay exponent $4$. (c) and (d): Order-parameter profile as a function of the distance from the surface, rescaled to the large-distance decay exponent $\Delta_\phi$. (e) and (f): Surface-bulk correlation functions of the field component $\varphi$, rescaled to the large-distance decay exponent $2+\Delta_\phi$.}
  \label{plots_surface}
\end{figure}
Next, we study the surface critical behavior.
As discussed above, in order to implement the normal UC, we simulated the $\phi^4$ model at the critical point with open BCs and a symmetry-breaking surface field.
Preliminary MC data suggested a reduction of corrections to scaling for a surface field $h_s=1.5\beta_s$. Our MC simulations reported below have thus been done at this value of $h_s$, for lattice sizes $L=32-192$.
To extract the normalization of the surface field component $\varphi$, we have computed its two-point function along the surface.
We show it in Fig.~\ref{plots_surface}(a), rescaled to its expected large-distance decay exponent $4$.
In this case finite-size corrections are rather small, such that we fit the data to \footnote{Even though $\vec{\varphi}$ has only a single component in the $N = 2$ case, we write it as a vector here to set the normalization for $N = 3$.}
\begin{equation}
  \<\vec{\varphi}({\bf x})\cdot\vec{\varphi}(0)\> = {\cal N}_\varphi {\bf x}^{-4}\left(1 + C {\bf x}^{-2}\right),
  \label{phi_ansatz}
\end{equation}
where, analogous to Eq.~(\ref{phi_bulk_ansatz}), the leading correction to scaling $\propto {\bf x}^{-2}$ originates from the expansion of the lattice operator in terms of continuum ones.
Fits to Eq.~(\ref{phi_ansatz}) deliver \cite{SM}
 \begin{equation}
  {\cal N}_\varphi = 0.328(3).
  \label{Nphi_xy}
\end{equation}

In Fig.~\ref{plots_surface}(c), we show
the magnetization profile $\langle \sigma(z) \rangle $ as a function of the distance from the surface $z$,
and rescaled to its asymptotic decay exponent $\Delta_\phi$.
For this quantity, scaling and finite-size corrections are relevant and we fit MC data to
\begin{equation}
  \langle\sigma(z)\rangle = M_\sigma (z + z_0)^{-\Delta_\phi}\Big[1+ B_\sigma \Big(\frac{z + z_0}{L}\Big)^3\Big],
  \label{sigma_ansatz}
\end{equation}
      where the perturbation of the surface action with the displacement operator ${\rm D}$ produces the replacement $z\rightarrow z+z_0$, with $z_0$ a nonuniversal constant, and
      the leading finite-size correction originates from the OPE (\ref{eq:BOPE}) \cite{SM};
      the latter is also known as distant wall correction \cite{FG-78,EKD-93,*EKD-93_erratum,ES-94}.
Fits to Eq.~(\ref{sigma_ansatz})
allow us to estimate \cite{SM}
\begin{equation}
    M_\sigma = 0.7540(3),\quad
    B_\sigma = 1.21(6),\quad
    z_0     = 1.018(6).
  \label{sigma_xy}
\end{equation}

In Fig.~\ref{plots_surface}(e) we show
the surface-bulk correlation function $\<\varphi(0)\varphi(0,z)\>$ of the field component $\varphi$, where one point is on the surface  and the other a distance $z$ away from the surface, so that the vector separating the two points is orthogonal to the surface; the correlations are rescaled to the large-distance decay exponent $2+\Delta_\phi$.
These correlations are affected by significant scaling corrections, while finite-size corrections, though not negligible, are smaller than in the case of $\langle \sigma(z) \rangle$.
Together with the relatively fast large-distance decay, $\sim z^{-2-\Delta_\phi}$, this makes the analysis of the surface-bulk correlations more involved.
A good ansatz for the MC data is
\begin{multline}
    \<\vec{\varphi}(0)\cdot\vec{\varphi}(0, z)\> = M_\varphi (z + z_0)^{-2-\Delta_\phi}\Big[1+ B_\varphi \left(\frac{z + z_0}{L}\right)^3\\
      + C (z + z_0) ^ {-2}\Big],
  \label{phiperp_ansatz2}
\end{multline}
  where we have included the corrections considered in Eq.~(\ref{phi_ansatz}) and Eq.~(\ref{sigma_ansatz}) \cite{SM}.
To avoid overfitting, in
fits
to Eq.~(\ref{phiperp_ansatz2}) we plug in the result for $z_0$ of Eq.~(\ref{sigma_xy}), varying its value within one error bar quoted there.
From the various fits we  estimate
\begin{equation}
    M_\varphi = 0.3146(8),\qquad
    B_\varphi = -0.7(2).
  \label{Mphi_xy}
\end{equation}

In the analysis of the MC data for the Heisenberg normal UC,
we proceed analogous to the case $N=2$, using the critical exponents $\Delta_\phi = 0.518920(25)$ and $\Delta_\epsilon=1.5948(2)$ \cite{Hasenbusch-20}.
To extract the normalization of the bulk field $\phi$,
we simulated the $\phi^4$ model for $N=3$, periodic BCs, and at the critical point, for lattice sizes $L=32-192$.
In Fig.~\ref{plots_bulk}(b) we show the two-point function of $\phi$, rescaled to the expected decay $x^{-2\Delta_\phi}$.
From fits of the correlations to Eq.~(\ref{phi_bulk_ansatz})
we obtain \cite{SM}
\begin{equation}
  \begin{split}
    {\cal N}_{\rm bulk} = 0.31230(15),\qquad
    B_\epsilon          = 2.432(7).
  \end{split}
  \label{Nbulk_heisenberg}
\end{equation}

Concerning the surface critical behavior,
preliminary MC
simulations of the model (\ref{model}) with $N=3$
suggested a reduction of subleading corrections for a surface field $h_s=1.4\beta_s$.
Here, we present results for this choice of $h_s$, and lattice sizes $L=32-192$.
In Fig.~\ref{plots_surface}(b) we show the surface correlations.
Fits of $\langle \vec{\varphi}({\bf x}) \cdot \vec{\varphi}(0)\rangle$ to Eq.~(\ref{phi_ansatz})
allow us to  estimate \cite{SM}
\begin{equation}
  {\cal N}_\varphi = 0.481(3).
  \label{Nphi_heisenberg}
\end{equation}
Fits of the order-parameter profile $\langle \sigma(z) \rangle$, shown in Fig.~\ref{plots_surface}(d),
deliver the following results:
\begin{equation}
    M_\sigma = 0.7062(2), \quad
    B_\sigma = 1.07(5),\quad
    z_0     = 1.031(4).
  \label{sigma_heisenberg}
\end{equation}
\vspace{-0.5cm}

In Fig.~\ref{plots_surface}(f) we show
the surface-bulk correlation function $\<\vec{\varphi}(0)\cdot \vec{\varphi}(0,z)\>$.
We fit it to Eq.~(\ref{phiperp_ansatz2}), employing the estimate of $z_0$ given in Eq.~(\ref{sigma_heisenberg}).
From the various fits
we obtain \cite{SM}
\begin{equation}
    M_\varphi = 0.4674(8),\qquad
    B_\varphi = -0.7(1).
  \label{Mphi_heisenberg}
\end{equation}

\prlsection{Discussion}
\begin{table}[t]
\centering
\begin{ruledtabular}
\begin{tabular}{c c c c c c}

$N$& $a_\sigma$ &$b_\textup{t}$& $b_{\rm D}$ &$\alpha$&  $\alpha_{\rm{eo}}$ \\
\hline
1 & 2.60(5) & & 0.244(8)& &\\

2 & 2.880(2) &  0.525(4) & &0.300(5) & 0.27(2)\cite{HDL-21}\\

3 & 3.136(2)& 0.529(3) & & 0.190(4) & 0.15(2)\cite{PT-20}\\
\end{tabular}
\end{ruledtabular}
\caption{Final results: universal amplitudes  $a_\sigma$, $b_{\rm t}$, $b_{\rm D}$ in Eq.~(\ref{eq:BOPE}) together with the corresponding value of $\alpha$, Eq.~(\ref{eq:alpha}). We also tabulate $\alpha_{\rm{eo}}$ found in Refs.~\cite{PT-20, HDL-21} by direct MC simulations of the extraordinary region. For $N =1$ we present the values of $a_\sigma$, $b_{\rm D}$ extracted from MC results of Refs.~\cite{PTD-10, Hasenbusch-10c}.}  
\label{tbl:final}
\end{table}
According to the discussion above of the normal UC and of the scaling forms in Ref.~\cite{SM},
the results of our scaling analysis of MC data
allow us to extract  universal amplitudes $a_\sigma$, $b_{\rm t}$ of the normal UC via
\begin{equation}
 a_\sigma = \frac{2^{\Delta_\phi}M_\sigma}{\sqrt{{\cal N}_{\rm bulk} / N}}, \;
 b_{\rm t} = \frac{2^{\Delta_\phi}M_\varphi}{4\sqrt{N-1}\sqrt{{\cal N}_{\rm bulk} / N} \sqrt{{\cal N}_\varphi}}.
\end{equation}
 $a_\sigma$ is  obtained from the amplitude of the order-parameter profile $M_\sigma$ (\ref{sigma_ansatz}) and the normalization of the bulk field ${\cal N}_{\rm bulk}$ (\ref{phi_bulk_ansatz}),
while $b_{\rm t}$ is obtained in terms of the amplitude of the surface-bulk correlations $M_\varphi$ (\ref{phiperp_ansatz2}), and the bulk and surface normalizations ${\cal N}_{\rm bulk}$, (\ref{phi_bulk_ansatz}), ${\cal N}_\varphi$, (\ref{phi_ansatz}). We collect our results for $a_\sigma$, $b_{\rm t}$ in Table \ref{tbl:final}, including
the value of $\alpha$ obtained from $a_\sigma$ and $b_{\rm t}$ via Eq.~(\ref{eq:alpha}).
For both $N = 2$ and $N = 3$, $\alpha > 0$, which indicates that the extraordinary-log UC exists, in accord with MC results of Refs.~\cite{PT-20, HDL-21}. Reference~\cite{Metlitski-20} predicts that $\alpha$ controls various universal properties of the extraordinary-log phase, including the exponent $q$ in Eq.~(\ref{eq:qlog}), which is related to $\alpha$ via Eq.~(\ref{eq:qalpha}). $\alpha$ found here agrees well with $\alpha$ extracted from direct MC simulations of the extraordinary region \cite{PT-20,HDL-21}, listed in Table \ref{tbl:final} as $\alpha_{\rm{eo}}$. This provides a highly nontrivial check of the theory in Ref.~\cite{Metlitski-20}. As pointed out in Ref.~\cite{PT-20}, the error bar on $\alpha_{\rm{eo}}$ should be taken with a grain of salt given the difficulty of fitting to the form in Eq.~(\ref{eq:qlog}) and the presence of subleading logarithmic corrections. Thus, we expect the method for determining $\alpha$ presented here to be more reliable than directly simulating the extraordinary-log phase.
We also present results for the coefficients $a_\sigma$, $b_{\rm D}$ in Eq.~(\ref{eq:BOPE}) for the normal UC of the Ising model ($N = 1$) \cite {SM,SDIsingSpectrum,Hasenbusch-10b,CPRV-02,Bray,Chang,Damay,Damay98,VicariA1p}, extracted from MC studies in Refs.~\cite{PTD-10, Hasenbusch-10c}.
 A numerical conformal bootstrap study of the normal UC of the O($N$) model with $N \ge 2$ was conducted in parallel to our work \cite{Padayasi}. Our results for  $a_\sigma$ and $b_{\rm t}$ are within the bounds produced by positive bootstrap and agree reasonably well with the approximate truncated bootstrap results. For $N = 1$, both $a_\sigma$ and $b_{\rm D}$ in Table \ref{tbl:final} agree within error bars with the truncated bootstrap findings of Ref.~\cite{GLMR-15}.

We conclude by outlining some possible future directions. It will be interesting to extend the calculations presented here to the O($N$) model with $N > 3$, with an eye to determining the critical value $N_c$ where the extraordinary-log UC disappears. $N = 4$ is a natural first target since bootstrap calculations \cite{Padayasi}, as well as previous MC simulations \cite{DengO4}, suggest that the extraordinary transition still exists in this case. Another extension is to study the free energy density for the normal UC in the geometry considered here, which combined with the coefficient $B_\sigma$ in Eq.~(\ref{sigma_ansatz}) and $a_\sigma$, allows one to determine the OPE coefficient $b_{\rm D}$ in Eq.~(\ref{eq:BOPE}), as well as the universal coefficient $C_{\rm D}$, characterizing the boundary OPE of the energy-momentum tensor $T_{zz} \stackrel{z \to 0}{\to} -\sqrt{C_{\rm D}} {\rm D}$ \cite{Cardy-90}. (In fact, this is how for the Ising model $b_{\rm D}$ in Table \ref{tbl:final} was obtained \footnote{In principle, it is also possible to determine $b_{\rm D}$ by a method similar to the one we used to find $b_{\rm t}$: one extracts the normalization of ${\rm D}$ from the connected two-point function of $\sigma$ on the boundary, and the coefficient $b_{\rm D}$ from the connected bulk-boundary two point function of $\sigma$. However, we found that due to the large scaling dimension $\hat{\Delta}_{\rm D} = 3$, the boundary two-point function $\langle \sigma({\bf x})\sigma({\rm 0})\rangle_{{\rm conn}}$ falls off very fast, so that the normalization is difficult to extract reliably.}.)
Further interesting avenues for future research would be to consider the $N\rightarrow 0$ limit \cite{DBLW-90,DBL-93,ZLTDB-90,Eisenriegler-book,HG-94, Cardy1997}, which describes the physics of dilute polymers \cite{deGennes-book,Cardy-book,PV-02}, and O($N$) loop models \cite{LDGB-12,DMNS-81}, which provide an extension of the standard O($N$) model to noninteger values of $N$.

\begin{acknowledgments}
  We are very grateful to Marco Meineri for discussions. M.M. thanks Ilya Gruzberg, Abijith Krishnan,  Marco Meineri and Jay Padaysi for a collaboration on a related project.
  We thank Jian-Ping Lv for useful communications.
  F.P.T. is funded by the Deutsche Forschungsgemeinschaft (DFG, German Research Foundation)--Project No. 414456783.
  We gratefully acknowledge the Gauss Centre for Supercomputing e.V. for funding this project by providing computing time through the John von Neumann Institute for Computing (NIC) on the GCS Supercomputer JUWELS at Jülich Supercomputing Centre (JSC) \cite{JUWELS}. M.M. is supported by the National Science Foundation under Grant No. DMR-1847861.
\end{acknowledgments}

\bibliography{francesco,max,sm}

\begin{thebibliography}{98}%
\makeatletter
\providecommand \@ifxundefined [1]{%
 \@ifx{#1\undefined}
}%
\providecommand \@ifnum [1]{%
 \ifnum #1\expandafter \@firstoftwo
 \else \expandafter \@secondoftwo
 \fi
}%
\providecommand \@ifx [1]{%
 \ifx #1\expandafter \@firstoftwo
 \else \expandafter \@secondoftwo
 \fi
}%
\providecommand \natexlab [1]{#1}%
\providecommand \enquote  [1]{``#1''}%
\providecommand \bibnamefont  [1]{#1}%
\providecommand \bibfnamefont [1]{#1}%
\providecommand \citenamefont [1]{#1}%
\providecommand \href@noop [0]{\@secondoftwo}%
\providecommand \href [0]{\begingroup \@sanitize@url \@href}%
\providecommand \@href[1]{\@@startlink{#1}\@@href}%
\providecommand \@@href[1]{\endgroup#1\@@endlink}%
\providecommand \@sanitize@url [0]{\catcode `\\12\catcode `\$12\catcode
  `\&12\catcode `\#12\catcode `\^12\catcode `\_12\catcode `\%12\relax}%
\providecommand \@@startlink[1]{}%
\providecommand \@@endlink[0]{}%
\providecommand \url  [0]{\begingroup\@sanitize@url \@url }%
\providecommand \@url [1]{\endgroup\@href {#1}{\urlprefix }}%
\providecommand \urlprefix  [0]{URL }%
\providecommand \Eprint [0]{\href }%
\providecommand \doibase [0]{https://doi.org/}%
\providecommand \selectlanguage [0]{\@gobble}%
\providecommand \bibinfo  [0]{\@secondoftwo}%
\providecommand \bibfield  [0]{\@secondoftwo}%
\providecommand \translation [1]{[#1]}%
\providecommand \BibitemOpen [0]{}%
\providecommand \bibitemStop [0]{}%
\providecommand \bibitemNoStop [0]{.\EOS\space}%
\providecommand \EOS [0]{\spacefactor3000\relax}%
\providecommand \BibitemShut  [1]{\csname bibitem#1\endcsname}%
\let\auto@bib@innerbib\@empty
\bibitem [{\citenamefont {{Cardy}}(1996)}]{Cardy-book}%
  \BibitemOpen
  \bibfield  {author} {\bibinfo {author} {\bibfnamefont {J.}~\bibnamefont
  {{Cardy}}},\ }\href@noop {} {\emph {\bibinfo {title} {{Scaling and
  Renormalization in Statistical Physics}}}}\ (\bibinfo  {publisher} {Cambridge
  University Press},\ \bibinfo {address} {Cambridge},\ \bibinfo {year}
  {1996})\BibitemShut {NoStop}%
\bibitem [{\citenamefont {Dosch}(2006)}]{Dosch-book}%
  \BibitemOpen
  \bibfield  {author} {\bibinfo {author} {\bibfnamefont {H.}~\bibnamefont
  {Dosch}},\ }\href {https://books.google.de/books?id=yZl0DgAAQBAJ} {\emph
  {\bibinfo {title} {Critical Phenomena at Surfaces and Interfaces: Evanescent
  X-Ray and Neutron Scattering}}},\ Springer Tracts in Modern Physics\
  (\bibinfo  {publisher} {Springer Berlin Heidelberg},\ \bibinfo {address}
  {Berlin},\ \bibinfo {year} {2006})\BibitemShut {NoStop}%
\bibitem [{\citenamefont {{Binder}}(1983)}]{Binder-83}%
  \BibitemOpen
  \bibfield  {author} {\bibinfo {author} {\bibfnamefont {K.}~\bibnamefont
  {{Binder}}},\ }\bibfield  {title} {\bibinfo {title} {Critical behavior at
  surfaces},\ }in\ \href@noop {} {\emph {\bibinfo {booktitle} {Phase
  Transitions and Critical Phenomena}}},\ Vol.~\bibinfo {volume} {8},\ \bibinfo
  {editor} {edited by\ \bibinfo {editor} {\bibfnamefont {C.}~\bibnamefont
  {{Domb}}}\ and\ \bibinfo {editor} {\bibfnamefont {J.~L.}\ \bibnamefont
  {{Lebowitz}}}}\ (\bibinfo  {publisher} {Academic Press},\ \bibinfo {address}
  {London},\ \bibinfo {year} {1983})\ p.~\bibinfo {pages} {1}\BibitemShut
  {NoStop}%
\bibitem [{\citenamefont {{Diehl}}(1986)}]{Diehl-86}%
  \BibitemOpen
  \bibfield  {author} {\bibinfo {author} {\bibfnamefont {H.~W.}\ \bibnamefont
  {{Diehl}}},\ }\bibfield  {title} {\bibinfo {title} {Field-theoretical
  approach to critical behaviour at surfaces},\ }in\ \href@noop {} {\emph
  {\bibinfo {booktitle} {Phase Transitions and Critical Phenomena}}},\
  Vol.~\bibinfo {volume} {10},\ \bibinfo {editor} {edited by\ \bibinfo {editor}
  {\bibfnamefont {C.}~\bibnamefont {{Domb}}}\ and\ \bibinfo {editor}
  {\bibfnamefont {J.~L.}\ \bibnamefont {{Lebowitz}}}}\ (\bibinfo  {publisher}
  {Academic Press},\ \bibinfo {address} {London},\ \bibinfo {year} {1986})\
  p.~\bibinfo {pages} {75}\BibitemShut {NoStop}%
\bibitem [{\citenamefont {{Pleimling}}(2004)}]{Pleimling-review}%
  \BibitemOpen
  \bibfield  {author} {\bibinfo {author} {\bibfnamefont {M.}~\bibnamefont
  {{Pleimling}}},\ }\bibfield  {title} {\bibinfo {title} {{Critical phenomena
  at perfect and non-perfect surfaces}},\ }\href
  {https://doi.org/10.1088/0305-4470/37/19/R01} {\bibfield  {journal} {\bibinfo
   {journal} {\JPAOLD}\ }\textbf {\bibinfo {volume} {37}},\ \bibinfo {pages}
  {R79} (\bibinfo {year} {2004})},\ \Eprint
  {https://arxiv.org/abs/cond-mat/0402574} {cond-mat/0402574} \BibitemShut
  {NoStop}%
\bibitem [{\citenamefont {Fisher}\ and\ \citenamefont
  {de~Gennes}(1978)}]{FG-78}%
  \BibitemOpen
  \bibfield  {author} {\bibinfo {author} {\bibfnamefont {M.~E.}\ \bibnamefont
  {Fisher}}\ and\ \bibinfo {author} {\bibfnamefont {P.-G.}\ \bibnamefont
  {de~Gennes}},\ }\bibfield  {title} {\bibinfo {title} {{Ph\'enom\`enes aux
  parois dans un m\'elange binaire critique}},\ }\href
  {http://gallica.bnf.fr/ark:/12148/bpt6k62353730/f61.image.r=fisher%20de%20gennes.langEN}
  {\bibfield  {journal} {\bibinfo  {journal} {C.~R.~Acad.~Sci.~Paris Ser.~B~}\
  }\textbf {\bibinfo {volume} {287}},\ \bibinfo {pages} {207} (\bibinfo {year}
  {1978})}\BibitemShut {NoStop}%
\bibitem [{\citenamefont {Krech}(1994)}]{Krech-94}%
  \BibitemOpen
  \bibfield  {author} {\bibinfo {author} {\bibfnamefont {M.}~\bibnamefont
  {Krech}},\ }\href {http://books.google.de/books?id=0ZuCngEACAAJ} {\emph
  {\bibinfo {title} {The Casimir Effect in Critical Systems}}}\ (\bibinfo
  {publisher} {World Scientific},\ \bibinfo {address} {London},\ \bibinfo
  {year} {1994})\BibitemShut {NoStop}%
\bibitem [{\citenamefont {{Krech}}(1999)}]{Krech-99}%
  \BibitemOpen
  \bibfield  {author} {\bibinfo {author} {\bibfnamefont {M.}~\bibnamefont
  {{Krech}}},\ }\bibfield  {title} {\bibinfo {title} {{Fluctuation-induced
  forces in critical fluids}},\ }\href
  {https://doi.org/10.1088/0953-8984/11/37/201} {\bibfield  {journal} {\bibinfo
   {journal} {\JPCM}\ }\textbf {\bibinfo {volume} {11}},\ \bibinfo {pages}
  {R391} (\bibinfo {year} {1999})},\ \Eprint
  {https://arxiv.org/abs/cond-mat/9909413} {cond-mat/9909413} \BibitemShut
  {NoStop}%
\bibitem [{\citenamefont {Brankov}\ \emph {et~al.}(2000)\citenamefont
  {Brankov}, \citenamefont {Danchev},\ and\ \citenamefont {Tonchev}}]{BDT-00}%
  \BibitemOpen
  \bibfield  {author} {\bibinfo {author} {\bibfnamefont {{\u{I}}.}~\bibnamefont
  {Brankov}}, \bibinfo {author} {\bibfnamefont {D.}~\bibnamefont {Danchev}},\
  and\ \bibinfo {author} {\bibfnamefont {N.}~\bibnamefont {Tonchev}},\ }\href
  {http://books.google.de/books?id=khHxnQEACAAJ} {\emph {\bibinfo {title}
  {Theory of Critical Phenomena in Finite-size Systems: Scaling and Quantum
  Effects}}},\ Series in modern condensed matter physics\ (\bibinfo
  {publisher} {World Scientific},\ \bibinfo {address} {Singapore},\ \bibinfo
  {year} {2000})\BibitemShut {NoStop}%
\bibitem [{\citenamefont {{Gambassi}}(2009)}]{Gambassi-09}%
  \BibitemOpen
  \bibfield  {author} {\bibinfo {author} {\bibfnamefont {A.}~\bibnamefont
  {{Gambassi}}},\ }\bibfield  {title} {\bibinfo {title} {{The Casimir effect:
  From quantum to critical fluctuations}},\ }\href
  {https://doi.org/10.1088/1742-6596/161/1/012037} {\bibfield  {journal}
  {\bibinfo  {journal} {J. Phys.: Conf. Ser.}\ }\textbf {\bibinfo {volume}
  {161}},\ \bibinfo {eid} {012037} (\bibinfo {year} {2009})},\ \Eprint
  {https://arxiv.org/abs/0812.0935} {arXiv:0812.0935 [cond-mat.stat-mech]}
  \BibitemShut {NoStop}%
\bibitem [{\citenamefont {{Gambassi}}\ and\ \citenamefont
  {{Dietrich}}(2011)}]{GD-11}%
  \BibitemOpen
  \bibfield  {author} {\bibinfo {author} {\bibfnamefont {A.}~\bibnamefont
  {{Gambassi}}}\ and\ \bibinfo {author} {\bibfnamefont {S.}~\bibnamefont
  {{Dietrich}}},\ }\bibfield  {title} {\bibinfo {title} {{Critical Casimir
  forces steered by patterned substrates}},\ }\href
  {https://doi.org/10.1039/c0sm00635a} {\bibfield  {journal} {\bibinfo
  {journal} {Soft Matter}\ }\textbf {\bibinfo {volume} {7}},\ \bibinfo {pages}
  {1247} (\bibinfo {year} {2011})},\ \Eprint {https://arxiv.org/abs/1011.1831}
  {arXiv:1011.1831 [cond-mat.soft]} \BibitemShut {NoStop}%
\bibitem [{\citenamefont {Macio{\l}ek}\ and\ \citenamefont
  {Dietrich}(2018)}]{MD-18}%
  \BibitemOpen
  \bibfield  {author} {\bibinfo {author} {\bibfnamefont {A.}~\bibnamefont
  {Macio{\l}ek}}\ and\ \bibinfo {author} {\bibfnamefont {S.}~\bibnamefont
  {Dietrich}},\ }\bibfield  {title} {\bibinfo {title} {{Collective behavior of
  colloids due to critical Casimir interactions}},\ }\href
  {https://doi.org/10.1103/RevModPhys.90.045001} {\bibfield  {journal}
  {\bibinfo  {journal} {\RMP}\ }\textbf {\bibinfo {volume} {90}},\ \bibinfo
  {eid} {045001} (\bibinfo {year} {2018})},\ \Eprint
  {https://arxiv.org/abs/1712.06678} {arXiv:1712.06678 [cond-mat.soft]}
  \BibitemShut {NoStop}%
\bibitem [{\citenamefont {McAvity}\ and\ \citenamefont
  {Osborn}(1995)}]{McAvity:1995zd}%
  \BibitemOpen
  \bibfield  {author} {\bibinfo {author} {\bibfnamefont {D.}~\bibnamefont
  {McAvity}}\ and\ \bibinfo {author} {\bibfnamefont {H.}~\bibnamefont
  {Osborn}},\ }\bibfield  {title} {\bibinfo {title} {{Conformal field theories
  near a boundary in general dimensions}},\ }\href
  {https://doi.org/10.1016/0550-3213(95)00476-9} {\bibfield  {journal}
  {\bibinfo  {journal} {Nucl. Phys.}\ }\textbf {\bibinfo {volume} {B455}},\
  \bibinfo {pages} {522} (\bibinfo {year} {1995})},\ \Eprint
  {https://arxiv.org/abs/cond-mat/9505127} {arXiv:cond-mat/9505127 [cond-mat]}
  \BibitemShut {NoStop}%
\bibitem [{\citenamefont {{Liendo}}\ \emph {et~al.}(2013)\citenamefont
  {{Liendo}}, \citenamefont {{Rastelli}},\ and\ \citenamefont {{van
  Rees}}}]{Liendo2013}%
  \BibitemOpen
  \bibfield  {author} {\bibinfo {author} {\bibfnamefont {P.}~\bibnamefont
  {{Liendo}}}, \bibinfo {author} {\bibfnamefont {L.}~\bibnamefont
  {{Rastelli}}},\ and\ \bibinfo {author} {\bibfnamefont {B.~C.}\ \bibnamefont
  {{van Rees}}},\ }\bibfield  {title} {\bibinfo {title} {{The bootstrap program
  for boundary CFT$_{ d }$}},\ }\href {https://doi.org/10.1007/JHEP07(2013)113}
  {\bibfield  {journal} {\bibinfo  {journal} {JHEP}\ }\textbf {\bibinfo
  {volume} {07}}\bibfield  {number} {\bibinfo  {number} { (2013)},\ \bibinfo
  {eid} {113}},\ }\Eprint {https://arxiv.org/abs/1210.4258} {arXiv:1210.4258
  [hep-th]} \BibitemShut {NoStop}%
\bibitem [{\citenamefont {Gliozzi}\ \emph {et~al.}(2015)\citenamefont
  {Gliozzi}, \citenamefont {Liendo}, \citenamefont {Meineri},\ and\
  \citenamefont {Rago}}]{Gliozzi2015}%
  \BibitemOpen
  \bibfield  {author} {\bibinfo {author} {\bibfnamefont {F.}~\bibnamefont
  {Gliozzi}}, \bibinfo {author} {\bibfnamefont {P.}~\bibnamefont {Liendo}},
  \bibinfo {author} {\bibfnamefont {M.}~\bibnamefont {Meineri}},\ and\ \bibinfo
  {author} {\bibfnamefont {A.}~\bibnamefont {Rago}},\ }\bibfield  {title}
  {\bibinfo {title} {{Boundary and Interface CFTs from the Conformal
  Bootstrap}},\ }\href {https://doi.org/10.1007/JHEP05(2015)036} {\bibfield
  {journal} {\bibinfo  {journal} {JHEP}\ }\textbf {\bibinfo {volume}
  {05}}\bibfield  {number} {\bibinfo  {number} { (2015)},\ \bibinfo {pages}
  {036}},\ }\Eprint {https://arxiv.org/abs/1502.07217} {arXiv:1502.07217
  [hep-th]} \BibitemShut {NoStop}%
\bibitem [{\citenamefont {Bill\'{o}}\ \emph {et~al.}(2016)\citenamefont
  {Bill\'{o}}, \citenamefont {Gon\c{c}alves}, \citenamefont {Lauria},\ and\
  \citenamefont {Meineri}}]{Billo:2016cpy}%
  \BibitemOpen
  \bibfield  {author} {\bibinfo {author} {\bibfnamefont {M.}~\bibnamefont
  {Bill\'{o}}}, \bibinfo {author} {\bibfnamefont {V.}~\bibnamefont
  {Gon\c{c}alves}}, \bibinfo {author} {\bibfnamefont {E.}~\bibnamefont
  {Lauria}},\ and\ \bibinfo {author} {\bibfnamefont {M.}~\bibnamefont
  {Meineri}},\ }\bibfield  {title} {\bibinfo {title} {{Defects in conformal
  field theory}},\ }\href {https://doi.org/10.1007/JHEP04(2016)091} {\bibfield
  {journal} {\bibinfo  {journal} {JHEP}\ }\textbf {\bibinfo {volume}
  {04}}\bibfield  {number} {\bibinfo  {number} { (2016)},\ \bibinfo {pages}
  {091}},\ }\Eprint {https://arxiv.org/abs/1601.02883} {arXiv:1601.02883
  [hep-th]} \BibitemShut {NoStop}%
\bibitem [{\citenamefont {Liendo}\ and\ \citenamefont
  {Meneghelli}(2017)}]{Liendo:2016ymz}%
  \BibitemOpen
  \bibfield  {author} {\bibinfo {author} {\bibfnamefont {P.}~\bibnamefont
  {Liendo}}\ and\ \bibinfo {author} {\bibfnamefont {C.}~\bibnamefont
  {Meneghelli}},\ }\bibfield  {title} {\bibinfo {title} {{Bootstrap equations
  for $ \mathcal{N} $ = 4 SYM with defects}},\ }\href
  {https://doi.org/10.1007/JHEP01(2017)122} {\bibfield  {journal} {\bibinfo
  {journal} {JHEP}\ }\textbf {\bibinfo {volume} {01}}\bibfield  {number}
  {\bibinfo  {number} { (2017)},\ \bibinfo {pages} {122}},\ }\Eprint
  {https://arxiv.org/abs/1608.05126} {arXiv:1608.05126 [hep-th]} \BibitemShut
  {NoStop}%
\bibitem [{\citenamefont {Lauria}\ \emph {et~al.}(2018)\citenamefont {Lauria},
  \citenamefont {Meineri},\ and\ \citenamefont {Trevisani}}]{Lauria:2017wav}%
  \BibitemOpen
  \bibfield  {author} {\bibinfo {author} {\bibfnamefont {E.}~\bibnamefont
  {Lauria}}, \bibinfo {author} {\bibfnamefont {M.}~\bibnamefont {Meineri}},\
  and\ \bibinfo {author} {\bibfnamefont {E.}~\bibnamefont {Trevisani}},\
  }\bibfield  {title} {\bibinfo {title} {{Radial coordinates for defect
  CFTs}},\ }\href {https://doi.org/10.1007/JHEP11(2018)148} {\bibfield
  {journal} {\bibinfo  {journal} {JHEP}\ }\textbf {\bibinfo {volume}
  {11}}\bibfield  {number} {\bibinfo  {number} { (2018)},\ \bibinfo {pages}
  {148}},\ }\Eprint {https://arxiv.org/abs/1712.07668} {arXiv:1712.07668
  [hep-th]} \BibitemShut {NoStop}%
\bibitem [{\citenamefont {Maz\'a\v{c}}\ \emph {et~al.}(2019)\citenamefont
  {Maz\'a\v{c}}, \citenamefont {Rastelli},\ and\ \citenamefont
  {Zhou}}]{Mazac:2018biw}%
  \BibitemOpen
  \bibfield  {author} {\bibinfo {author} {\bibfnamefont {D.}~\bibnamefont
  {Maz\'a\v{c}}}, \bibinfo {author} {\bibfnamefont {L.}~\bibnamefont
  {Rastelli}},\ and\ \bibinfo {author} {\bibfnamefont {X.}~\bibnamefont
  {Zhou}},\ }\bibfield  {title} {\bibinfo {title} {{An analytic approach to
  BCFT$_{d}$}},\ }\href {https://doi.org/10.1007/JHEP12(2019)004} {\bibfield
  {journal} {\bibinfo  {journal} {JHEP}\ }\textbf {\bibinfo {volume}
  {12}}\bibfield  {number} {\bibinfo  {number} { (2019)},\ \bibinfo {pages}
  {004}},\ }\Eprint {https://arxiv.org/abs/1812.09314} {arXiv:1812.09314
  [hep-th]} \BibitemShut {NoStop}%
\bibitem [{\citenamefont {Kaviraj}\ and\ \citenamefont
  {Paulos}(2020)}]{Kaviraj:2018tfd}%
  \BibitemOpen
  \bibfield  {author} {\bibinfo {author} {\bibfnamefont {A.}~\bibnamefont
  {Kaviraj}}\ and\ \bibinfo {author} {\bibfnamefont {M.~F.}\ \bibnamefont
  {Paulos}},\ }\bibfield  {title} {\bibinfo {title} {{The functional bootstrap
  for boundary CFT}},\ }\href {https://doi.org/10.1007/JHEP04(2020)135}
  {\bibfield  {journal} {\bibinfo  {journal} {JHEP}\ }\textbf {\bibinfo
  {volume} {04}}\bibfield  {number} {\bibinfo  {number} { (2020)},\ \bibinfo
  {pages} {135}},\ }\Eprint {https://arxiv.org/abs/1812.04034}
  {arXiv:1812.04034 [hep-th]} \BibitemShut {NoStop}%
\bibitem [{\citenamefont {Dey}\ \emph {et~al.}(2020)\citenamefont {Dey},
  \citenamefont {Hansen},\ and\ \citenamefont {Shpot}}]{Dey:2020lwp}%
  \BibitemOpen
  \bibfield  {author} {\bibinfo {author} {\bibfnamefont {P.}~\bibnamefont
  {Dey}}, \bibinfo {author} {\bibfnamefont {T.}~\bibnamefont {Hansen}},\ and\
  \bibinfo {author} {\bibfnamefont {M.}~\bibnamefont {Shpot}},\ }\bibfield
  {title} {\bibinfo {title} {{Operator expansions, layer susceptibility and
  two-point functions in BCFT}},\ }\href
  {https://doi.org/10.1007/JHEP12(2020)051} {\bibfield  {journal} {\bibinfo
  {journal} {JHEP}\ }\textbf {\bibinfo {volume} {12}}\bibfield  {number}
  {\bibinfo  {number} { (2020)},\ \bibinfo {pages} {051}},\ }\Eprint
  {https://arxiv.org/abs/2006.11253} {arXiv:2006.11253 [hep-th]} \BibitemShut
  {NoStop}%
\bibitem [{\citenamefont {Behan}\ \emph {et~al.}(2020)\citenamefont {Behan},
  \citenamefont {Di~Pietro}, \citenamefont {Lauria},\ and\ \citenamefont
  {Van~Rees}}]{Behan:2020nsf}%
  \BibitemOpen
  \bibfield  {author} {\bibinfo {author} {\bibfnamefont {C.}~\bibnamefont
  {Behan}}, \bibinfo {author} {\bibfnamefont {L.}~\bibnamefont {Di~Pietro}},
  \bibinfo {author} {\bibfnamefont {E.}~\bibnamefont {Lauria}},\ and\ \bibinfo
  {author} {\bibfnamefont {B.~C.}\ \bibnamefont {Van~Rees}},\ }\bibfield
  {title} {\bibinfo {title} {{Bootstrapping boundary-localized interactions}},\
  }\href {https://doi.org/10.1007/JHEP12(2020)182} {\bibfield  {journal}
  {\bibinfo  {journal} {JHEP}\ }\textbf {\bibinfo {volume} {12}}\bibfield
  {number} {\bibinfo  {number} { (2020)},\ \bibinfo {pages} {182}},\ }\Eprint
  {https://arxiv.org/abs/2009.03336} {arXiv:2009.03336 [hep-th]} \BibitemShut
  {NoStop}%
\bibitem [{\citenamefont {Gimenez-Grau}\ \emph {et~al.}(2021)\citenamefont
  {Gimenez-Grau}, \citenamefont {Liendo},\ and\ \citenamefont {van
  Vliet}}]{Gimenez-Grau:2020jvf}%
  \BibitemOpen
  \bibfield  {author} {\bibinfo {author} {\bibfnamefont {A.}~\bibnamefont
  {Gimenez-Grau}}, \bibinfo {author} {\bibfnamefont {P.}~\bibnamefont
  {Liendo}},\ and\ \bibinfo {author} {\bibfnamefont {P.}~\bibnamefont {van
  Vliet}},\ }\bibfield  {title} {\bibinfo {title} {{Superconformal boundaries
  in $4-\epsilon$ dimensions}},\ }\href
  {https://doi.org/10.1007/JHEP04(2021)167} {\bibfield  {journal} {\bibinfo
  {journal} {JHEP}\ }\textbf {\bibinfo {volume} {04}}\bibfield  {number}
  {\bibinfo  {number} { (2021)},\ \bibinfo {pages} {167}},\ }\Eprint
  {https://arxiv.org/abs/2012.00018} {arXiv:2012.00018 [hep-th]} \BibitemShut
  {NoStop}%
\bibitem [{\citenamefont {Grover}\ and\ \citenamefont
  {Vishwanath}()}]{TarungaplessSPT}%
  \BibitemOpen
  \bibfield  {author} {\bibinfo {author} {\bibfnamefont {T.}~\bibnamefont
  {Grover}}\ and\ \bibinfo {author} {\bibfnamefont {A.}~\bibnamefont
  {Vishwanath}},\ }\bibfield  {title} {\bibinfo {title} {Quantum criticality in
  topological insulators and superconductors: Emergence of strongly coupled
  majoranas and supersymmetry},\ }\href@noop {} {\bibfield  {journal} {\bibinfo
   {journal} {{}}\ }}\Eprint {https://arxiv.org/abs/1206.1332} {arXiv:1206.1332
  [cond-mat.str-el]} \BibitemShut {NoStop}%
\bibitem [{\citenamefont {Barkeshli}\ \emph {et~al.}(2015)\citenamefont
  {Barkeshli}, \citenamefont {Mulligan},\ and\ \citenamefont
  {Fisher}}]{MaissamCFL}%
  \BibitemOpen
  \bibfield  {author} {\bibinfo {author} {\bibfnamefont {M.}~\bibnamefont
  {Barkeshli}}, \bibinfo {author} {\bibfnamefont {M.}~\bibnamefont
  {Mulligan}},\ and\ \bibinfo {author} {\bibfnamefont {M.~P.~A.}\ \bibnamefont
  {Fisher}},\ }\bibfield  {title} {\bibinfo {title} {Particle-hole symmetry and
  the composite fermi liquid},\ }\href
  {https://doi.org/10.1103/physrevb.92.165125} {\bibfield  {journal} {\bibinfo
  {journal} {\prb}\ }\textbf {\bibinfo {volume} {92}},\ \bibinfo {pages}
  {165125} (\bibinfo {year} {2015})},\ \Eprint
  {https://arxiv.org/abs/1502.05404} {arXiv:1502.05404 [cond-mat.str-el]}
  \BibitemShut {NoStop}%
\bibitem [{\citenamefont {Barkeshli}\ and\ \citenamefont
  {Qi}(2014)}]{MaissamGenon}%
  \BibitemOpen
  \bibfield  {author} {\bibinfo {author} {\bibfnamefont {M.}~\bibnamefont
  {Barkeshli}}\ and\ \bibinfo {author} {\bibfnamefont {X.-L.}\ \bibnamefont
  {Qi}},\ }\bibfield  {title} {\bibinfo {title} {Synthetic topological qubits
  in conventional bilayer quantum hall systems},\ }\href
  {https://doi.org/10.1103/physrevx.4.041035} {\bibfield  {journal} {\bibinfo
  {journal} {\PRX}\ }\textbf {\bibinfo {volume} {4}},\ \bibinfo {pages}
  {041035} (\bibinfo {year} {2014})},\ \Eprint
  {https://arxiv.org/abs/1302.2673} {arXiv:1302.2673 [cond-mat.mes-hall]}
  \BibitemShut {NoStop}%
\bibitem [{\citenamefont {Cano}\ \emph {et~al.}(2015)\citenamefont {Cano},
  \citenamefont {Cheng}, \citenamefont {Barkeshli}, \citenamefont {Clarke},\
  and\ \citenamefont {Nayak}}]{MaissamMajorana}%
  \BibitemOpen
  \bibfield  {author} {\bibinfo {author} {\bibfnamefont {J.}~\bibnamefont
  {Cano}}, \bibinfo {author} {\bibfnamefont {M.}~\bibnamefont {Cheng}},
  \bibinfo {author} {\bibfnamefont {M.}~\bibnamefont {Barkeshli}}, \bibinfo
  {author} {\bibfnamefont {D.~J.}\ \bibnamefont {Clarke}},\ and\ \bibinfo
  {author} {\bibfnamefont {C.}~\bibnamefont {Nayak}},\ }\bibfield  {title}
  {\bibinfo {title} {Chirality-protected majorana zero modes at the gapless
  edge of abelian quantum hall states},\ }\href
  {https://doi.org/10.1103/physrevb.92.195152} {\bibfield  {journal} {\bibinfo
  {journal} {\prb}\ }\textbf {\bibinfo {volume} {92}},\ \bibinfo {pages}
  {195152} (\bibinfo {year} {2015})},\ \Eprint
  {https://arxiv.org/abs/1505.07825} {arXiv:1505.07825 [cond-mat.str-el]}
  \BibitemShut {NoStop}%
\bibitem [{\citenamefont {Scaffidi}\ \emph {et~al.}(2017)\citenamefont
  {Scaffidi}, \citenamefont {Parker},\ and\ \citenamefont
  {Vasseur}}]{ScaffidiGapless}%
  \BibitemOpen
  \bibfield  {author} {\bibinfo {author} {\bibfnamefont {T.}~\bibnamefont
  {Scaffidi}}, \bibinfo {author} {\bibfnamefont {D.~E.}\ \bibnamefont
  {Parker}},\ and\ \bibinfo {author} {\bibfnamefont {R.}~\bibnamefont
  {Vasseur}},\ }\bibfield  {title} {\bibinfo {title} {Gapless
  symmetry-protected topological order},\ }\href
  {https://doi.org/10.1103/PhysRevX.7.041048} {\bibfield  {journal} {\bibinfo
  {journal} {Phys. Rev. X}\ }\textbf {\bibinfo {volume} {7}},\ \bibinfo {pages}
  {041048} (\bibinfo {year} {2017})},\ \Eprint
  {https://arxiv.org/abs/1705.01557} {arXiv:1705.01557 [cond-mat.str-el]}
  \BibitemShut {NoStop}%
\bibitem [{\citenamefont {Parker}\ \emph {et~al.}(2018)\citenamefont {Parker},
  \citenamefont {Scaffidi},\ and\ \citenamefont {Vasseur}}]{ParkerGapless}%
  \BibitemOpen
  \bibfield  {author} {\bibinfo {author} {\bibfnamefont {D.~E.}\ \bibnamefont
  {Parker}}, \bibinfo {author} {\bibfnamefont {T.}~\bibnamefont {Scaffidi}},\
  and\ \bibinfo {author} {\bibfnamefont {R.}~\bibnamefont {Vasseur}},\
  }\bibfield  {title} {\bibinfo {title} {Topological luttinger liquids from
  decorated domain walls},\ }\href {https://doi.org/10.1103/physrevb.97.165114}
  {\bibfield  {journal} {\bibinfo  {journal} {\prb}\ }\textbf {\bibinfo
  {volume} {97}},\ \bibinfo {pages} {165114} (\bibinfo {year} {2018})},\
  \Eprint {https://arxiv.org/abs/1711.09106} {arXiv:1711.09106
  [cond-mat.str-el]} \BibitemShut {NoStop}%
\bibitem [{\citenamefont {Verresen}\ \emph {et~al.}(2021)\citenamefont
  {Verresen}, \citenamefont {Thorngren}, \citenamefont {Jones},\ and\
  \citenamefont {Pollmann}}]{RyangaplessSPT}%
  \BibitemOpen
  \bibfield  {author} {\bibinfo {author} {\bibfnamefont {R.}~\bibnamefont
  {Verresen}}, \bibinfo {author} {\bibfnamefont {R.}~\bibnamefont {Thorngren}},
  \bibinfo {author} {\bibfnamefont {N.~G.}\ \bibnamefont {Jones}},\ and\
  \bibinfo {author} {\bibfnamefont {F.}~\bibnamefont {Pollmann}},\ }\bibfield
  {title} {\bibinfo {title} {Gapless topological phases and symmetry-enriched
  quantum criticality},\ }\href {https://doi.org/10.1103/PhysRevX.11.041059}
  {\bibfield  {journal} {\bibinfo  {journal} {Phys. Rev. X}\ }\textbf {\bibinfo
  {volume} {11}},\ \bibinfo {pages} {041059} (\bibinfo {year} {2021})},\
  \Eprint {https://arxiv.org/abs/1905.06969} {arXiv:1905.06969
  [cond-mat.str-el]} \BibitemShut {NoStop}%
\bibitem [{\citenamefont {Verresen}()}]{RubengaplessSPT}%
  \BibitemOpen
  \bibfield  {author} {\bibinfo {author} {\bibfnamefont {R.}~\bibnamefont
  {Verresen}},\ }\bibfield  {title} {\bibinfo {title} {Topology and edge states
  survive quantum criticality between topological insulators},\ }\href@noop {}
  {\bibfield  {journal} {\bibinfo  {journal} {{}}\ }}\Eprint
  {https://arxiv.org/abs/2003.05453} {arXiv:2003.05453 [cond-mat.str-el]}
  \BibitemShut {NoStop}%
\bibitem [{\citenamefont {Thorngren}\ \emph {et~al.}(2021)\citenamefont
  {Thorngren}, \citenamefont {Vishwanath},\ and\ \citenamefont
  {Verresen}}]{RubenIntGapless}%
  \BibitemOpen
  \bibfield  {author} {\bibinfo {author} {\bibfnamefont {R.}~\bibnamefont
  {Thorngren}}, \bibinfo {author} {\bibfnamefont {A.}~\bibnamefont
  {Vishwanath}},\ and\ \bibinfo {author} {\bibfnamefont {R.}~\bibnamefont
  {Verresen}},\ }\bibfield  {title} {\bibinfo {title} {Intrinsically gapless
  topological phases},\ }\href {https://doi.org/10.1103/PhysRevB.104.075132}
  {\bibfield  {journal} {\bibinfo  {journal} {Phys. Rev. B}\ }\textbf {\bibinfo
  {volume} {104}},\ \bibinfo {pages} {075132} (\bibinfo {year} {2021})},\
  \Eprint {https://arxiv.org/abs/2008.06638} {arXiv:2008.06638
  [cond-mat.str-el]} \BibitemShut {NoStop}%
\bibitem [{\citenamefont {{Suzuki}}\ and\ \citenamefont
  {{Sato}}(2012)}]{SS-12}%
  \BibitemOpen
  \bibfield  {author} {\bibinfo {author} {\bibfnamefont {T.}~\bibnamefont
  {{Suzuki}}}\ and\ \bibinfo {author} {\bibfnamefont {M.}~\bibnamefont
  {{Sato}}},\ }\bibfield  {title} {\bibinfo {title} {{Gapless edge states and
  their stability in two-dimensional quantum magnets}},\ }\href
  {https://doi.org/10.1103/PhysRevB.86.224411} {\bibfield  {journal} {\bibinfo
  {journal} {\prb}\ }\textbf {\bibinfo {volume} {86}},\ \bibinfo {eid} {224411}
  (\bibinfo {year} {2012})},\ \Eprint {https://arxiv.org/abs/1209.3097}
  {arXiv:1209.3097 [cond-mat.str-el]} \BibitemShut {NoStop}%
\bibitem [{\citenamefont {Zhang}\ and\ \citenamefont {Wang}(2017)}]{ZW-17}%
  \BibitemOpen
  \bibfield  {author} {\bibinfo {author} {\bibfnamefont {L.}~\bibnamefont
  {Zhang}}\ and\ \bibinfo {author} {\bibfnamefont {F.}~\bibnamefont {Wang}},\
  }\bibfield  {title} {\bibinfo {title} {{Unconventional Surface Critical
  Behavior Induced by a Quantum Phase Transition from the Two-Dimensional
  Affleck-Kennedy-Lieb-Tasaki Phase to a N{\'e}el-Ordered Phase}},\ }\href
  {https://doi.org/10.1103/PhysRevLett.118.087201} {\bibfield  {journal}
  {\bibinfo  {journal} {\PRL}\ }\textbf {\bibinfo {volume} {118}},\ \bibinfo
  {eid} {087201} (\bibinfo {year} {2017})},\ \Eprint
  {https://arxiv.org/abs/1611.06477} {arXiv:1611.06477 [cond-mat.str-el]}
  \BibitemShut {NoStop}%
\bibitem [{\citenamefont {Ding}\ \emph {et~al.}(2018)\citenamefont {Ding},
  \citenamefont {Zhang},\ and\ \citenamefont {Guo}}]{DZG-18}%
  \BibitemOpen
  \bibfield  {author} {\bibinfo {author} {\bibfnamefont {C.}~\bibnamefont
  {Ding}}, \bibinfo {author} {\bibfnamefont {L.}~\bibnamefont {Zhang}},\ and\
  \bibinfo {author} {\bibfnamefont {W.}~\bibnamefont {Guo}},\ }\bibfield
  {title} {\bibinfo {title} {{Engineering Surface Critical Behavior of $(2 +1
  )$-Dimensional O(3) Quantum Critical Points}},\ }\href
  {https://doi.org/10.1103/PhysRevLett.120.235701} {\bibfield  {journal}
  {\bibinfo  {journal} {\PRL}\ }\textbf {\bibinfo {volume} {120}},\ \bibinfo
  {eid} {235701} (\bibinfo {year} {2018})},\ \Eprint
  {https://arxiv.org/abs/1801.10035} {arXiv:1801.10035 [cond-mat.str-el]}
  \BibitemShut {NoStop}%
\bibitem [{\citenamefont {Weber}\ \emph {et~al.}(2018)\citenamefont {Weber},
  \citenamefont {Parisen~Toldin},\ and\ \citenamefont {Wessel}}]{WPTW-18}%
  \BibitemOpen
  \bibfield  {author} {\bibinfo {author} {\bibfnamefont {L.}~\bibnamefont
  {Weber}}, \bibinfo {author} {\bibfnamefont {F.}~\bibnamefont
  {Parisen~Toldin}},\ and\ \bibinfo {author} {\bibfnamefont {S.}~\bibnamefont
  {Wessel}},\ }\bibfield  {title} {\bibinfo {title} {Nonordinary edge
  criticality of two-dimensional quantum critical magnets},\ }\href
  {https://doi.org/10.1103/PhysRevB.98.140403} {\bibfield  {journal} {\bibinfo
  {journal} {\prb}\ }\textbf {\bibinfo {volume} {98}},\ \bibinfo {eid}
  {140403(R)} (\bibinfo {year} {2018})},\ \Eprint
  {https://arxiv.org/abs/1804.06820} {arXiv:1804.06820 [cond-mat.str-el]}
  \BibitemShut {NoStop}%
\bibitem [{\citenamefont {Weber}\ and\ \citenamefont {Wessel}(2019)}]{WW-19}%
  \BibitemOpen
  \bibfield  {author} {\bibinfo {author} {\bibfnamefont {L.}~\bibnamefont
  {Weber}}\ and\ \bibinfo {author} {\bibfnamefont {S.}~\bibnamefont {Wessel}},\
  }\bibfield  {title} {\bibinfo {title} {Nonordinary criticality at the edges
  of planar spin-1 heisenberg antiferromagnets},\ }\href
  {https://doi.org/10.1103/PhysRevB.100.054437} {\bibfield  {journal} {\bibinfo
   {journal} {\prb}\ }\textbf {\bibinfo {volume} {100}},\ \bibinfo {eid}
  {054437} (\bibinfo {year} {2019})},\ \Eprint
  {https://arxiv.org/abs/1906.07051} {arXiv:1906.07051 [cond-mat.str-el]}
  \BibitemShut {NoStop}%
\bibitem [{\citenamefont {Jian}\ \emph {et~al.}(2021)\citenamefont {Jian},
  \citenamefont {Xu}, \citenamefont {Wu},\ and\ \citenamefont {Xu}}]{JXWX-20}%
  \BibitemOpen
  \bibfield  {author} {\bibinfo {author} {\bibfnamefont {C.-M.}\ \bibnamefont
  {Jian}}, \bibinfo {author} {\bibfnamefont {Y.}~\bibnamefont {Xu}}, \bibinfo
  {author} {\bibfnamefont {X.-C.}\ \bibnamefont {Wu}},\ and\ \bibinfo {author}
  {\bibfnamefont {C.}~\bibnamefont {Xu}},\ }\bibfield  {title} {\bibinfo
  {title} {{Continuous N{\'e}el-VBS quantum phase transition in non-local
  one-dimensional systems with SO(3) symmetry}},\ }\href
  {https://doi.org/10.21468/SciPostPhys.10.2.033} {\bibfield  {journal}
  {\bibinfo  {journal} {SciPost Phys.}\ }\textbf {\bibinfo {volume} {10}},\
  \bibinfo {eid} {033} (\bibinfo {year} {2021})},\ \Eprint
  {https://arxiv.org/abs/2004.07852} {arXiv:2004.07852 [cond-mat.str-el]}
  \BibitemShut {NoStop}%
\bibitem [{\citenamefont {Zhu}\ \emph {et~al.}(2021)\citenamefont {Zhu},
  \citenamefont {Ding}, \citenamefont {Zhang},\ and\ \citenamefont
  {Guo}}]{ZDZG-20}%
  \BibitemOpen
  \bibfield  {author} {\bibinfo {author} {\bibfnamefont {W.}~\bibnamefont
  {Zhu}}, \bibinfo {author} {\bibfnamefont {C.}~\bibnamefont {Ding}}, \bibinfo
  {author} {\bibfnamefont {L.}~\bibnamefont {Zhang}},\ and\ \bibinfo {author}
  {\bibfnamefont {W.}~\bibnamefont {Guo}},\ }\bibfield  {title} {\bibinfo
  {title} {{Surface critical behavior of coupled Haldane chains}},\ }\href
  {https://doi.org/10.1103/PhysRevB.103.024412} {\bibfield  {journal} {\bibinfo
   {journal} {\prb}\ }\textbf {\bibinfo {volume} {103}},\ \bibinfo {eid}
  {024412} (\bibinfo {year} {2021})},\ \Eprint
  {https://arxiv.org/abs/2010.10920} {arXiv:2010.10920 [cond-mat.str-el]}
  \BibitemShut {NoStop}%
\bibitem [{\citenamefont {Weber}\ and\ \citenamefont {Wessel}(2021)}]{WW-20}%
  \BibitemOpen
  \bibfield  {author} {\bibinfo {author} {\bibfnamefont {L.}~\bibnamefont
  {Weber}}\ and\ \bibinfo {author} {\bibfnamefont {S.}~\bibnamefont {Wessel}},\
  }\bibfield  {title} {\bibinfo {title} {{Spin versus bond correlations along
  dangling edges of quantum critical magnets}},\ }\href
  {https://doi.org/10.1103/PhysRevB.103.L020406} {\bibfield  {journal}
  {\bibinfo  {journal} {\PRB}\ }\textbf {\bibinfo {volume} {103}},\ \bibinfo
  {pages} {L020406} (\bibinfo {year} {2021})},\ \Eprint
  {https://arxiv.org/abs/2010.15691} {arXiv:2010.15691 [cond-mat.str-el]}
  \BibitemShut {NoStop}%
\bibitem [{\citenamefont {Ding}\ \emph {et~al.}(2021)\citenamefont {Ding},
  \citenamefont {Zhu}, \citenamefont {Guo},\ and\ \citenamefont
  {Zhang}}]{DZGZ-21}%
  \BibitemOpen
  \bibfield  {author} {\bibinfo {author} {\bibfnamefont {C.}~\bibnamefont
  {Ding}}, \bibinfo {author} {\bibfnamefont {W.}~\bibnamefont {Zhu}}, \bibinfo
  {author} {\bibfnamefont {W.}~\bibnamefont {Guo}},\ and\ \bibinfo {author}
  {\bibfnamefont {L.}~\bibnamefont {Zhang}},\ }\bibfield  {title} {\bibinfo
  {title} {{Special Transition and Extraordinary Phase on the Surface of a
  (2+1)-Dimensional Quantum Heisenberg Antiferromagnet}},\ }\href
  {https://ui.adsabs.harvard.edu/abs/2021arXiv211004762D} {\bibfield  {journal}
  {\bibinfo  {journal} {{}}\ ,\ \bibinfo {pages} {arXiv:2110.04762}} (\bibinfo
  {year} {2021})},\ \Eprint {https://arxiv.org/abs/2110.04762}
  {arXiv:2110.04762 [cond-mat.str-el]} \BibitemShut {NoStop}%
\bibitem [{\citenamefont {Metlitski}(2022)}]{Metlitski-20}%
  \BibitemOpen
  \bibfield  {author} {\bibinfo {author} {\bibfnamefont {M.~A.}\ \bibnamefont
  {Metlitski}},\ }\bibfield  {title} {\bibinfo {title} {{Boundary criticality
  of the O(N) model in d = 3 critically revisited}},\ }\href
  {https://doi.org/10.21468/SciPostPhys.12.4.131} {\bibfield  {journal}
  {\bibinfo  {journal} {Scipost}\ }\textbf {\bibinfo {volume} {12}},\ \bibinfo
  {pages} {131} (\bibinfo {year} {2022})},\ \Eprint
  {https://arxiv.org/abs/2009.05119} {arXiv:2009.05119 [cond-mat.str-el]}
  \BibitemShut {NoStop}%
\bibitem [{\citenamefont {Parisen~Toldin}(2021)}]{PT-20}%
  \BibitemOpen
  \bibfield  {author} {\bibinfo {author} {\bibfnamefont {F.}~\bibnamefont
  {Parisen~Toldin}},\ }\bibfield  {title} {\bibinfo {title} {{Boundary critical
  behavior of the three-dimensional Heisenberg universality class}},\ }\href
  {https://doi.org/10.1103/PhysRevLett.126.135701} {\bibfield  {journal}
  {\bibinfo  {journal} {\prl}\ }\textbf {\bibinfo {volume} {126}},\ \bibinfo
  {pages} {135701} (\bibinfo {year} {2021})},\ \Eprint
  {https://arxiv.org/abs/2012.00039} {arXiv:2012.00039 [cond-mat.stat-mech]}
  \BibitemShut {NoStop}%
\bibitem [{\citenamefont {Hu}\ \emph {et~al.}(2021)\citenamefont {Hu},
  \citenamefont {Deng},\ and\ \citenamefont {Lv}}]{HDL-21}%
  \BibitemOpen
  \bibfield  {author} {\bibinfo {author} {\bibfnamefont {M.}~\bibnamefont
  {Hu}}, \bibinfo {author} {\bibfnamefont {Y.}~\bibnamefont {Deng}},\ and\
  \bibinfo {author} {\bibfnamefont {J.-P.}\ \bibnamefont {Lv}},\ }\bibfield
  {title} {\bibinfo {title} {{Extraordinary-Log Surface Phase Transition in the
  Three-Dimensional X Y Model}},\ }\href
  {https://doi.org/10.1103/PhysRevLett.127.120603} {\bibfield  {journal}
  {\bibinfo  {journal} {\prl}\ }\textbf {\bibinfo {volume} {127}},\ \bibinfo
  {eid} {120603} (\bibinfo {year} {2021})},\ \Eprint
  {https://arxiv.org/abs/2104.05152} {arXiv:2104.05152 [cond-mat.stat-mech]}
  \BibitemShut {NoStop}%
\bibitem [{\citenamefont {{Pelissetto}}\ and\ \citenamefont
  {{Vicari}}(2002)}]{PV-02}%
  \BibitemOpen
  \bibfield  {author} {\bibinfo {author} {\bibfnamefont {A.}~\bibnamefont
  {{Pelissetto}}}\ and\ \bibinfo {author} {\bibfnamefont {E.}~\bibnamefont
  {{Vicari}}},\ }\bibfield  {title} {\bibinfo {title} {{Critical phenomena and
  renormalization-group theory}},\ }\href
  {https://doi.org/10.1016/S0370-1573(02)00219-3} {\bibfield  {journal}
  {\bibinfo  {journal} {\physrep}\ }\textbf {\bibinfo {volume} {368}},\
  \bibinfo {pages} {549} (\bibinfo {year} {2002})},\ \Eprint
  {https://arxiv.org/abs/cond-mat/0012164} {cond-mat/0012164} \BibitemShut
  {NoStop}%
\bibitem [{\citenamefont {Krech}(2000)}]{Krech-00}%
  \BibitemOpen
  \bibfield  {author} {\bibinfo {author} {\bibfnamefont {M.}~\bibnamefont
  {Krech}},\ }\bibfield  {title} {\bibinfo {title} {Surface scaling behavior of
  isotropic heisenberg systems: Critical exponents, structure factor, and
  profiles},\ }\href {https://doi.org/10.1103/PhysRevB.62.6360} {\bibfield
  {journal} {\bibinfo  {journal} {\prb}\ }\textbf {\bibinfo {volume} {62}},\
  \bibinfo {pages} {6360} (\bibinfo {year} {2000})},\ \Eprint
  {https://arxiv.org/abs/cond-mat/0006448} {arXiv:cond-mat/0006448
  [cond-mat.stat-mech]} \BibitemShut {NoStop}%
\bibitem [{\citenamefont {{Deng}}\ \emph {et~al.}(2005)\citenamefont {{Deng}},
  \citenamefont {{Bl{\"o}te}},\ and\ \citenamefont {{Nightingale}}}]{DBN-05}%
  \BibitemOpen
  \bibfield  {author} {\bibinfo {author} {\bibfnamefont {Y.}~\bibnamefont
  {{Deng}}}, \bibinfo {author} {\bibfnamefont {H.~W.~J.}\ \bibnamefont
  {{Bl{\"o}te}}},\ and\ \bibinfo {author} {\bibfnamefont {M.~P.}\ \bibnamefont
  {{Nightingale}}},\ }\bibfield  {title} {\bibinfo {title} {{Surface and bulk
  transitions in three-dimensional O(n) models}},\ }\href
  {https://doi.org/10.1103/PhysRevE.72.016128} {\bibfield  {journal} {\bibinfo
  {journal} {\pre}\ }\textbf {\bibinfo {volume} {72}},\ \bibinfo {eid} {016128}
  (\bibinfo {year} {2005})},\ \Eprint {https://arxiv.org/abs/cond-mat/0504173}
  {cond-mat/0504173 [cond-mat.stat-mech]} \BibitemShut {NoStop}%
\bibitem [{Note1()}]{Note1}%
  \BibitemOpen
  \bibinfo {note} {Earlier a hint of the special transition for $N = 3$ was
  seen in Ref.~\cite {DBN-05}}\BibitemShut {NoStop}%
\bibitem [{\citenamefont {{Bray}}\ and\ \citenamefont {{Moore}}(1977)}]{BM-77}%
  \BibitemOpen
  \bibfield  {author} {\bibinfo {author} {\bibfnamefont {A.~J.}\ \bibnamefont
  {{Bray}}}\ and\ \bibinfo {author} {\bibfnamefont {M.~A.}\ \bibnamefont
  {{Moore}}},\ }\bibfield  {title} {\bibinfo {title} {{Critical behaviour of
  semi-infinite systems}},\ }\href
  {https://doi.org/10.1088/0305-4470/10/11/021} {\bibfield  {journal} {\bibinfo
   {journal} {\JPAOLD}\ }\textbf {\bibinfo {volume} {10}},\ \bibinfo {pages}
  {1927} (\bibinfo {year} {1977})}\BibitemShut {NoStop}%
\bibitem [{\citenamefont {Burkhardt}\ and\ \citenamefont
  {Cardy}(1987)}]{BC-87}%
  \BibitemOpen
  \bibfield  {author} {\bibinfo {author} {\bibfnamefont {T.~W.}\ \bibnamefont
  {Burkhardt}}\ and\ \bibinfo {author} {\bibfnamefont {J.~L.}\ \bibnamefont
  {Cardy}},\ }\bibfield  {title} {\bibinfo {title} {{Surface critical behaviour
  and local operators with boundary-induced critical profiles}},\ }\href
  {https://doi.org/10.1088/0305-4470/20/4/010} {\bibfield  {journal} {\bibinfo
  {journal} {\JPAOLD}\ }\textbf {\bibinfo {volume} {20}},\ \bibinfo {pages}
  {L233} (\bibinfo {year} {1987})}\BibitemShut {NoStop}%
\bibitem [{\citenamefont {{Parisen Toldin}}\ and\ \citenamefont
  {{Dietrich}}()}]{PTD-10}%
  \BibitemOpen
  \bibfield  {author} {\bibinfo {author} {\bibfnamefont {F.}~\bibnamefont
  {{Parisen Toldin}}}\ and\ \bibinfo {author} {\bibfnamefont {S.}~\bibnamefont
  {{Dietrich}}},\ }\bibfield  {title} {\bibinfo {title} {{Critical Casimir
  forces and adsorption profiles in the presence of a chemically structured
  substrate}},\ }\href {https://doi.org/10.1088/1742-5468/2010/11/P11003}
  {\bibfield  {journal} {\bibinfo  {journal} {JSTAT}\ }\bibfield  {number}
  {\bibinfo  {number} { (2010)},\ \bibinfo {eid} {P11003}},\ }\Eprint
  {https://arxiv.org/abs/1007.3913} {arXiv:1007.3913 [cond-mat.stat-mech]}
  \BibitemShut {NoStop}%
\bibitem [{\citenamefont {{Campostrini}}\ \emph
  {et~al.}(2002{\natexlab{a}})\citenamefont {{Campostrini}}, \citenamefont
  {{Hasenbusch}}, \citenamefont {{Pelissetto}}, \citenamefont {{Rossi}},\ and\
  \citenamefont {{Vicari}}}]{CHPRV-02}%
  \BibitemOpen
  \bibfield  {author} {\bibinfo {author} {\bibfnamefont {M.}~\bibnamefont
  {{Campostrini}}}, \bibinfo {author} {\bibfnamefont {M.}~\bibnamefont
  {{Hasenbusch}}}, \bibinfo {author} {\bibfnamefont {A.}~\bibnamefont
  {{Pelissetto}}}, \bibinfo {author} {\bibfnamefont {P.}~\bibnamefont
  {{Rossi}}},\ and\ \bibinfo {author} {\bibfnamefont {E.}~\bibnamefont
  {{Vicari}}},\ }\bibfield  {title} {\bibinfo {title} {{Critical exponents and
  equation of state of the three-dimensional Heisenberg universality class}},\
  }\href {https://doi.org/10.1103/PhysRevB.65.144520} {\bibfield  {journal}
  {\bibinfo  {journal} {\prb}\ }\textbf {\bibinfo {volume} {65}},\ \bibinfo
  {eid} {144520} (\bibinfo {year} {2002}{\natexlab{a}})},\ \Eprint
  {https://arxiv.org/abs/cond-mat/0110336} {cond-mat/0110336} \BibitemShut
  {NoStop}%
\bibitem [{\citenamefont {Campostrini}\ \emph {et~al.}(2006)\citenamefont
  {Campostrini}, \citenamefont {Hasenbusch}, \citenamefont {Pelissetto},\ and\
  \citenamefont {Vicari}}]{CHPV-06}%
  \BibitemOpen
  \bibfield  {author} {\bibinfo {author} {\bibfnamefont {M.}~\bibnamefont
  {Campostrini}}, \bibinfo {author} {\bibfnamefont {M.}~\bibnamefont
  {Hasenbusch}}, \bibinfo {author} {\bibfnamefont {A.}~\bibnamefont
  {Pelissetto}},\ and\ \bibinfo {author} {\bibfnamefont {E.}~\bibnamefont
  {Vicari}},\ }\bibfield  {title} {\bibinfo {title} {{Theoretical estimates of
  the critical exponents of the superfluid transition in $^4$He by lattice
  methods}},\ }\href {https://doi.org/10.1103/PhysRevB.74.144506} {\bibfield
  {journal} {\bibinfo  {journal} {\prb}\ }\textbf {\bibinfo {volume} {74}},\
  \bibinfo {eid} {144506} (\bibinfo {year} {2006})},\ \Eprint
  {https://arxiv.org/abs/cond-mat/0605083} {cond-mat/0605083} \BibitemShut
  {NoStop}%
\bibitem [{\citenamefont {Hasenbusch}(2019)}]{Hasenbusch-19}%
  \BibitemOpen
  \bibfield  {author} {\bibinfo {author} {\bibfnamefont {M.}~\bibnamefont
  {Hasenbusch}},\ }\bibfield  {title} {\bibinfo {title} {{Monte Carlo study of
  an improved clock model in three dimensions}},\ }\href
  {https://doi.org/10.1103/PhysRevB.100.224517} {\bibfield  {journal} {\bibinfo
   {journal} {\prb}\ }\textbf {\bibinfo {volume} {100}},\ \bibinfo {eid}
  {224517} (\bibinfo {year} {2019})},\ \Eprint
  {https://arxiv.org/abs/1910.05916} {arXiv:1910.05916 [cond-mat.stat-mech]}
  \BibitemShut {NoStop}%
\bibitem [{\citenamefont {Parisen~Toldin}(2022)}]{PT-21}%
  \BibitemOpen
  \bibfield  {author} {\bibinfo {author} {\bibfnamefont {F.}~\bibnamefont
  {Parisen~Toldin}},\ }\bibfield  {title} {\bibinfo {title} {{Finite-Size
  Scaling at fixed Renormalization-Group invariant}},\ }\href
  {https://doi.org/10.1103/PhysRevE.105.034137} {\bibfield  {journal} {\bibinfo
   {journal} {\PRE}\ }\textbf {\bibinfo {volume} {105}},\ \bibinfo {pages}
  {034137} (\bibinfo {year} {2022})},\ \Eprint
  {https://arxiv.org/abs/2112.00392} {arXiv:2112.00392 [cond-mat.stat-mech]}
  \BibitemShut {NoStop}%
\bibitem [{\citenamefont {Hasenbusch}(2020)}]{Hasenbusch-20}%
  \BibitemOpen
  \bibfield  {author} {\bibinfo {author} {\bibfnamefont {M.}~\bibnamefont
  {Hasenbusch}},\ }\bibfield  {title} {\bibinfo {title} {Monte carlo study of a
  generalized icosahedral model on the simple cubic lattice},\ }\href
  {https://doi.org/10.1103/PhysRevB.102.024406} {\bibfield  {journal} {\bibinfo
   {journal} {\prb}\ }\textbf {\bibinfo {volume} {102}},\ \bibinfo {eid}
  {024406} (\bibinfo {year} {2020})},\ \Eprint
  {https://arxiv.org/abs/2005.04448} {arXiv:2005.04448 [cond-mat.stat-mech]}
  \BibitemShut {NoStop}%
\bibitem [{\citenamefont {{Hasenbusch}}()}]{Hasenbusch-09b}%
  \BibitemOpen
  \bibfield  {author} {\bibinfo {author} {\bibfnamefont {M.}~\bibnamefont
  {{Hasenbusch}}},\ }\bibfield  {title} {\bibinfo {title} {{The thermodynamic
  Casimir effect in the neighbourhood of the {$\lambda$}-transition: a Monte
  Carlo study of an improved three-dimensional lattice model}},\ }\href
  {https://doi.org/10.1088/1742-5468/2009/07/P07031} {\bibfield  {journal}
  {\bibinfo  {journal} {JSTAT}\ }\bibfield  {number} {\bibinfo  {number} {
  (2010)},\ \bibinfo {eid} {P07031}},\ }\Eprint
  {https://arxiv.org/abs/0905.2096} {arXiv:0905.2096 [cond-mat.stat-mech]}
  \BibitemShut {NoStop}%
\bibitem [{\citenamefont {{Hasenbusch}}(2010)}]{Hasenbusch-10c}%
  \BibitemOpen
  \bibfield  {author} {\bibinfo {author} {\bibfnamefont {M.}~\bibnamefont
  {{Hasenbusch}}},\ }\bibfield  {title} {\bibinfo {title} {{Thermodynamic
  Casimir effect for films in the three-dimensional Ising universality class:
  Symmetry-breaking boundary conditions}},\ }\href
  {https://doi.org/10.1103/PhysRevB.82.104425} {\bibfield  {journal} {\bibinfo
  {journal} {\prb}\ }\textbf {\bibinfo {volume} {82}},\ \bibinfo {eid} {104425}
  (\bibinfo {year} {2010})},\ \Eprint {https://arxiv.org/abs/1005.4749}
  {arXiv:1005.4749 [cond-mat.stat-mech]} \BibitemShut {NoStop}%
\bibitem [{\citenamefont {{Hasenbusch}}(2011)}]{Hasenbusch-11}%
  \BibitemOpen
  \bibfield  {author} {\bibinfo {author} {\bibfnamefont {M.}~\bibnamefont
  {{Hasenbusch}}},\ }\bibfield  {title} {\bibinfo {title} {{Thermodynamic
  Casimir force: A Monte Carlo study of the crossover between the ordinary and
  the normal surface universality class}},\ }\href
  {https://doi.org/10.1103/PhysRevB.83.134425} {\bibfield  {journal} {\bibinfo
  {journal} {\prb}\ }\textbf {\bibinfo {volume} {83}},\ \bibinfo {eid} {134425}
  (\bibinfo {year} {2011})},\ \Eprint {https://arxiv.org/abs/1012.4986}
  {arXiv:1012.4986 [cond-mat.stat-mech]} \BibitemShut {NoStop}%
\bibitem [{\citenamefont {Hasenbusch}(2011)}]{Hasenbusch-11b}%
  \BibitemOpen
  \bibfield  {author} {\bibinfo {author} {\bibfnamefont {M.}~\bibnamefont
  {Hasenbusch}},\ }\bibfield  {title} {\bibinfo {title} {{Monte Carlo study of
  surface critical phenomena: The special point}},\ }\href
  {https://doi.org/10.1103/PhysRevB.84.134405} {\bibfield  {journal} {\bibinfo
  {journal} {\prb}\ }\textbf {\bibinfo {volume} {84}},\ \bibinfo {eid} {134405}
  (\bibinfo {year} {2011})},\ \Eprint {https://arxiv.org/abs/1108.2425}
  {arXiv:1108.2425 [cond-mat.stat-mech]} \BibitemShut {NoStop}%
\bibitem [{\citenamefont {{Hasenbusch}}(2012)}]{Hasenbusch-12}%
  \BibitemOpen
  \bibfield  {author} {\bibinfo {author} {\bibfnamefont {M.}~\bibnamefont
  {{Hasenbusch}}},\ }\bibfield  {title} {\bibinfo {title} {{Thermodynamic
  Casimir effect: Universality and corrections to scaling}},\ }\href
  {https://doi.org/10.1103/PhysRevB.85.174421} {\bibfield  {journal} {\bibinfo
  {journal} {\prb}\ }\textbf {\bibinfo {volume} {85}},\ \bibinfo {eid} {174421}
  (\bibinfo {year} {2012})},\ \Eprint {https://arxiv.org/abs/1202.6206}
  {arXiv:1202.6206 [cond-mat.stat-mech]} \BibitemShut {NoStop}%
\bibitem [{\citenamefont {{Parisen Toldin}}\ \emph {et~al.}(2013)\citenamefont
  {{Parisen Toldin}}, \citenamefont {{Tr{\"o}ndle}},\ and\ \citenamefont
  {{Dietrich}}}]{PTTD-13}%
  \BibitemOpen
  \bibfield  {author} {\bibinfo {author} {\bibfnamefont {F.}~\bibnamefont
  {{Parisen Toldin}}}, \bibinfo {author} {\bibfnamefont {M.}~\bibnamefont
  {{Tr{\"o}ndle}}},\ and\ \bibinfo {author} {\bibfnamefont {S.}~\bibnamefont
  {{Dietrich}}},\ }\bibfield  {title} {\bibinfo {title} {{Critical Casimir
  forces between homogeneous and chemically striped surfaces}},\ }\href
  {https://doi.org/10.1103/PhysRevE.88.052110} {\bibfield  {journal} {\bibinfo
  {journal} {\pre}\ }\textbf {\bibinfo {volume} {88}},\ \bibinfo {eid} {052110}
  (\bibinfo {year} {2013})},\ \Eprint {https://arxiv.org/abs/1303.6104}
  {arXiv:1303.6104 [cond-mat.stat-mech]} \BibitemShut {NoStop}%
\bibitem [{\citenamefont {{Parisen Toldin}}(2015)}]{PT-13}%
  \BibitemOpen
  \bibfield  {author} {\bibinfo {author} {\bibfnamefont {F.}~\bibnamefont
  {{Parisen Toldin}}},\ }\bibfield  {title} {\bibinfo {title} {{Critical
  Casimir force in the presence of random local adsorption preference}},\
  }\href {https://doi.org/10.1103/PhysRevE.91.032105} {\bibfield  {journal}
  {\bibinfo  {journal} {\PRE}\ }\textbf {\bibinfo {volume} {91}},\ \bibinfo
  {pages} {032105} (\bibinfo {year} {2015})},\ \Eprint
  {https://arxiv.org/abs/1308.5220} {arXiv:1308.5220 [cond-mat.stat-mech]}
  \BibitemShut {NoStop}%
\bibitem [{\citenamefont {{Parisen Toldin}}\ \emph {et~al.}(2015)\citenamefont
  {{Parisen Toldin}}, \citenamefont {{Tr{\"o}ndle}},\ and\ \citenamefont
  {{Dietrich}}}]{PTTD-14}%
  \BibitemOpen
  \bibfield  {author} {\bibinfo {author} {\bibfnamefont {F.}~\bibnamefont
  {{Parisen Toldin}}}, \bibinfo {author} {\bibfnamefont {M.}~\bibnamefont
  {{Tr{\"o}ndle}}},\ and\ \bibinfo {author} {\bibfnamefont {S.}~\bibnamefont
  {{Dietrich}}},\ }\bibfield  {title} {\bibinfo {title} {{Line contribution to
  the critical Casimir force between a homogeneous and a chemically stepped
  surface}},\ }\href {https://doi.org/10.1088/0953-8984/27/21/214010}
  {\bibfield  {journal} {\bibinfo  {journal} {\JPCM}\ }\textbf {\bibinfo
  {volume} {27}},\ \bibinfo {eid} {214010} (\bibinfo {year} {2015})},\ \Eprint
  {https://arxiv.org/abs/1409.5536} {arXiv:1409.5536 [cond-mat.stat-mech]}
  \BibitemShut {NoStop}%
\bibitem [{\citenamefont {{Parisen Toldin}}\ \emph {et~al.}(2017)\citenamefont
  {{Parisen Toldin}}, \citenamefont {{Assaad}},\ and\ \citenamefont
  {{Wessel}}}]{PTAW-17}%
  \BibitemOpen
  \bibfield  {author} {\bibinfo {author} {\bibfnamefont {F.}~\bibnamefont
  {{Parisen Toldin}}}, \bibinfo {author} {\bibfnamefont {F.~F.}\ \bibnamefont
  {{Assaad}}},\ and\ \bibinfo {author} {\bibfnamefont {S.}~\bibnamefont
  {{Wessel}}},\ }\bibfield  {title} {\bibinfo {title} {{Critical behavior in
  the presence of an order-parameter pinning field}},\ }\href
  {https://doi.org/10.1103/PhysRevB.95.014401} {\bibfield  {journal} {\bibinfo
  {journal} {\prb}\ }\textbf {\bibinfo {volume} {95}},\ \bibinfo {eid} {014401}
  (\bibinfo {year} {2017})},\ \Eprint {https://arxiv.org/abs/1607.04270}
  {arXiv:1607.04270 [cond-mat.stat-mech]} \BibitemShut {NoStop}%
\bibitem [{\citenamefont {{Wolff}}(1989)}]{Wolff-89}%
  \BibitemOpen
  \bibfield  {author} {\bibinfo {author} {\bibfnamefont {U.}~\bibnamefont
  {{Wolff}}},\ }\bibfield  {title} {\bibinfo {title} {{Collective Monte Carlo
  updating for spin systems}},\ }\href
  {https://doi.org/10.1103/PhysRevLett.62.361} {\bibfield  {journal} {\bibinfo
  {journal} {\PRL}\ }\textbf {\bibinfo {volume} {62}},\ \bibinfo {pages} {361}
  (\bibinfo {year} {1989})}\BibitemShut {NoStop}%
\bibitem [{\citenamefont {{Cardy}}(1990)}]{Cardy-90}%
  \BibitemOpen
  \bibfield  {author} {\bibinfo {author} {\bibfnamefont {J.~L.}\ \bibnamefont
  {{Cardy}}},\ }\bibfield  {title} {\bibinfo {title} {{Universal critical-point
  amplitudes in parallel-plate geometries}},\ }\href
  {https://doi.org/10.1103/PhysRevLett.65.1443} {\bibfield  {journal} {\bibinfo
   {journal} {\PRL}\ }\textbf {\bibinfo {volume} {65}},\ \bibinfo {pages}
  {1443} (\bibinfo {year} {1990})}\BibitemShut {NoStop}%
\bibitem [{\citenamefont {Padayasi}\ \emph {et~al.}()\citenamefont {Padayasi},
  \citenamefont {Krishnan}, \citenamefont {Metlitski}, \citenamefont
  {Gruzberg},\ and\ \citenamefont {Meineri}}]{Padayasi}%
  \BibitemOpen
  \bibfield  {author} {\bibinfo {author} {\bibfnamefont {J.}~\bibnamefont
  {Padayasi}}, \bibinfo {author} {\bibfnamefont {A.}~\bibnamefont {Krishnan}},
  \bibinfo {author} {\bibfnamefont {M.}~\bibnamefont {Metlitski}}, \bibinfo
  {author} {\bibfnamefont {I.}~\bibnamefont {Gruzberg}},\ and\ \bibinfo
  {author} {\bibfnamefont {M.}~\bibnamefont {Meineri}},\ }\bibfield  {title}
  {\bibinfo {title} {The extraordinary boundary transition in the 3d o(n) model
  via conformal bootstrap},\ }\href@noop {} {\bibfield  {journal} {\bibinfo
  {journal} {{}}\ }}\Eprint {https://arxiv.org/abs/2111.03071}
  {arXiv:2111.03071 [cond-mat.str-el]} \BibitemShut {NoStop}%
\bibitem [{SM()}]{SM}%
  \BibitemOpen
  \href@noop {} {}\bibinfo {note} {See Supplemental Material for a discussion
  of the scaling forms of various observables of the lattice model, technical
  details on the fits of statistically correlated data, and results of the fits
  of MC data.}\BibitemShut {Stop}%
\bibitem [{\citenamefont {Chester}\ \emph {et~al.}(2020)\citenamefont
  {Chester}, \citenamefont {Landry}, \citenamefont {Liu}, \citenamefont
  {Poland}, \citenamefont {Simmons-Duffin}, \citenamefont {Su},\ and\
  \citenamefont {Vichi}}]{CLLPSDSV-19}%
  \BibitemOpen
  \bibfield  {author} {\bibinfo {author} {\bibfnamefont {S.~M.}\ \bibnamefont
  {Chester}}, \bibinfo {author} {\bibfnamefont {W.}~\bibnamefont {Landry}},
  \bibinfo {author} {\bibfnamefont {J.}~\bibnamefont {Liu}}, \bibinfo {author}
  {\bibfnamefont {D.}~\bibnamefont {Poland}}, \bibinfo {author} {\bibfnamefont
  {D.}~\bibnamefont {Simmons-Duffin}}, \bibinfo {author} {\bibfnamefont
  {N.}~\bibnamefont {Su}},\ and\ \bibinfo {author} {\bibfnamefont
  {A.}~\bibnamefont {Vichi}},\ }\bibfield  {title} {\bibinfo {title} {{Carving
  out OPE space and precise O(2) model critical exponents}},\ }\href
  {https://doi.org/10.1007/JHEP06(2020)142} {\bibfield  {journal} {\bibinfo
  {journal} {JHEP}\ }\textbf {\bibinfo {volume} {06}}\bibfield  {number}
  {\bibinfo  {number} { (2020)},\ \bibinfo {eid} {142}},\ }\Eprint
  {https://arxiv.org/abs/1912.03324} {arXiv:1912.03324 [hep-th]} \BibitemShut
  {NoStop}%
\bibitem [{\citenamefont {Young}(2015)}]{Young_notes}%
  \BibitemOpen
  \bibfield  {author} {\bibinfo {author} {\bibfnamefont {A.~P.}\ \bibnamefont
  {Young}},\ }\href {https://doi.org/10.1007/978-3-319-19051-8} {\emph
  {\bibinfo {title} {{Everything You Wanted to Know About Data Analysis and
  Fitting but Were Afraid to Ask}}}},\ SpringerBriefs in Physics\ (\bibinfo
  {publisher} {Springer International Publishing},\ \bibinfo {year} {2015})\
  \Eprint {https://arxiv.org/abs/arXiv:1210.3781} {arXiv:1210.3781}
  \BibitemShut {NoStop}%
\bibitem [{\citenamefont {Michael}(1994)}]{Michael-94}%
  \BibitemOpen
  \bibfield  {author} {\bibinfo {author} {\bibfnamefont {C.}~\bibnamefont
  {Michael}},\ }\bibfield  {title} {\bibinfo {title} {Fitting correlated
  data},\ }\href {https://doi.org/10.1103/PhysRevD.49.2616} {\bibfield
  {journal} {\bibinfo  {journal} {\prd}\ }\textbf {\bibinfo {volume} {49}},\
  \bibinfo {pages} {2616} (\bibinfo {year} {1994})},\ \Eprint
  {https://arxiv.org/abs/hep-lat/9310026} {arXiv:hep-lat/9310026 [hep-lat]}
  \BibitemShut {NoStop}%
\bibitem [{\citenamefont {Seibert}(1994)}]{Seibert-94}%
  \BibitemOpen
  \bibfield  {author} {\bibinfo {author} {\bibfnamefont {D.}~\bibnamefont
  {Seibert}},\ }\bibfield  {title} {\bibinfo {title} {Undesirable effects of
  covariance matrix techniques for error analysis},\ }\href
  {https://doi.org/10.1103/PhysRevD.49.6240} {\bibfield  {journal} {\bibinfo
  {journal} {\prd}\ }\textbf {\bibinfo {volume} {49}},\ \bibinfo {pages} {6240}
  (\bibinfo {year} {1994})},\ \Eprint {https://arxiv.org/abs/hep-lat/9305014}
  {arXiv:hep-lat/9305014 [hep-lat]} \BibitemShut {NoStop}%
\bibitem [{Note2()}]{Note2}%
  \BibitemOpen
  \bibinfo {note} {Even though $\protect \vec {\varphi }$ has only a single
  component in the $N = 2$ case, we write it as a vector here to set the
  normalization for $N = 3$.}\BibitemShut {Stop}%
\bibitem [{\citenamefont {{Eisenriegler}}\ \emph
  {et~al.}(1993{\natexlab{a}})\citenamefont {{Eisenriegler}}, \citenamefont
  {{Krech}},\ and\ \citenamefont {{Dietrich}}}]{EKD-93}%
  \BibitemOpen
  \bibfield  {author} {\bibinfo {author} {\bibfnamefont {E.}~\bibnamefont
  {{Eisenriegler}}}, \bibinfo {author} {\bibfnamefont {M.}~\bibnamefont
  {{Krech}}},\ and\ \bibinfo {author} {\bibfnamefont {S.}~\bibnamefont
  {{Dietrich}}},\ }\bibfield  {title} {\bibinfo {title} {{Absence of
  hyperuniversality in critical films}},\ }\href
  {https://doi.org/10.1103/PhysRevLett.70.619} {\bibfield  {journal} {\bibinfo
  {journal} {\prl}\ }\textbf {\bibinfo {volume} {70}},\ \bibinfo {pages} {619}
  (\bibinfo {year} {1993}{\natexlab{a}})}\BibitemShut {NoStop}%
\bibitem [{\citenamefont {{Eisenriegler}}\ \emph
  {et~al.}(1993{\natexlab{b}})\citenamefont {{Eisenriegler}}, \citenamefont
  {{Krech}},\ and\ \citenamefont {{Dietrich}}}]{EKD-93_erratum}%
  \BibitemOpen
  \bibfield  {author} {\bibinfo {author} {\bibfnamefont {E.}~\bibnamefont
  {{Eisenriegler}}}, \bibinfo {author} {\bibfnamefont {M.}~\bibnamefont
  {{Krech}}},\ and\ \bibinfo {author} {\bibfnamefont {S.}~\bibnamefont
  {{Dietrich}}},\ }\bibfield  {title} {\bibinfo {title} {{Erratum: ``Absence of
  hyperuniversality in critical films'' [Phys. Rev. Lett. 70, 619 (1993)]}},\
  }\href {https://doi.org/10.1103/PhysRevLett.70.2051.2} {\bibfield  {journal}
  {\bibinfo  {journal} {\prl}\ }\textbf {\bibinfo {volume} {70}},\ \bibinfo
  {pages} {2051(E)} (\bibinfo {year} {1993}{\natexlab{b}})}\BibitemShut
  {NoStop}%
\bibitem [{\citenamefont {{Eisenriegler}}\ and\ \citenamefont
  {{Stapper}}(1994)}]{ES-94}%
  \BibitemOpen
  \bibfield  {author} {\bibinfo {author} {\bibfnamefont {E.}~\bibnamefont
  {{Eisenriegler}}}\ and\ \bibinfo {author} {\bibfnamefont {M.}~\bibnamefont
  {{Stapper}}},\ }\bibfield  {title} {\bibinfo {title} {{Critical behavior near
  a symmetry-breaking surface and the stress tensor}},\ }\href
  {https://doi.org/10.1103/PhysRevB.50.10009} {\bibfield  {journal} {\bibinfo
  {journal} {\prb}\ }\textbf {\bibinfo {volume} {50}},\ \bibinfo {pages}
  {10009} (\bibinfo {year} {1994})}\BibitemShut {NoStop}%
\bibitem [{\citenamefont {Simmons-Duffin}(2017)}]{SDIsingSpectrum}%
  \BibitemOpen
  \bibfield  {author} {\bibinfo {author} {\bibfnamefont {D.}~\bibnamefont
  {Simmons-Duffin}},\ }\bibfield  {title} {\bibinfo {title} {{The Lightcone
  Bootstrap and the Spectrum of the 3d Ising CFT}},\ }\href
  {https://doi.org/10.1007/JHEP03(2017)086} {\bibfield  {journal} {\bibinfo
  {journal} {JHEP}\ }\textbf {\bibinfo {volume} {03}}\bibfield  {number}
  {\bibinfo  {number} { (2017)},\ \bibinfo {pages} {086}},\ }\Eprint
  {https://arxiv.org/abs/1612.08471} {arXiv:1612.08471 [hep-th]} \BibitemShut
  {NoStop}%
\bibitem [{\citenamefont {Hasenbusch}(2010)}]{Hasenbusch-10b}%
  \BibitemOpen
  \bibfield  {author} {\bibinfo {author} {\bibfnamefont {M.}~\bibnamefont
  {Hasenbusch}},\ }\bibfield  {title} {\bibinfo {title} {{Universal amplitude
  ratios in the three-dimensional Ising universality class}},\ }\href
  {https://doi.org/10.1103/PhysRevB.82.174434} {\bibfield  {journal} {\bibinfo
  {journal} {\prb}\ }\textbf {\bibinfo {volume} {82}},\ \bibinfo {eid} {174434}
  (\bibinfo {year} {2010})},\ \Eprint {https://arxiv.org/abs/1004.4983}
  {arXiv:1004.4983 [cond-mat.stat-mech]} \BibitemShut {NoStop}%
\bibitem [{\citenamefont {{Campostrini}}\ \emph
  {et~al.}(2002{\natexlab{b}})\citenamefont {{Campostrini}}, \citenamefont
  {{Pelissetto}}, \citenamefont {{Rossi}},\ and\ \citenamefont
  {{Vicari}}}]{CPRV-02}%
  \BibitemOpen
  \bibfield  {author} {\bibinfo {author} {\bibfnamefont {M.}~\bibnamefont
  {{Campostrini}}}, \bibinfo {author} {\bibfnamefont {A.}~\bibnamefont
  {{Pelissetto}}}, \bibinfo {author} {\bibfnamefont {P.}~\bibnamefont
  {{Rossi}}},\ and\ \bibinfo {author} {\bibfnamefont {E.}~\bibnamefont
  {{Vicari}}},\ }\bibfield  {title} {\bibinfo {title} {{25th-order
  high-temperature expansion results for three-dimensional Ising-like systems
  on the simple-cubic lattice}},\ }\href
  {https://doi.org/10.1103/PhysRevE.65.066127} {\bibfield  {journal} {\bibinfo
  {journal} {\pre}\ }\textbf {\bibinfo {volume} {65}},\ \bibinfo {eid} {066127}
  (\bibinfo {year} {2002}{\natexlab{b}})},\ \Eprint
  {https://arxiv.org/abs/cond-mat/0201180} {cond-mat/0201180} \BibitemShut
  {NoStop}%
\bibitem [{\citenamefont {Bray}(1976)}]{Bray}%
  \BibitemOpen
  \bibfield  {author} {\bibinfo {author} {\bibfnamefont {A.~J.}\ \bibnamefont
  {Bray}},\ }\bibfield  {title} {\bibinfo {title} {Dispersion-theory approach
  to the correlation function for critical scattering},\ }\href
  {https://doi.org/10.1103/PhysRevB.14.1248} {\bibfield  {journal} {\bibinfo
  {journal} {Phys. Rev. B}\ }\textbf {\bibinfo {volume} {14}},\ \bibinfo
  {pages} {1248} (\bibinfo {year} {1976})}\BibitemShut {NoStop}%
\bibitem [{\citenamefont {Chang}\ \emph {et~al.}(1979)\citenamefont {Chang},
  \citenamefont {Burstyn},\ and\ \citenamefont {Sengers}}]{Chang}%
  \BibitemOpen
  \bibfield  {author} {\bibinfo {author} {\bibfnamefont {R.~F.}\ \bibnamefont
  {Chang}}, \bibinfo {author} {\bibfnamefont {H.}~\bibnamefont {Burstyn}},\
  and\ \bibinfo {author} {\bibfnamefont {J.~V.}\ \bibnamefont {Sengers}},\
  }\bibfield  {title} {\bibinfo {title} {Correlation function near the critical
  mixing point of a binary liquid},\ }\href
  {https://doi.org/10.1103/PhysRevA.19.866} {\bibfield  {journal} {\bibinfo
  {journal} {Phys. Rev. A}\ }\textbf {\bibinfo {volume} {19}},\ \bibinfo
  {pages} {866} (\bibinfo {year} {1979})}\BibitemShut {NoStop}%
\bibitem [{\citenamefont {Damay}\ \emph {et~al.}(1989)\citenamefont {Damay},
  \citenamefont {Leclercq},\ and\ \citenamefont {Chieux}}]{Damay}%
  \BibitemOpen
  \bibfield  {author} {\bibinfo {author} {\bibfnamefont {P.}~\bibnamefont
  {Damay}}, \bibinfo {author} {\bibfnamefont {F.}~\bibnamefont {Leclercq}},\
  and\ \bibinfo {author} {\bibfnamefont {P.}~\bibnamefont {Chieux}},\
  }\bibfield  {title} {\bibinfo {title} {Critical scattering function in a
  binary fluid mixture: A study of sodium-deuteroammonia solution at the
  critical concentration by small-angle neutron scattering},\ }\href
  {https://doi.org/10.1103/PhysRevB.40.4696} {\bibfield  {journal} {\bibinfo
  {journal} {Phys. Rev. B}\ }\textbf {\bibinfo {volume} {40}},\ \bibinfo
  {pages} {4696} (\bibinfo {year} {1989})}\BibitemShut {NoStop}%
\bibitem [{\citenamefont {Damay}\ \emph {et~al.}(1998)\citenamefont {Damay},
  \citenamefont {Leclercq}, \citenamefont {Magli}, \citenamefont {Formisano},\
  and\ \citenamefont {Lindner}}]{Damay98}%
  \BibitemOpen
  \bibfield  {author} {\bibinfo {author} {\bibfnamefont {P.}~\bibnamefont
  {Damay}}, \bibinfo {author} {\bibfnamefont {F.}~\bibnamefont {Leclercq}},
  \bibinfo {author} {\bibfnamefont {R.}~\bibnamefont {Magli}}, \bibinfo
  {author} {\bibfnamefont {F.}~\bibnamefont {Formisano}},\ and\ \bibinfo
  {author} {\bibfnamefont {P.}~\bibnamefont {Lindner}},\ }\bibfield  {title}
  {\bibinfo {title} {Universal critical-scattering function: An experimental
  approach},\ }\href {https://doi.org/10.1103/PhysRevB.58.12038} {\bibfield
  {journal} {\bibinfo  {journal} {Phys. Rev. B}\ }\textbf {\bibinfo {volume}
  {58}},\ \bibinfo {pages} {12038} (\bibinfo {year} {1998})}\BibitemShut
  {NoStop}%
\bibitem [{\citenamefont {Mart\'{\i}n-Mayor}\ \emph {et~al.}(2002)\citenamefont
  {Mart\'{\i}n-Mayor}, \citenamefont {Pelissetto},\ and\ \citenamefont
  {Vicari}}]{VicariA1p}%
  \BibitemOpen
  \bibfield  {author} {\bibinfo {author} {\bibfnamefont {V.}~\bibnamefont
  {Mart\'{\i}n-Mayor}}, \bibinfo {author} {\bibfnamefont {A.}~\bibnamefont
  {Pelissetto}},\ and\ \bibinfo {author} {\bibfnamefont {E.}~\bibnamefont
  {Vicari}},\ }\bibfield  {title} {\bibinfo {title} {Critical structure factor
  in ising systems},\ }\href {https://doi.org/10.1103/PhysRevE.66.026112}
  {\bibfield  {journal} {\bibinfo  {journal} {Phys. Rev. E}\ }\textbf {\bibinfo
  {volume} {66}},\ \bibinfo {pages} {026112} (\bibinfo {year}
  {2002})}\BibitemShut {NoStop}%
\bibitem [{\citenamefont {{Gliozzi}}\ \emph {et~al.}(2015)\citenamefont
  {{Gliozzi}}, \citenamefont {{Liendo}}, \citenamefont {{Meineri}},\ and\
  \citenamefont {{Rago}}}]{GLMR-15}%
  \BibitemOpen
  \bibfield  {author} {\bibinfo {author} {\bibfnamefont {F.}~\bibnamefont
  {{Gliozzi}}}, \bibinfo {author} {\bibfnamefont {P.}~\bibnamefont {{Liendo}}},
  \bibinfo {author} {\bibfnamefont {M.}~\bibnamefont {{Meineri}}},\ and\
  \bibinfo {author} {\bibfnamefont {A.}~\bibnamefont {{Rago}}},\ }\bibfield
  {title} {\bibinfo {title} {{Boundary and interface CFTs from the conformal
  bootstrap}},\ }\href {https://doi.org/10.1007/JHEP05(2015)036} {\bibfield
  {journal} {\bibinfo  {journal} {JHEP}\ }\textbf {\bibinfo {volume}
  {05}}\bibfield  {number} {\bibinfo  {number} { (2015)},\ \bibinfo {eid}
  {036}},\ }\Eprint {https://arxiv.org/abs/1502.07217} {arXiv:1502.07217
  [hep-th]} \BibitemShut {NoStop}%
\bibitem [{\citenamefont {Deng}(2006)}]{DengO4}%
  \BibitemOpen
  \bibfield  {author} {\bibinfo {author} {\bibfnamefont {Y.}~\bibnamefont
  {Deng}},\ }\bibfield  {title} {\bibinfo {title} {{Bulk and surface phase
  transitions in the three-dimensional O$(4)$ spin model}},\ }\href
  {https://doi.org/10.1103/PhysRevE.73.056116} {\bibfield  {journal} {\bibinfo
  {journal} {Phys. Rev. E}\ }\textbf {\bibinfo {volume} {73}},\ \bibinfo
  {pages} {056116} (\bibinfo {year} {2006})}\BibitemShut {NoStop}%
\bibitem [{Note3()}]{Note3}%
  \BibitemOpen
  \bibinfo {note} {In principle, it is also possible to determine $b_{\protect
  \rm D}$ by a method similar to the one we used to find $b_{\protect \rm t}$:
  one extracts the normalization of ${\protect \rm D}$ from the connected
  two-point function of $\sigma $ on the boundary, and the coefficient
  $b_{\protect \rm D}$ from the connected bulk-boundary two point function of
  $\sigma $. However, we found that due to the large scaling dimension
  $\protect \hat {\Delta }_{\protect \rm D} = 3$, the boundary two-point
  function $\langle \sigma ({\protect \bf x})\sigma ({\protect \rm 0})\rangle
  _{{\protect \rm conn}}$ falls off very fast, so that the normalization is
  difficult to extract reliably.}\BibitemShut {Stop}%
\bibitem [{\citenamefont {De'Bell}\ \emph {et~al.}(1990)\citenamefont
  {De'Bell}, \citenamefont {Lookman},\ and\ \citenamefont
  {Whittington}}]{DBLW-90}%
  \BibitemOpen
  \bibfield  {author} {\bibinfo {author} {\bibfnamefont {K.}~\bibnamefont
  {De'Bell}}, \bibinfo {author} {\bibfnamefont {T.}~\bibnamefont {Lookman}},\
  and\ \bibinfo {author} {\bibfnamefont {S.~G.}\ \bibnamefont {Whittington}},\
  }\bibfield  {title} {\bibinfo {title} {{Analysis of exact enumeration data
  for self-avoiding walks attached to a surface}},\ }\href
  {https://doi.org/10.1103/PhysRevA.41.682} {\bibfield  {journal} {\bibinfo
  {journal} {\pra}\ }\textbf {\bibinfo {volume} {41}},\ \bibinfo {pages} {682}
  (\bibinfo {year} {1990})}\BibitemShut {NoStop}%
\bibitem [{\citenamefont {De'Bell}\ and\ \citenamefont
  {Lookman}(1993)}]{DBL-93}%
  \BibitemOpen
  \bibfield  {author} {\bibinfo {author} {\bibfnamefont {K.}~\bibnamefont
  {De'Bell}}\ and\ \bibinfo {author} {\bibfnamefont {T.}~\bibnamefont
  {Lookman}},\ }\bibfield  {title} {\bibinfo {title} {{Surface phase
  transitions in polymer systems}},\ }\href
  {https://doi.org/10.1103/RevModPhys.65.87} {\bibfield  {journal} {\bibinfo
  {journal} {\RMP}\ }\textbf {\bibinfo {volume} {65}},\ \bibinfo {pages} {87}
  (\bibinfo {year} {1993})}\BibitemShut {NoStop}%
\bibitem [{\citenamefont {Zhao}\ \emph {et~al.}(1990)\citenamefont {Zhao},
  \citenamefont {Lookman},\ and\ \citenamefont {De'Bell}}]{ZLTDB-90}%
  \BibitemOpen
  \bibfield  {author} {\bibinfo {author} {\bibfnamefont {D.}~\bibnamefont
  {Zhao}}, \bibinfo {author} {\bibfnamefont {T.}~\bibnamefont {Lookman}},\ and\
  \bibinfo {author} {\bibfnamefont {K.}~\bibnamefont {De'Bell}},\ }\bibfield
  {title} {\bibinfo {title} {Crossover behavior for self-avoiding walks
  interacting with a surface},\ }\href
  {https://doi.org/10.1103/PhysRevA.42.4591} {\bibfield  {journal} {\bibinfo
  {journal} {\pra}\ }\textbf {\bibinfo {volume} {42}},\ \bibinfo {pages} {4591}
  (\bibinfo {year} {1990})}\BibitemShut {NoStop}%
\bibitem [{\citenamefont {Eisenriegler}(1993)}]{Eisenriegler-book}%
  \BibitemOpen
  \bibfield  {author} {\bibinfo {author} {\bibfnamefont {E.}~\bibnamefont
  {Eisenriegler}},\ }\href {https://doi.org/10.1142/1354} {\emph {\bibinfo
  {title} {{Polymers Near Surfaces}}}}\ (\bibinfo  {publisher} {World
  Scientific},\ \bibinfo {address} {Singapore},\ \bibinfo {year}
  {1993})\BibitemShut {NoStop}%
\bibitem [{\citenamefont {Hegger}\ and\ \citenamefont
  {Grassberger}(1994)}]{HG-94}%
  \BibitemOpen
  \bibfield  {author} {\bibinfo {author} {\bibfnamefont {R.}~\bibnamefont
  {Hegger}}\ and\ \bibinfo {author} {\bibfnamefont {P.}~\bibnamefont
  {Grassberger}},\ }\bibfield  {title} {\bibinfo {title} {{Chain polymers near
  an adsorbing surface}},\ }\href {https://doi.org/10.1088/0305-4470/27/12/015}
  {\bibfield  {journal} {\bibinfo  {journal} {\JPAOLD}\ }\textbf {\bibinfo
  {volume} {27}},\ \bibinfo {pages} {4069} (\bibinfo {year}
  {1994})}\BibitemShut {NoStop}%
\bibitem [{\citenamefont {Batchelor}\ and\ \citenamefont
  {Cardy}(1997)}]{Cardy1997}%
  \BibitemOpen
  \bibfield  {author} {\bibinfo {author} {\bibfnamefont {M.~T.}\ \bibnamefont
  {Batchelor}}\ and\ \bibinfo {author} {\bibfnamefont {J.}~\bibnamefont
  {Cardy}},\ }\bibfield  {title} {\bibinfo {title} {Extraordinary transition in
  the two-dimensional o(n) model},\ }\href
  {https://doi.org/10.1016/s0550-3213(97)00533-6} {\bibfield  {journal}
  {\bibinfo  {journal} {\NPB}\ }\textbf {\bibinfo {volume} {506}},\ \bibinfo
  {pages} {553–564} (\bibinfo {year} {1997})}\BibitemShut {NoStop}%
\bibitem [{\citenamefont {de~Gennes}(1979)}]{deGennes-book}%
  \BibitemOpen
  \bibfield  {author} {\bibinfo {author} {\bibfnamefont {P.-G.}\ \bibnamefont
  {de~Gennes}},\ }\href@noop {} {\emph {\bibinfo {title} {{Scaling Concepts in
  Polymer Physics}}}}\ (\bibinfo  {publisher} {Cornell University Press},\
  \bibinfo {address} {Ithaca, NY},\ \bibinfo {year} {1979})\BibitemShut
  {NoStop}%
\bibitem [{\citenamefont {Liu}\ \emph {et~al.}(2012)\citenamefont {Liu},
  \citenamefont {Deng}, \citenamefont {Garoni},\ and\ \citenamefont
  {Bl{\"o}te}}]{LDGB-12}%
  \BibitemOpen
  \bibfield  {author} {\bibinfo {author} {\bibfnamefont {Q.}~\bibnamefont
  {Liu}}, \bibinfo {author} {\bibfnamefont {Y.}~\bibnamefont {Deng}}, \bibinfo
  {author} {\bibfnamefont {T.~M.}\ \bibnamefont {Garoni}},\ and\ \bibinfo
  {author} {\bibfnamefont {H.~W.~J.}\ \bibnamefont {Bl{\"o}te}},\ }\bibfield
  {title} {\bibinfo {title} {{The O(n) loop model on a three-dimensional
  lattice}},\ }\href {https://doi.org/10.1016/j.nuclphysb.2012.01.026}
  {\bibfield  {journal} {\bibinfo  {journal} {\NPB}\ }\textbf {\bibinfo
  {volume} {859}},\ \bibinfo {pages} {107} (\bibinfo {year} {2012})},\ \Eprint
  {https://arxiv.org/abs/1112.5647} {arXiv:1112.5647 [cond-mat.stat-mech]}
  \BibitemShut {NoStop}%
\bibitem [{\citenamefont {Domany}\ \emph {et~al.}(1981)\citenamefont {Domany},
  \citenamefont {Mukamel}, \citenamefont {Nienhuis},\ and\ \citenamefont
  {Schwimmer}}]{DMNS-81}%
  \BibitemOpen
  \bibfield  {author} {\bibinfo {author} {\bibfnamefont {E.}~\bibnamefont
  {Domany}}, \bibinfo {author} {\bibfnamefont {D.}~\bibnamefont {Mukamel}},
  \bibinfo {author} {\bibfnamefont {B.}~\bibnamefont {Nienhuis}},\ and\
  \bibinfo {author} {\bibfnamefont {A.}~\bibnamefont {Schwimmer}},\ }\bibfield
  {title} {\bibinfo {title} {{Duality relations and equivalences for models
  with O(N) and cubic symmetry}},\ }\href
  {https://doi.org/10.1016/0550-3213(81)90559-9} {\bibfield  {journal}
  {\bibinfo  {journal} {\NPB}\ }\textbf {\bibinfo {volume} {190}},\ \bibinfo
  {pages} {279} (\bibinfo {year} {1981})}\BibitemShut {NoStop}%
\bibitem [{\citenamefont {{J\"{u}lich Supercomputing Centre}}(2019)}]{JUWELS}%
  \BibitemOpen
  \bibfield  {author} {\bibinfo {author} {\bibnamefont {{J\"{u}lich
  Supercomputing Centre}}},\ }\bibfield  {title} {\bibinfo {title} {{JUWELS:
  Modular Tier-0/1 Supercomputer at the J\"{u}lich Supercomputing Centre}},\
  }\href {https://doi.org/10.17815/jlsrf-5-171} {\bibfield  {journal} {\bibinfo
   {journal} {{J. Large-Scale Res. Facil.}}\ }\textbf {\bibinfo {volume} {5}},\
  \bibinfo {pages} {A135} (\bibinfo {year} {2019})}\BibitemShut {NoStop}%
\end{thebibliography}%

\clearpage


\onecolumngrid
  \parbox[c][3em][t]{\textwidth}{\centering \large\bf Supplemental Material}
\smallskip
\twocolumngrid

\setcounter{equation}{0}
\renewcommand{\theHequation}{S.\arabic{equation}}
\renewcommand{\theequation}{S.\arabic{equation}}

\setcounter{figure}{0}
\renewcommand{\theHfigure}{S.\arabic{figure}}
\renewcommand{\thefigure}{S.\arabic{figure}}

\setcounter{table}{0}
\renewcommand{\theHtable}{S.\Roman{table}}
\renewcommand{\thetable}{S.\Roman{table}}

\section{Scaling forms}
\label{app:corr}

In this section, we discuss the  scaling forms of various observables  for the normal boundary universality class of the O($N$) model that we use to fit our data. We always assume that the model is tuned to the bulk critical point. Unless otherwise noted, all the operators in this section are those of the continuum conformal field theory; lattice fields are denoted with an explicit subscript ``lat''.  We will denote bulk operators by ${\cal O}$ and boundary operators (with the exception of protected operators) by $\hat{{\cal O}}$ and the corresponding scaling dimensions by $\Delta_{O}$ and $\hat{\Delta}_{{\cal O}}$. We normalize bulk operators so that in infinite space $\langle {\cal O}_1(x) {\cal O}_2(y)\rangle = \frac{\delta_{12}}{(x-y)^{2 \Delta_O}}$, and boundary operators so that in half-infinite space $\langle \hat{{\cal O}}_1({\bf x}) \hat{{\cal O}}_2({\bf y})\rangle = \frac{\delta_{12}}{({\bf x}-{\bf y})^{2 \hat{\Delta}_O}}$. 

We begin with the bulk OPE of the leading O($N$) vector $\phi^a$ with itself: $\phi^a \times \phi^b$. This OPE will contain operators transforming in the singlet, traceless symmetric  and antisymmetric tensor representations of O($N$). After taking the O($N$) trace, only the O($N$) singlet operators survive: 
\begin{equation} \phi^a(x) \phi^a(0) = \frac{N}{x^{2 \Delta_{\phi}}}\left(1+ \lambda_{\phi \phi \epsilon} x^{\Delta_\epsilon} \epsilon(0) + \ldots\right), \,\, x \to 0.\end{equation}
Here, $\epsilon$ is the lowest dimension  singlet --- the relevant perturbation of the O($N$) model, often called the thermal or the energy operator, and $\lambda_{\phi \phi \epsilon}$ is an OPE coefficient. We have dropped all higher dimension operators in the OPE. When considering the system in a periodic box of size  $L$, we expect $\langle \epsilon(x) \rangle = \frac{u_\epsilon}{L^{\Delta_{\epsilon}}}$, with $u_\epsilon$ --- a universal number. Thus, in a periodic box the two point function including the leading finite size scaling correction is:
\begin{equation} \langle \phi^a(x) \phi^a(0)\rangle \approx \frac{N}{x^{2 \Delta_{\phi}}}\left[1+ \lambda_{\phi \phi \epsilon} u_\epsilon \left(\frac{x}{L}\right)^{\Delta_\epsilon}\right], \quad x \ll L. \label{eq:phibulkconf} \end{equation}

We also expect corrections to scaling. We are working with a model where the leading irrelevant O($N$) scalar $\epsilon'$ has been tuned away. Thus, corrections to scaling will come from a variety of sources: 
\begin{enumerate}
\item The next O($N$) scalar and angular momentum $\ell = 0$ perturbation $\epsilon''$---the  dimension of this is not known precisely for $N \ge 2$. For $N = 1$  numerical bootstrap gives $\Delta_{\epsilon''} \approx 6.9$ \cite{SDIsingSpectrum}, i.e. the correction to scaling exponent $\omega'' \approx 3.9$. These corrections will be ignored below. \label{en:epspp}
\item The leading O($N$) scalar and angular momentum $\ell = 4$ perturbation (which is allowed by the cubic anisotropy of the lattice) - this is estimated to have scaling dimension $\Delta \approx 5$, so $\omega_{NR} \approx 2$ \cite{SDIsingSpectrum, Hasenbusch-19, Hasenbusch-20}.\label{en:NR}
\item The expansion of the lattice operator in terms of continuum operators, 
\begin{equation} \phi^a_{{\rm lat}} \approx \sqrt{{\cal N}_{\rm bulk}/N} (1 + c \, \Box ) \phi^a + \ldots, \end{equation} 
where we've dropped higher descendants of $\phi^a$, as well as higher dimension O($N$) vector primaries (it is generally thought that they have dimension $\Delta > 3 > \Delta_\phi + 2$). \label{en:op}
\end{enumerate}
Combining the effects \ref{en:NR} and \ref{en:op} we obtain our scaling ansatz (\ref{phi_bulk_ansatz}). 
Note that we have only included  corrections to scaling to the leading term in the $x \ll L$ limit in (\ref{phi_bulk_ansatz}). 

We next proceed to the theory in the presence of a normal boundary. We begin by taking the expectation value of the $\sigma$ OPE in (\ref{eq:BOPE}).
In a periodic slab geometry,  $\langle D \rangle = u_{\mathrm{D}} L^{-3}$, so we obtain
\begin{equation} \langle \sigma(z) \rangle \approx  \frac{a_\sigma}{(2 z)^{\Delta_{\phi}}}  \left[1 + \frac{8 b_{\mathrm{D}} u_{\mathrm{D}}}{a_\sigma} \left(\frac{z}{L}\right)^3\right], \quad z \ll L.  \label{eq:sigmascal} \end{equation}
What are the corrections to scaling to the above form? The corrections \ref{en:NR} and \ref{en:op} discussed for the bulk correlator are still present. In addition, there are corrections due to  irrelevant perturbations on the boundary: the lowest of these is the displacement ${\mathrm D}$, whose effect is to replace $z \to z+ z_0$  in Eq.~(\ref{eq:sigmascal}). Here $z_0$ is a non-universal constant with units of length. Indeed, we have the boundary OPE,  $T^{zz} \sim -\sqrt{C_\mathrm{D}} {\rm D}$, $z \to 0$, where $T^{\mu \nu}$ is the stress-tensor and $C_\mathrm{D}$ - a universal constant. Thus, perturbing the boundary with the displacement changes the action by $\delta S = z_0 \int d^2 {\bf x} \left(T^{zz}(z=0, {\bf x}) +  T^{zz}(z=L, {\bf x})\right)$, where we've assumed reflection symmetry $z \to L - z$. Then
\begin{equation} \langle \sigma(z) \rangle_{S + \delta S} = \langle \sigma(z+z_0) \rangle_{S + \delta S'} \label{eq:SSp}\end{equation}
with $\delta S' =  2 z_0 \int d^2 {\bf x} \,  T^{zz}(z=L, {\bf x})$. Here, we've moved one insertion of the stress-tensor from $z = 0$ to $z = L$. With the perturbation $\delta S'$ localized on the $z  = L$ boundary, for $z \ll L $ we can now use the OPE (\ref{eq:BOPE}) on the right-hand-side of (\ref{eq:SSp}). We will have $\langle D \rangle_{S + \delta S'} = u_{\mathrm{D}} L^{-3} (1 + O(1/L))$. Thus, including $1/z$ corrections to the second term in brackets in Eq.~(\ref{eq:sigmascal}), but not $1/L$ corrections (which are smaller in the $z \ll L$ limit), we obtain Eq.~(\ref{eq:sigmascal}) with $z \to z+z_0$. We will ignore corrections to scaling from irrelevant boundary operators above $D$: according to large-$N$ analysis, the lowest of these has $\hat{\Delta} = 5$ \cite{Padayasi}, which will lead to $\omega = 3$. Thus, we obtain the ansatz
\begin{multline}
  \langle \sigma_{\rm{lat}}(z) \rangle \approx  M_\sigma (z+z_0)^{-\Delta_{\phi}}  \Bigg[1 + B_\sigma \left(\frac{z+z_0}{L}\right)^3 \\
  + C (z+ z_0)^{-2}\Bigg]
\end{multline}
with 
 \begin{equation} M_\sigma = 2^{-\Delta_\phi} a_\sigma\sqrt{{\cal N}_{\rm bulk}/N}. \end{equation}
  Note that the replacement $z \to z + z_0$ in the $C$ term is for esthetic purposes --- we do not claim to know the corrections to such accuracy.

We next discuss the boundary two-point function $\langle \varphi^i_{\rm{lat}}({\bf x}) \varphi^i_{\rm{lat}}(0)\rangle$. The leading contribution to the boundary lattice field is from the tilt operator ${\rm t}^i$:
\begin{equation} \varphi^i_{\rm{lat}}({\bf x}) = \sqrt{{\cal N}_{\varphi}/(N-1)} (1 + c \,\Box_{\bf x}) {\rm t}^i(\bf x) + \ldots \,. \end{equation}
We neglect higher dimension boundary primaries in the above expansion: in the large-$N$ limit these start with $\hat{\Delta} = 5$ \cite{Padayasi}.
Now, we have the OPE
\begin{equation} {\rm t}^i({\bf x}) {\rm t}^i(0) = (N-1) {\bf x}^{-4} (1 + \lambda_{{\rm t t D}} {\bf x}^3 {\rm D}(0) + \ldots ) \label{ttDOPE}\end{equation}
so for $x \ll L$, we expect
\begin{equation} \langle {\rm t}^i({\bf x}) {\rm t}^i(0) \rangle \approx  (N-1) {\bf x}^{-4} \left[1 + \lambda_{{\rm t t D}} u_{{\rm D}} \left(\frac{{\bf x}}{L}\right)^3 \right]. \label{ttcorrL} \end{equation}
What are the effects of the displacement perturbation on the above correlator? We may regularize this perturbation as  $\delta S = z_0 \int d^2{\bf x} (T^{zz}({\bf x}, z = \epsilon) +T^{zz}({\bf x}, z = L-\epsilon))$ with $\epsilon \to 0^+$. Then in computing $\langle {\rm t}^i({\bf x}) {\rm t}^i(0) \rangle$ we may freely slide one of the $T^{zz}$ insertions to the other boundary so that $\delta S \to 2 z_0 \int d^2 {\bf x} \,  T^{zz}(z=L, {\bf x})$. Now using the OPE (\ref{ttDOPE})  the only change to (\ref{ttcorrL}) is a $1/L$ correction to the second term in brackets coming from a $1/L$ correction to $\langle D\rangle$. We ignore this correction below. Collecting other sources of corrections to scaling:
\begin{equation}  \langle \varphi^i_{\rm{lat}}({\bf x}) \varphi^i_{\rm{lat}}(0)\rangle = {\cal N}_{\varphi} {\bf x}^{-4} \left[1 + B_{\varphi D}  \left(\frac{{\bf x}}{L}\right)^3 + C {\bf x}^{-2}\right].\end{equation}

Finally, we discuss the bulk-boundary correlation function $\langle \varphi^i_{\rm{lat}}(0,z) \varphi^i_{\rm{lat}}(0)\rangle$. With one factor of ${\rm t}^i(0)$ already on the boundary, we may further fuse $\varphi^i(0,z)$ onto the boundary to get for $z \to 0$
\begin{equation} \varphi^i(0,z)  {\rm t}^i(0) = 2^{2 - \Delta_\varphi} (N-1) b_{\rm t} z^{-2 - \Delta_\varphi}(1 + \beta \, z^3 {\rm D}(0) + \ldots).\end{equation}
The constant in front is fixed by taking the expectation value in a semi-infinite geometry and using Eq.~(\ref{tvarphi}). We do not expect the  constant $\beta$ to be related to any of the OPE coefficients already introduced. Taking the expectation value,
\begin{multline}
  \langle \varphi^i(0,z)  {\rm t}^i(0)\rangle \\
  \approx 2^{2 - \Delta_\varphi} (N-1) b_{\rm t} z^{-2 - \Delta_\varphi}\left[1 + \beta u_{\rm D} \left(\frac{z}{L}\right)^3 \right].
\end{multline}
By a familiar argument, including corrections from the boundary displacement perturbation results in $z \to z+ z_0$ in the above expression up to a $1/L$ correction to the second term in the brackets. Finally, including other sources of corrections to scaling we obtain Eq.~(\ref{phiperp_ansatz2}) with
 \begin{equation} M_{\varphi} = \sqrt{{\cal N}_{\varphi} {\cal N}_{{\rm bulk}} (N-1)/N}\,2^{2-\Delta_\varphi} b_{\rm t}.\end{equation}

\section{The normal universality class in the Ising model}

In this section, we use the results of Refs.~\cite{PTD-10,Hasenbusch-10c} to extract the values of universal amplitudes $a_\sigma$, $b_{\rm D}$ for the normal universality class of the 3d Ising model.

We begin with $a_\sigma$. Using the notation of Ref.~\cite{PTD-10},
\beq \langle \sigma_{{\rm lat}}(z)\rangle = B_M c_{\pm} (z/\xi_{\pm,0,{\rm gap}})^{-\Delta_\sigma}, \quad\quad z \to 0. \label{phizlat}\eeq
The non-universal constant $B_M$ is defined so that in the ordered phase in infinite volume
\beq \langle \sigma_{{\rm lat}} \rangle  = B_M (-t)^{\Delta_{\sigma} \nu} \label{philatbulk} \eeq
with $t = (T-T_c)/T_c$ and $\nu$ --- the correlation length exponent. The non-universal numbers $\xi_{\pm,0,{\rm gap}}$  are defined via
\beq \xi_{\pm,{\rm gap}} = {\xi_{\pm,0,{\rm gap}}} |t|^{-\nu},\label{xit}\eeq
where $\xi_{\pm,{\rm gap}}$ is the correlation length governing exponential decay of $\sigma$ correlation function at large distance in the high and low temperature phases. The constants $c_{\pm}$ in Eq.~(\ref{phizlat}) are universal. Now, let $\sigma(x)$ be a continuum field normalized in the bulk at the critical point to $\langle \sigma(x) \sigma(0) \rangle = x^{-2 \Delta_{\sigma}}$. We have in the ordered phase in the bulk
\beq \langle \sigma \rangle = u \xi_{-, {\rm gap}}^{-\Delta_{\sigma}}, \label{phicontbulk} \eeq
where $u$ is a universal constant that we will be able to relate to other known universal constants below. Now, with $\sigma_{{\rm lat}} = \sqrt{N_b} \sigma$, combining (\ref{philatbulk}), (\ref{phicontbulk}), (\ref{xit}), we have 
\beq B_M = \sqrt{N_b} u \xi_{-,0,{\rm gap}}^{-\Delta_\sigma}. \label{B}\eeq
At the critical point for the normal class,
\beq \langle \sigma(z) \rangle = \frac{a_{\sigma}}{(2z)^{\Delta_\sigma}}.\eeq
Matching this to (\ref{phizlat}) and using (\ref{B}), 
\beq a_\sigma = 2^{\Delta_\sigma} u \,c_-.\eeq
Now, from (\ref{phizlat}),   $c_+/c_- =  U_{\xi, {\rm gap}}^{-\Delta_{\sigma}}$, where the universal constant $U_{\xi,{\rm gap}} = \frac{\xi_{+,0,{\rm gap}}}{\xi_{-,0,{\rm gap}}}$,
\beq a_\sigma = (2 U_{\xi,{\rm gap}})^{\Delta_\sigma} u c_+ .\label{asigmaucp}\eeq

We next relate $u$ to known universal constants. Unless otherwise stated, we use the notation of Ref.~\cite{PV-02}.
 In the high-temperature phase,
 \begin{multline}
   G_{{\rm lat}}(q) = \int d^d x e^{-i \vec{q} \cdot \vec{x}} \langle \sigma_{{\rm lat}}(x) \sigma_{{\rm lat}}(0)\rangle \\
   \stackrel{q\xi_{+,2} \gg 1}{\to} \frac{G_{{\rm lat}}(q = 0) A^+_1}{(q \xi_{+,2})^{d-2 \Delta_\sigma}},
 \end{multline}
where $\xi_{+,2}$ is the second moment correlation length and $A^+_1$ is a universal number. Further,
\beq G_{{\rm lat}}(q = 0) = C^+ t^{-\nu(d - 2 \Delta_{\sigma})} = C^+ (\xi_{+,{\rm gap}}/\xi_{+,0,{\rm gap}})^{d-2\Delta_{\sigma}}. \eeq 
with $C^+$ -- a non-universal coefficient. Thus,
\beq G_{{\rm lat}}(q) = C^+ A^+_1 \left(\frac{\xi_{+,{\rm gap}}}{\xi_{+,0,{\rm gap}}} \frac{1}{q \xi_{+,2}}\right)^{d-2\Delta_\sigma}, \quad q\xi_{+,2} \gg 1.\eeq
Matching to the bulk normalization of the two-point function of the  continuum field $\sigma(x)$, we obtain
\beq N_b = \frac{C^+ A^+_1 \Gamma(\Delta_\sigma)}{\pi^{d/2} \Gamma(d/2-\Delta_\sigma)} \left(\frac{\xi_{+,{\rm gap}}}{2 \xi_{+,2} \xi_{+,0,{\rm gap}}} \right)^{d-2\Delta_\sigma}.\eeq
Then from Eq.~(\ref{B}),
\beq u = \pi^{d/4} 2^{d/2-\Delta_\sigma} \left(\frac{\Gamma(d/2-\Delta_\sigma)Q_c}{\Gamma(\Delta_\sigma) A^+_1}\right)^{1/2} \left(\frac{Q^-_\xi}{U_{\xi}}\right)^{\Delta_\sigma}\eeq
with 
\beq Q_c = \frac{B^2_M \xi^d_{+,0,2}}{C^+}, \quad U_\xi = \frac{\xi_{+,0,2}}{\xi_{-,0,2}}, \quad Q^-_\xi = \frac{\xi_{-,0,{\rm gap}}}{\xi_{-,0,2}}\eeq
where
\beq \xi_{\pm,2} = \xi_{\pm, 0,2} |t|^{-\nu}.\eeq
Combining this with (\ref{asigmaucp}),
\beq a_\sigma =(4 \pi)^{d/4} \left(\frac{\Gamma(d/2-\Delta_\sigma) Q_c}{\Gamma(\Delta_\sigma) A^+_1}\right)^{1/2}  (Q^+_{\xi})^{\Delta_\sigma} c_+\eeq
with
\beq Q^+_\xi = \frac{\xi_{+,0,{\rm gap}}}{\xi_{+,0,2}}.\eeq
Ref.~\cite{PTD-10} reports $c_+ = 0.844(6)$. We use $\Delta_\sigma = 0.5181489(10)$ \cite{SDIsingSpectrum}, $Q_c = 0.3293(2)$ {\cite{Hasenbusch-10b}},
$Q_\xi^+ = 1.000200(3)$ \cite{CPRV-02}.
The greatest uncertainty is on the value of $A^+_1$. The value obtained by the $\epsilon$-expansion at 3-loops is $A^+_1 = 0.922$ \cite{Bray}. This appears consistent with experimental results  $A^+_1 = 0.96(4)$, $A^+_1 = 0.95(4)$ \cite{Chang} (two different parameterizations were used), $A^+_1 = 0.91-0.94$ \cite{Damay}, $A^+_1 = 0.915(21)$ \cite{Damay98}, as well as roughly consistent with Monte Carlo \cite{VicariA1p}. We will use the value $A^+_1 \approx 0.915(21)$ from the experimental paper \cite{Damay98}. The value of $A^+_1$ has the greatest uncertainty out of all the ingredients in the calculation.
Combining everything together, we obtain
\beq a_\sigma = 2.60(5).\eeq

The truncated bootstrap result is $a_\sigma = 2.5994(8)$ \cite{GLMR-15}, which is spot on.

We next consider the OPE coefficient $b_{\rm D}$:
\bea T_{zz}({\rm{x}}, z) &\sim& -\sqrt{C_{{\rm D}}} {\rm D}({\rm{x}}) + \ldots, \nonumber\\
\sigma({\rm{x}}, z) &\sim& \frac{a_{\sigma}}{(2 z)^{\Delta_{\sigma}}} + b_{\rm D} (2z)^{d- \Delta_\sigma}  {\rm D}({\rm{x}}) + \ldots\,.
\eea
Here and below $T^{\mu \nu} = -\frac{2}{\sqrt{g}} \frac{\delta S}{\delta g_{\mu \nu}}$ is the standard stress-energy tensor.
As shown by Cardy \cite{Cardy-90},
\beq \frac{b_{\rm D}}{\Delta_\sigma a_\sigma }  = \frac{1}{S_d \sqrt{C_{\rm D}}} \label{mua} \eeq  
with $S_d  = \frac{2 \pi^{d/2}}{\Gamma(d/2)}$. (Note that our normalization and sign conventions are different from those in Ref.~\cite{Cardy-90}).

 Taking the expectation value in a film geometry of thickness $L$, we have
\bea \langle T_{zz}(z) \rangle &=& - \sqrt{C_{\rm D}} \langle {\rm D} \rangle,  \nonumber\\
\langle \sigma(z) \rangle &=& \frac{a_{\sigma}}{(2 z)^{\Delta_{\sigma}}} \left(1 +  \frac{b_{\rm D}}{a_{\sigma}}    (2 z)^{d} \langle {\rm D} \rangle + \ldots \right) \nonumber \\
&=& \frac{a_\sigma}{(2 z)^{\Delta_{\sigma}}} \left(1 + B_{\sigma} \left(\frac z L\right)^d + \ldots \right)\eea
with 
\beq B_{\sigma} = \frac{\Delta_{\sigma}}{S_d \sqrt{C}_{\rm D}} (2 L)^d \langle {\rm D}\rangle. \eeq

Finally, with the free energy $F = - \log Z$, where  $Z$ is  the partition function,  let the universal contribution to the free energy of the film per unit surface area, $F/L^{d-1}_{\parallel}$, be
\beq f = \frac{A}{L^{d-1}} \eeq
with $A$ - a universal coefficient. Then 
\beq \frac{d f}{d L} = -\langle T_{zz}\rangle,\eeq
i.e.
\beq A = -\frac{\sqrt{C_{\rm D}} \langle {\rm D} \rangle L^d }{d-1}, \eeq
so
\beq \frac{B_\sigma}{A} = - \frac{2^d (d-1) \Delta_{\sigma}}{S_d C_{\rm D}}. \eeq
Now, for $++$ boundary conditions on the two surfaces of the slab, Ref.~\cite{PTD-10} gives $B_{\sigma} = 1.40(1)$, while Ref.~\cite{Hasenbusch-10c} gives $(d-1) A = -0.820(15)$, so
\beq C_{\rm D} = 0.193(5), \quad b_{\rm D} = 0.244(8).\eeq
In comparison, truncated bootstrap gives $b_{\rm D} = 0.25064(6)$. Thus,  MC and truncated bootstrap values of $b_{\rm D}$  agree within error-bars.

\section{Fits of statistically correlated data}
\label{app:correlated_fits}
In this work we determine the amplitudes of various one- and two-point functions by fitting the space and size dependence of the MC results. The determination of the best fit parameters is done with the $\chi^2$ minimization.
In every fit, all points at a given lattice size originate from the same MC run, hence they are statistically correlated. Taking such a correlation into account is essential in order to reliably estimate the uncertainty of the fitted parameters. In principle, the fit procedure can be naturally adapted to take a statistical covariance of data into account, by employing the inverse of the covariance matrix in the formula for $\chi^2$. Although resampling methods like the Jackknife \cite{Young_notes} used here allow to estimate the covariance matrix, it is well known that its inversion is numerically unstable, and can lead to wrong results in the final estimates \cite{Michael-94,Seibert-94}. Therefore, as we did in Ref.~\cite{PTD-10}, to correctly estimate the uncertainty of the fit parameters, we resort to using the Jackknife resampling method in the computation of the $\chi^2$ itself.
Specifically, given a set of MC estimates $\{O_i\}$, we consider the Jackknife bin estimate $\{O^J_i\}$, calculated using all MC data except the discarded MC bin $J$. Given a fit to a function $f(\{O_i\}, \{p_i\})$, depending on parameters $\{p_i\}$ that we want to determine, we minimize
\begin{equation}
  \left(\chi^2\right)^J = \sum_i \left(\frac{O^J_i - f(\{O_i^J\}, \{p_i\})}{\sigma_i^2}\right)^2,
\label{chi_square}
\end{equation}
where $\sigma_i$ is an estimate of the uncertainty of $O_i$, computed with the usual Jackknife \cite{Young_notes}.
For $J=1\ldots N_{\rm bin}$, the minimization of Eq.~(\ref{chi_square}) delivers $N_{\rm bin}$ Jackknife estimates of $\{p_i^J\}$, from which the uncertainty is computed with the usual Jackknife formula. In the analysis done below, we have chosen $N_{\rm bin}=100$.
We have further employed the fitted values $\{p_i\}$ obtained using the mean-value estimates of $\{O_i\}$ to subtract the Jackknife bias $\propto 1/N_{\rm bin}$ in the estimates of $\{p_i\}$; as expected, such a correction is negligible with respect to the statistical error bar.
We have also used the dispersion over the Jackknife bins of the minimum $(\chi^2)^J$ to estimate the standard deviation of the minimum $\chi^2$.

\section{Fits of one- and two-point functions}
\label{app:fits}
\subsection{$N=2$}
\label{app:fits:2}
\subsubsection{Bulk correlations}
\label{app:fits:2:bulk}

\begin{table}[b]
  \begin{ruledtabular}
    \begin{tabular}{llcccc}
    $(x/L)_{\rm max}$ & $L_{\rm min}$ & ${\cal N}_{\rm bulk}$ & $B_\epsilon$ & $C$ & $\chi^2/{\rm d.o.f.}$ \\
      \hline
      &   32    &  0.281034(69)   &  2.8017(39)   &  0.3457(30)   &  10.3(2.3)   \\
      &   48    &  0.281004(68)   &  2.8055(44)   &  0.3478(29)   &  9.7(2.4)   \\
$1/4$ &   64    &  0.280957(64)   &  2.8126(50)   &  0.3513(27)   &  7.6(2.3)   \\
      &   96    &  0.280922(57)   &  2.8206(63)   &  0.3541(23)   &  5.4(2.0)   \\
      &  128    &  0.280954(54)   &  2.8190(75)   &  0.3532(21)   &  5.2(2.4)   \\
\hline
      &   32    &  0.281180(75)   &  2.7702(58)   &  0.3404(31)   &  12.8(2.6)   \\
      &   48    &  0.281164(76)   &  2.7728(64)   &  0.3412(32)   &  12.9(2.7)   \\
$1/8$ &   64    &  0.281090(76)   &  2.7870(76)   &  0.3452(32)   &  11.8(3.0)   \\
      &   96    &  0.280991(72)   &  2.8081(93)   &  0.3506(30)   &  9.6(3.0)   \\
      &  128    &  0.280960(66)   &  2.819(10)   &  0.3527(27)   &  9.5(3.4)   \\
    \end{tabular}
  \end{ruledtabular}
  \caption{Fits of the bulk two-point function for $N=2$ to Eq.~(\ref{phi_bulk_ansatz}) as a function of the minimum lattice size $L_{\rm min}$, and of the maximum value of $(x/L)$ taken into account.
    For all fits we consider MC data for $x\ge x_{\rm min} = 4$.
    The quoted error bars are the sum of the statistical uncertainty originating from the fit, and the dependence of the results on varying $\Delta_\phi=0.519088(22)$, $\Delta_\epsilon=1.51136(22)$ \cite{CLLPSDSV-19} within one error bar.}
  \label{fits_bulk_xy_xmin4}
\end{table}

\begin{table}[h]
  \begin{ruledtabular}
    \begin{tabular}{llcccc}
    $(x/L)_{\rm max}$ & $L_{\rm min}$ & ${\cal N}_{\rm bulk}$ & $B_\epsilon$ & $C$ & $\chi^2/{\rm d.o.f.}$ \\
      \hline
      &   32    &  0.28132(11)   &  2.7936(44)   &  0.291(11)   &  4.4(1.4)   \\
      &   48    &  0.28131(11)   &  2.7941(49)   &  0.292(12)   &  4.4(1.4)   \\
$1/4$ &   64    &  0.28126(11)   &  2.7984(56)   &  0.301(11)   &  3.8(1.6)   \\
      &   96    &  0.281181(99)   &  2.8060(67)   &  0.3132(98)   &  2.8(1.5)   \\
      &  128    &  0.281216(89)   &  2.8025(75)   &  0.3119(86)   &  2.6(1.9)   \\
      \hline
      &   48    &  0.28155(12)   &  2.7567(82)   &  0.277(11)   &  1.12(99)   \\
$1/8$ &   64    &  0.28154(13)   &  2.7580(99)   &  0.278(12)   &  1.1(1.0)   \\
      &   96    &  0.28148(14)   &  2.765(13)   &  0.284(13)   &  1.0(1.2)   \\
      &  128    &  0.28151(14)   &  2.761(15)   &  0.281(14)   &  1.2(1.4)   \\
    \end{tabular}
  \end{ruledtabular}
  \caption{Same as Table \ref{fits_bulk_xy_xmin4} for $x_{\rm min}=6$.}
  \label{fits_bulk_xy_xmin6}
\end{table}

As mentioned in the main text,
we simulated the $\phi^4$ model with periodic BCs at the critical point, for lattice sizes $L=32-192$.
The rescaled two-point function shown in
Fig.~\ref{plots_bulk}(a) suggests that the universal part of the correlations is found for $x\gtrsim 4$.
In Tables \ref{fits_bulk_xy_xmin4} and \ref{fits_bulk_xy_xmin6}, we show results of fits to Eq.~(\ref{phi_bulk_ansatz}), as a function of a minimum value $x_{\rm min}$ of $x$, the maximum value $(x/L)_{\rm max}$ of $x/L$ and the minimum lattice size $L$ taken into account.
In order to evaluate the possible systematic error due to the uncertainty in the critical exponents $\Delta_\phi$ and $\Delta_\epsilon$, in the fits we vary the value of $\Delta_\phi=0.519088(22)$ and $\Delta_\epsilon=1.51136(22)$ \cite{CLLPSDSV-19} within one error bar. The quoted error is the sum (in absolute value) of the statistical error bar stemming from the minimization of the $\chi^2/{\rm d.o.f.}$ (d.o.f. denotes the degrees of freedom), and the variation of the fitted value due to the uncertainty in $\Delta_\phi$ and $\Delta_\epsilon$.
Within the high accuracy of our MC data, the last contribution is not completely negligible, but it is typically smaller, or of the same order, as the statistical error bars.
Fits for $x_{\rm min}=4$ (Table \ref{fits_bulk_xy_xmin4}) exhibit a large $\chi^2/{\rm d.o.f.}$, which does not improve when $(x/L)_{\rm max}$ is decreased to $1/8$. The $\chi^2/{\rm d.o.f.}$ is significantly improved when $x_{\rm min}=6$ (Table \ref{fits_bulk_xy_xmin6}), but still somewhat large for $(x/L)_{\rm max} = 1/4$.
A good $\chi^2/{\rm d.o.f.}$ is eventually found for $x_{\rm min}=6$ and $(x/L)_{\rm max} = 1/8$. In this case the results are also very stable on increasing $L_{\rm min}$ and allow us to determine the estimates in Eq.~(\ref{Nbulk_xy}).

\subsubsection{Surface correlations}
\label{app:fits:2:surface}
\begin{table}[t]
  \begin{ruledtabular}
    \begin{tabular}{llccc}
    $({\bf x}/L)_{\rm max}$ & $L_{\rm min}$ & ${\cal N}_\varphi$ & $C$ & $\chi^2/{\rm d.o.f.}$ \\
      \hline
            & 32   &   0.33938(42)   &   0.192(20)   &   4.72(66)   \\
            & 48   &   0.33935(42)   &   0.194(20)   &   4.74(68)   \\
      $1/8$ & 64   &   0.33924(43)   &   0.199(20)   &   4.94(72)   \\
            & 96   &   0.33906(47)   &   0.207(22)   &   5.02(74)   \\
            & 128  &   0.33929(56)   &   0.196(26)   &   5.02(77)   \\
      \hline
             & 48   &   0.33950(45)   &   0.187(21)   &   7.7(1.2)   \\
      $1/12$ & 64   &   0.33951(45)   &   0.186(21)   &   8.0(1.2)   \\
             & 96   &   0.33917(47)   &   0.202(23)   &   8.0(1.2)   \\
             & 128  &   0.33937(55)   &   0.192(26)   &   8.0(1.3)   \\
    \end{tabular}
  \end{ruledtabular}
  \caption{Fits of the surface two-point function for $N=2$ to Eq.~(\ref{phi_ansatz}) as a function of the minimum lattice size $L_{\rm min}$, and of the maximum value of $({\bf x}/L)$ taken into account. For all fits we consider MC data for ${\bf x}\ge {\bf x}_{\rm min} = 4$.}
  \label{fits_phi_xy_xmin4}
\end{table}

\begin{table}[b]
  \begin{ruledtabular}
    \begin{tabular}{llccc}
    $({\bf x}/L)_{\rm max}$ & $L_{\rm min}$ & ${\cal N}_{\rm bulk}$ & $C$ & $\chi^2/{\rm d.o.f.}$ \\
      \hline
            & 48   &   0.3284(19)   &   1.20(22)   &   0.82(32)   \\
      $1/8$ & 64   &   0.3285(19)   &   1.17(22)   &   0.80(32)   \\
            & 96   &   0.3282(19)   &   1.20(22)   &   0.83(32)   \\
            & 128   &   0.3269(21)   &   1.38(26)   &   0.80(34)   \\
      \hline
      $1/12$ & 96   &   0.3278(19)   &   1.25(23)   &   0.74(43)   \\
             & 128   &   0.3272(21)   &   1.36(26)   &   0.77(45)   \\
    \end{tabular}
  \end{ruledtabular}
  \caption{Same as Table \ref{fits_phi_xy_xmin4} for ${\bf x}_{\rm min}=6$.}
  \label{fits_phi_xy_xmin6}
\end{table}

As discussed in the main text, in order to implement the normal UC, we simulated the $\phi^4$ model at the critical point with open BCs and a symmetry-breaking surface field $h_s=1.5\beta_s$
for lattice sizes $L=32-192$.
In Fig.~\ref{plots_surface}(a) we show the two-point function of the field component $\varphi$ along the surface, rescaled to the expected large-distance decay exponent $4$.
We observe a quick increase of the error bars on increasing the distance ${\bf x}$. This is due to the relatively fast decay ${\bf x}^{-4}$ of the correlations, which requires
an increasing computational effort to estimate the amplitude of the correlations $\<\varphi({\bf x})\varphi(0)\>{\bf x}^4$.
In consideration of the fact that the size dependence in Fig.~\ref{plots_surface}(a) appears to be rather small,
we have fitted the MC data to Eq.~(\ref{phi_ansatz}), where finite-size corrections are neglected.
Fig.~\ref{plots_surface}(a) suggests that the universal part of the correlations is found for ${\bf x}\gtrsim 4$. In Tables \ref{fits_phi_xy_xmin4} and \ref{fits_phi_xy_xmin6} we report fit results, as a function of a minimum value ${\bf x}_{\rm min}$ of ${\bf x}$, the maximum value $({\bf x}/L)_{\rm max}$ of ${\bf x}/L$ and the minimum lattice size $L_{\rm min}$ taken into account.
Due to the aforementioned rapid increase of statistical error bars, compared to the fits to Eq.~(\ref{phi_bulk_ansatz}) we consider in this case smaller values of $({\bf x}/L)_{\rm max}$.
Fits for ${\bf x}_{\rm min} = 4$ reported in Table \ref{fits_phi_xy_xmin4} exhibit a large value of $\chi^2/{\rm d.o.f.}$.
A good value of $\chi^2/{\rm d.o.f}$ is instead found for  ${\bf x}_{\rm min} = 6$ (Table \ref{fits_phi_xy_xmin6}), and the corresponding fitted values are stable on decreasing $({\bf x}/L)_{\rm max}$ and increasing $L_{\rm min}$.
Accordingly,
a conservative estimate of ${\cal N}_\varphi$, compatible with all results of Table \ref{fits_phi_xy_xmin6} is given in Eq.~(\ref{Nphi_xy}).

\subsubsection{Magnetization profile}
\label{app:fits:2:profile}
\begin{table*}[t]
  \begin{ruledtabular}
    \begin{tabular}{llcccc}
      $(z/L)_{\rm max}$ & $L_{\rm min}$ & $M_\sigma$ & $B_\sigma$ & $z_0$ & $\chi^2/{\rm d.o.f.}$ \\
      \hline
      &   32    &  0.754574(93)   &  1.2150(42)   &  1.03263(92)   &  9.0(1.5)   \\
      &   48    &  0.754534(94)   &  1.2258(57)   &  1.03196(93)   &  8.3(1.4)   \\
$1/4$ &   64    &  0.754514(94)   &  1.2304(73)   &  1.03156(93)   &  8.5(1.5)   \\
      &   96    &  0.754513(93)   &  1.2240(89)   &  1.03145(91)   &  8.8(1.6)   \\
      &  128    &  0.754528(96)   &  1.203(11)   &  1.03157(94)   &  8.9(1.6)   \\
      \hline
      &   32    &  0.754685(91)   &  1.126(15)   &  1.03394(86)   &  11.6(2.2)   \\
      &   48    &  0.754680(95)   &  1.130(19)   &  1.03388(91)   &  11.8(2.2)   \\
$1/8$ &   64    &  0.754689(98)   &  1.115(25)   &  1.03397(96)   &  12.0(2.3)   \\
      &   96    &  0.75477(10)   &  1.012(37)   &  1.0351(10)   &  10.3(2.7)   \\
      &  128    &  0.754807(99)   &  0.925(43)   &  1.03557(98)   &  10.2(2.8)   \\
    \end{tabular}
  \end{ruledtabular}
  \caption{Fits of the order-parameter profile for $N=2$ to Eq.~(\ref{sigma_ansatz}) as a function of the minimum lattice size $L_{\rm min}$, and of the maximum value of $(z/L)$ taken into account.
    For all fits we consider MC data for $z\ge z_{\rm min} = 4$.
    The quoted error bars are the sum of the statistical uncertainty originating from the fit, and the dependence of the results on varying $\Delta_\phi=0.519088(22)$ \cite{CLLPSDSV-19} within one error bar.}
  \label{fits_sigma_xy_xmin4}
\end{table*}

\begin{table*}
  \begin{ruledtabular}
    \begin{tabular}{llcccc}
      $(z/L)_{\rm max}$ & $L_{\rm min}$ & $M_\sigma$ & $B_\sigma$ & $z_0$ & $\chi^2/{\rm d.o.f.}$ \\
      \hline
      &   32    &  0.75428(12)   &  1.2319(44)   &  1.0251(17)   &  4.29(88)   \\
      &   48    &  0.75419(12)   &  1.2478(61)   &  1.0232(18)   &  2.68(72)   \\
$1/4$ &   64    &  0.75412(13)   &  1.2613(79)   &  1.0216(18)   &  1.71(58)   \\
      &   96    &  0.75410(12)   &  1.2630(95)   &  1.0211(18)   &  1.75(62)   \\
      &  128    &  0.75412(12)   &  1.249(12)   &  1.0212(17)   &  1.31(55)   \\
      \hline
      &   48    &  0.75430(12)   &  1.171(20)   &  1.0247(17)   &  1.73(80)   \\
$1/8$ &   64    &  0.75427(13)   &  1.185(27)   &  1.0243(18)   &  1.66(76)   \\
      &   96    &  0.75430(15)   &  1.161(45)   &  1.0248(22)   &  1.55(88)   \\
      &  128    &  0.75431(15)   &  1.139(54)   &  1.0247(22)   &  1.4(1.0)   \\
    \end{tabular}
  \end{ruledtabular}
  \caption{Same as Table \ref{fits_sigma_xy_xmin4} for $z_{\rm min} = 6$.}
  \label{fits_sigma_xy_xmin6}
\end{table*}

\begin{table*}
  \begin{ruledtabular}
    \begin{tabular}{llcccc}
      $(z/L)_{\rm max}$ & $L_{\rm min}$ & $M_\sigma$ & $B_\sigma$ & $z_0$ & $\chi^2/{\rm d.o.f.}$ \\
      \hline
      &   32    &  0.75413(15)   &  1.2464(54)   &  1.0209(28)   &  2.75(65)   \\
      &   48    &  0.75406(16)   &  1.2560(66)   &  1.0191(30)   &  1.89(58)   \\
$1/4$ &   64    &  0.75394(16)   &  1.2720(88)   &  1.0159(31)   &  0.81(37)   \\
      &   96    &  0.75389(16)   &  1.279(11)   &  1.0144(32)   &  0.70(35)   \\
      &  128    &  0.75392(16)   &  1.268(13)   &  1.0146(31)   &  0.32(22)   \\
      \hline
      &   64    &  0.75409(17)   &  1.203(31)   &  1.0187(31)   &  0.34(31)   \\
$1/8$ &   96    &  0.75410(20)   &  1.198(49)   &  1.0189(38)   &  0.34(34)   \\
      &  128    &  0.75407(22)   &  1.203(66)   &  1.0181(43)   &  0.23(34)   \\
    \end{tabular}
  \end{ruledtabular}
  \caption{Same as Table \ref{fits_sigma_xy_xmin4} for $z_{\rm min} = 8$.}
  \label{fits_sigma_xy_xmin8}
\end{table*}

The magnetization profile $\langle \sigma(z)\rangle$ shown in Fig.~\ref{plots_surface}(c), for a surface field $h_s=1.5\beta_s$, and lattice sizes $L=32-192$, displays clearly scaling corrections and finite-size effects.
In line with the discussion of the scaling forms, we have fitted the MC to Eq.~(\ref{sigma_ansatz}), where we take into account the leading scaling correction $\propto z^{-1}$ and the leadinig finite-size term.
In Tables \ref{fits_sigma_xy_xmin4}-\ref{fits_sigma_xy_xmin8} we report results of fits to Eq.~(\ref{sigma_ansatz}), as a function of a minimum value $z_{\rm min}$ of $z$, the maximum value $(z/L)_{\rm max}$ of $z/L$ and the minimum lattice size $L_{\rm min}$ taken into account.
The impact of varying $\Delta_\phi=0.519088(22)$ \cite{CLLPSDSV-19} within one error bar mainly affects the precision of $M_\sigma$.
Fits for $z_{\rm min}=4$ (Table \ref{fits_sigma_xy_xmin4}) exhibit a large $\chi^2/{\rm d.o.f.}$, which for $z_{\rm min}=6$ (Table \ref{fits_sigma_xy_xmin6}) is substantially reduced, and is compatible with $1$ within one estimated standard deviation.
As a further check, in Table \ref{fits_sigma_xy_xmin8} we consider fits for $z_{\rm min}=8$: in this case we have $\chi^2/{\rm d.o.f.} < 1$ for $L_{\rm min}\ge 64$.
The fit results differ only slightly from those of Table \ref{fits_sigma_xy_xmin6}.
Judging conservatively the variation in the fit results, we arrive to the estimates of Eq.~(\ref{sigma_xy}).

\subsubsection{Surface-bulk correlations}
\label{app:fits:2:surface-bulk}
The rescaled surface-bulk correlation function $\<\varphi(0)\varphi(0,z)\>$
shown in Fig.~\ref{plots_surface}(e) clearly displays scaling corrections, as well as a finite-size dependence.
Compared to the order-parameter profile, Fig.~\ref{plots_surface}(c), in Fig.~\ref{plots_surface}(e) scaling corrections appear optically larger, whereas finite-size corrections are smaller.
We have attempted to fit the surface-bulk correlation to
\begin{equation}
  \<\varphi(0)\varphi(0,z)\> = M_\varphi (z + z_0)^{-2-\Delta_\phi}\left[1+ B_\varphi \left(\frac{z + z_0}{L}\right)^3\right],
  \label{phiperp_ansatz}
\end{equation}
where we have included the leading corrections only.
Fits to Eq.~(\ref{phiperp_ansatz}) (not reported here) give rather unstable results and a large value of $\chi^2/{\rm d.o.f.}$.
In particular, we observe a rather poor determination of the finite-size correction $B_\varphi$, and a fitted value of $z_0$ which deviates from the result of Eq.~(\ref{sigma_xy}).
On increasing the value of $z_{\rm min}$ the fitted value of $z_0$ slowly increases, though remaining incompatible with Eq.~(\ref{sigma_xy}).
At the same time, the minimum $\chi^2/{\rm d.o.f.}$ reduces, although it remains somewhat large $\chi^2/{\rm d.o.f.} \sim 1.5$ for $z_{\rm min}=8$.
In line with the observations above on Fig.~\ref{plots_surface}(e),
we interpret these observations as a sign that subleading scaling corrections are numerically relevant, and that the finite-size term is small.
Guided by these consideration, we fit the MC data to Eq.~(\ref{phiperp_ansatz2})
using the result for $z_0$ given in Eq.~(\ref{sigma_xy}), and varying its value within one error bar quoted there.
Here, the uncertainty of $\Delta_\phi=0.519088(22)$ \cite{CLLPSDSV-19} gives a negligible contribution to the error bars of the fitted parameters.

\begin{table*}
  \begin{ruledtabular}
    \begin{tabular}{llcccc}
      $(z/L)_{\rm max}$ & $L_{\rm min}$ & $M_\varphi$ & $B_\varphi$ & $C$ & $\chi^2/{\rm d.o.f.}$ \\
      \hline
      &   32    &  0.31163(36)   &&  3.182(72)   &  11.9(1.2)   \\
      &   48    &  0.31169(36)   &&  3.178(72)   &  10.5(1.2)   \\
$1/4$ &   64    &  0.31177(36)   &&  3.172(73)   &  10.2(1.2)   \\
      &   96    &  0.31188(36)   &&  3.162(72)   &  10.3(1.3)   \\
      &  128    &  0.31196(37)   &&  3.154(74)   &  11.6(1.5)   \\
      \hline
      &   32    &  0.31186(36)   &  -0.451(52)   &  3.162(72)   &  10.4(1.2)   \\
      &   48    &  0.31178(37)   &  -0.203(91)   &  3.170(72)   &  10.4(1.2)   \\
$1/4$ &   64    &  0.31171(37)   &  0.16(15)   &  3.177(71)   &  10.3(1.2)   \\
      &   96    &  0.31166(38)   &  0.93(27)   &  3.182(72)   &  9.7(1.2)   \\
      &  128    &  0.31172(39)   &  1.63(42)   &  3.177(73)   &  10.7(1.3)   \\
      \hline
      &   32    &  0.31169(36)   &&  3.177(71)   &  22.5(2.5)   \\
      &   48    &  0.31168(36)   &&  3.179(72)   &  22.5(2.5)   \\
$1/8$ &   64    &  0.31172(36)   &&  3.176(72)   &  21.9(2.5)   \\
      &   96    &  0.31184(36)   &&  3.167(72)   &  21.7(2.7)   \\
      &  128    &  0.31193(37)   &&  3.158(74)   &  24.3(3.3)   \\
      \hline
      &   32    &  0.31181(37)   &  -0.58(10)   &  3.169(71)   &  21.9(2.5)   \\
      &   48    &  0.31181(37)   &  -0.59(13)   &  3.169(72)   &  22.3(2.6)   \\
$1/8$ &   64    &  0.31168(38)   &  0.19(24)   &  3.180(72)   &  22.4(2.6)   \\
      &   96    &  0.31133(41)   &  3.06(52)   &  3.211(73)   &  19.2(2.5)   \\
      &  128    &  0.31123(43)   &  6.11(87)   &  3.219(74)   &  19.1(2.6)   \\
    \end{tabular}
  \end{ruledtabular}
  \caption{Fits of the surface-bulk correlations for $N=2$ to Eq.~(\ref{phiperp_ansatz2}) as a function of the minimum lattice size $L_{\rm min}$, and of the maximum value of $(z/L)$ taken into account.
    For all fits we use $z_0=1.018(6)$ (Eq.~(\ref{sigma_xy})) and consider MC data for $z\ge z_{\rm min} = 4$.
    The quoted error bars are the sum of the uncertainty stemming from the fit, and the spread of the results on varying $z_0$ within one error bar quoted in Eq.~(\ref{sigma_xy}).
    An absent $B_\varphi$ indicates that we fixed it to $B_\varphi=0$, see main text.}
  \label{fits_phiperp_xy_zmin4}
\end{table*}

\begin{table*}
  \begin{ruledtabular}
    \begin{tabular}{llcccc}
      $(z/L)_{\rm max}$ & $L_{\rm min}$ & $M_\varphi$ & $B_\varphi$ & $C$ & $\chi^2/{\rm d.o.f.}$ \\
      \hline
      &   32    &  0.31368(38)   &&  2.78(11)   &  3.39(48)   \\
      &   48    &  0.31369(38)   &&  2.78(11)   &  2.28(42)   \\
$1/4$ &   64    &  0.31378(38)   &&  2.77(11)   &  1.67(38)   \\
      &   96    &  0.31395(37)   &&  2.75(11)   &  1.20(30)   \\
      &  128    &  0.31421(40)   &&  2.71(11)   &  1.06(31)   \\
      \hline
      &   32    &  0.31426(38)   &  -0.615(52)   &  2.70(11)   &  1.02(27)   \\
      &   48    &  0.31428(38)   &  -0.638(81)   &  2.69(10)   &  1.02(27)   \\
$1/4$ &   64    &  0.31428(39)   &  -0.63(14)   &  2.70(10)   &  1.05(28)   \\
      &   96    &  0.31421(42)   &  -0.47(25)   &  2.71(11)   &  1.06(29)   \\
      &  128    &  0.31436(43)   &  -0.39(39)   &  2.68(11)   &  1.02(31)   \\
      \hline
      &   48    &  0.31407(38)   &&  2.72(10)   &  2.12(57)   \\
$1/8$ &   64    &  0.31403(38)   &&  2.73(10)   &  1.93(58)   \\
      &   96    &  0.31406(38)   &&  2.73(10)   &  1.37(53)   \\
      &  128    &  0.31426(40)   &&  2.70(11)   &  0.96(49)   \\
      \hline
      &   48    &  0.31434(39)   &  -0.91(15)   &  2.69(10)   &  1.11(43)   \\
$1/8$ &   64    &  0.31443(40)   &  -1.11(23)   &  2.68(10)   &  1.07(41)   \\
      &   96    &  0.31440(47)   &  -1.01(53)   &  2.68(11)   &  1.14(45)   \\
      &  128    &  0.31447(52)   &  -0.83(87)   &  2.66(12)   &  0.92(48)   \\
    \end{tabular}
  \end{ruledtabular}
  \caption{Same as Table \ref{fits_phiperp_xy_zmin4} for $z_{\rm min} = 6$.}
  \label{fits_phiperp_xy_zmin6}
\end{table*}

\begin{table*}
  \begin{ruledtabular}
    \begin{tabular}{llcccc}
      $(z/L)_{\rm max}$ & $L_{\rm min}$ & $M_\varphi$ & $B_\varphi$ & $C$ & $\chi^2/{\rm d.o.f.}$ \\
      \hline
      &   32    &  0.31378(45)   &&  2.75(18)   &  2.29(41)   \\
      &   48    &  0.31373(44)   &&  2.78(18)   &  1.96(39)   \\
$1/4$ &   64    &  0.31378(45)   &&  2.79(18)   &  1.47(36)   \\
      &   96    &  0.31394(46)   &&  2.77(18)   &  1.07(29)   \\
      &  128    &  0.31423(50)   &&  2.71(19)   &  1.00(31)   \\
      \hline
      &   32    &  0.31457(46)   &  -0.592(64)   &  2.59(18)   &  0.84(23)   \\
      &   48    &  0.31466(46)   &  -0.645(84)   &  2.57(18)   &  0.82(23)   \\
$1/4$ &   64    &  0.31466(49)   &  -0.67(14)   &  2.57(18)   &  0.84(24)   \\
      &   96    &  0.31453(52)   &  -0.56(24)   &  2.62(18)   &  0.90(26)   \\
      &  128    &  0.31456(55)   &  -0.46(38)   &  2.62(19)   &  0.96(30)   \\
      \hline
      &   64    &  0.31436(46)   &&  2.63(17)   &  1.21(48)   \\
$1/8$ &   96    &  0.31428(46)   &&  2.67(17)   &  0.96(48)   \\
      &  128    &  0.31439(49)   &&  2.66(18)   &  0.77(44)   \\
      \hline
      &   64    &  0.31483(50)   &  -1.15(30)   &  2.56(17)   &  0.59(31)   \\
$1/8$ &   96    &  0.31493(60)   &  -1.32(57)   &  2.53(19)   &  0.60(31)   \\
      &  128    &  0.31487(75)   &  -1.2(1.1)   &  2.55(23)   &  0.69(38)   \\
    \end{tabular}
  \end{ruledtabular}
  \caption{Same as Table \ref{fits_phiperp_xy_zmin4} for $z_{\rm min} = 8$.}
  \label{fits_phiperp_xy_zmin8}
\end{table*}

In Tables \ref{fits_phiperp_xy_zmin4}-\ref{fits_phiperp_xy_zmin8} we report results of fits to Eq.~(\ref{phiperp_ansatz2}).
Analogous to the previous sections, we consider in the fits MC data for $z \ge z_{\rm min}$, $(z/L) \le (z/L)_{\rm max}$, and $L \ge L_{\rm min}$.
We first consider fits where the finite-size correction is ignored, i.e., we fix $B_\varphi=0$ in Eq.~(\ref{phiperp_ansatz2}).
Fits for $z_{\rm min} = 4$ (Table \ref{fits_phiperp_xy_zmin4}) exhibit a large value of $\chi^2/{\rm d.o.f.}$.
The quality of the fits is substantially improved when we restrict the data to $z_{\rm min} = 6$ (Table \ref{fits_phiperp_xy_zmin6}), although the value of $\chi^2/{\rm d.o.f.}$ is still somewhat large.
The inclusion of the finite-size term $B_\varphi (z/L)^3$ in the fits further improves the fits, giving a good $\chi^2/{\rm d.o.f.}$.
In agreement with the observations above, the fitted value of $B_\varphi$ is small, and appears to be difficult to resolve with precision.
The fitted amplitude $M_\varphi$ is instead very stable and precise.
As an additional check of the robustness of the results, in Table \ref{fits_phiperp_xy_zmin8} we repeat the fits for $z_{\rm min} = 8$, obtaining results in line with those found for $z_{\rm min} = 6$.
By considering the variation of the fit results in Tables \ref{fits_phiperp_xy_zmin6} and \ref{fits_phiperp_xy_zmin8}, we extract the estimates of Eq.~(\ref{Mphi_xy}).

\subsection{$N=3$}
\label{app:fits:3}
\subsubsection{Bulk correlations}
\label{app:fits:3:bulk}
\begin{table}
  \begin{ruledtabular}
    \begin{tabular}{llcccc}
    $(x/L)_{\rm max}$ & $L_{\rm min}$ & ${\cal N}_{\rm bulk}$ & $B_\epsilon$ & $C$ & $\chi^2/{\rm d.o.f.}$ \\
      \hline
      &   32    &  0.311698(62)   &  2.4795(33)   &  0.3484(21)   &  22.0(3.0)   \\
      &   48    &  0.311660(59)   &  2.4849(37)   &  0.3507(19)   &  19.2(2.9)   \\
$1/4$ &   64    &  0.311631(57)   &  2.4901(41)   &  0.3527(18)   &  16.3(2.8)   \\
      &   96    &  0.311617(54)   &  2.4944(50)   &  0.3539(17)   &  14.0(2.8)   \\
      &  128    &  0.311636(51)   &  2.4980(63)   &  0.3537(15)   &  12.0(2.7)   \\
      \hline
      &   32    &  0.311817(63)   &  2.4500(43)   &  0.3443(21)   &  31.3(4.2)   \\
      &   48    &  0.311793(63)   &  2.4546(48)   &  0.3454(21)   &  31.0(4.3)   \\
$1/8$ &   64    &  0.311742(62)   &  2.4660(55)   &  0.3480(20)   &  29.5(4.5)   \\
      &   96    &  0.311676(61)   &  2.4823(67)   &  0.3512(20)   &  27.7(4.8)   \\
      &  128    &  0.311622(55)   &  2.5035(82)   &  0.3542(17)   &  24.3(4.8)   \\
    \end{tabular}
  \end{ruledtabular}
  \caption{Fits of the bulk two-point function for $N=3$ to Eq.~(\ref{phi_bulk_ansatz}) as a function of the minimum lattice size $L_{\rm min}$, and of the maximum value of $(x/L)$ taken into account.
    For all fits we consider MC data for $x\ge x_{\rm min} = 4$.
    The quoted error bars are the sum of the statistical uncertainty originating from the fit, and the dependence of the results on varying $\Delta_\phi = 0.518920(25)$ and $\Delta_\epsilon=1.5948(2)$ \cite{Hasenbusch-20} within one error bar.}
  \label{fits_bulk_heisenberg_xmin4}
\end{table}

\begin{table}
  \begin{ruledtabular}
    \begin{tabular}{llcccc}
      $(x/L)_{\rm max}$ & $L_{\rm min}$ & ${\cal N}_{\rm bulk}$ & $B_\epsilon$ & $C$ & $\chi^2/{\rm d.o.f.}$ \\
      \hline
      &   32    &  0.311987(92)   &  2.4714(32)   &  0.3000(73)   &  7.6(1.3)   \\
      &   48    &  0.311974(90)   &  2.4724(36)   &  0.3019(71)   &  7.4(1.4)   \\
$1/4$ &   64    &  0.311941(87)   &  2.4752(40)   &  0.3068(68)   &  6.5(1.5)   \\
      &   96    &  0.311917(83)   &  2.4775(48)   &  0.3114(64)   &  5.5(1.6)   \\
      &  128    &  0.311921(76)   &  2.4793(60)   &  0.3139(56)   &  4.1(1.7)   \\
      \hline
      &   48    &  0.312214(96)   &  2.4324(56)   &  0.2846(70)   &  1.3(1.0)   \\
$1/8$ &   64    &  0.312207(99)   &  2.4333(63)   &  0.2853(74)   &  1.3(1.0)   \\
      &   96    &  0.31220(10)   &  2.4348(82)   &  0.2863(80)   &  1.4(1.2)   \\
      &  128    &  0.31217(10)   &  2.439(10)   &  0.2894(79)   &  1.5(1.5)   \\
    \end{tabular}
  \end{ruledtabular}
  \caption{Same as Table \ref{fits_bulk_heisenberg_xmin4} for $x_{\rm min}=6$.}
  \label{fits_bulk_heisenberg_xmin6}
\end{table}

\begin{table}
  \begin{ruledtabular}
    \begin{tabular}{llcccc}
      $(x/L)_{\rm max}$ & $L_{\rm min}$ & ${\cal N}_{\rm bulk}$ & $B_\epsilon$ & $C$ & $\chi^2/{\rm d.o.f.}$ \\
      \hline
      &   32    &  0.31195(13)   &  2.4770(37)   &  0.297(18)   &  5.9(1.2)   \\
      &   48    &  0.31197(13)   &  2.4758(40)   &  0.293(19)   &  5.9(1.2)   \\
$1/4$ &   64    &  0.31195(13)   &  2.4770(45)   &  0.299(19)   &  5.6(1.3)   \\
      &   96    &  0.31191(12)   &  2.4788(55)   &  0.310(19)   &  5.0(1.5)   \\
      &  128    &  0.31188(11)   &  2.4820(64)   &  0.325(16)   &  3.9(1.6)   \\
      \hline
      &   64    &  0.31229(14)   &  2.4318(79)   &  0.260(18)   &  0.42(63)   \\
$1/8$ &   96    &  0.31230(15)   &  2.4305(97)   &  0.258(20)   &  0.41(61)   \\
      &  128    &  0.31231(17)   &  2.428(14)   &  0.256(24)   &  0.46(70)   \\
    \end{tabular}
  \end{ruledtabular}
  \caption{Same as Table \ref{fits_bulk_heisenberg_xmin4} for $x_{\rm min}=8$.}
  \label{fits_bulk_heisenberg_xmin8}
\end{table}

We proceed here analogous to the $XY$ case.
The rescaled bulk two-point shown in Fig.~\ref{plots_bulk}(b) suggests that its universal part is found for $x\gtrsim 4$.
In Tables \ref{fits_bulk_heisenberg_xmin4}-\ref{fits_bulk_heisenberg_xmin8} we show fits to the Eq.~(\ref{phi_bulk_ansatz}), where we also consider the variation of $\Delta_\phi = 0.518920(25)$ and $\Delta_\epsilon=1.5948(2)$ \cite{Hasenbusch-20} within one error bar.
Fits for $x_{\rm min} = 4$ (Table \ref{fits_bulk_heisenberg_xmin4}) display a large $\chi^2/{\rm d.o.f.}$.
For $x_{\rm min} = 6$ (Table \ref{fits_bulk_heisenberg_xmin6}) we observe a significant decrease of $\chi^2/{\rm d.o.f.}$, which becomes compatible with $1$ within one standard deviation for $(x/L)_{\rm max} = 1/8$; further decreasing $(x/L)_{\rm max}$ does not alter the minimum $\chi^2/{\rm d.o.f.}$.
As a further check, we considered $x_{\rm min} = 8$.
Corresponding fit results shown in Table \ref{fits_bulk_heisenberg_xmin8} exhibits a small $\chi^2/{\rm d.o.f.}$ for $(x/L)_{\rm max} = 1/8$ and the fitted values of ${\cal N}_{\rm bulk}$ are in agreement with the results for $x_{\rm min} = 6$ and $(x/L)_{\rm max} = 1/8$.
Based on these fits, we arrive to the estimates of Eq.~(\ref{Nbulk_heisenberg}).

\subsubsection{Surface correlations}
\label{app:fits:3:surface}
We observe in the surface correlations of the field component $\varphi$ shown in Fig.~\ref{plots_surface}(b) a rapid increase of the error bars on increasing the distance ${\bf x}$.
Similar to the $N=2$ case (Fig.~\ref{plots_surface}(a)), this is
due to the fast decay of correlations.

\begin{table}[t]
  \begin{ruledtabular}
    \begin{tabular}{llccc}
      $({\bf x}/L)_{\rm max}$ & $L_{\rm min}$ & ${\cal N}_\varphi$ & $C$ & $\chi^2/{\rm d.o.f.}$ \\
      \hline
            & 32    &    0.50134(45)    &    0.207(15)    &    11.6(1.0)    \\
            & 48    &    0.50132(45)    &    0.209(15)    &    11.8(1.0)    \\
      $1/8$ & 64    &    0.50112(47)    &    0.215(15)    &    12.2(1.1)    \\
            & 96    &    0.50076(49)    &    0.226(16)    &    12.1(1.1)    \\
            & 128   &    0.50024(51)    &    0.243(17)    &    12.7(1.3)    \\
      \hline
             & 48    &    0.50161(46)    &    0.198(15)    &    19.3(1.7)    \\
      $1/12$ & 64    &    0.50161(46)    &    0.199(15)    &    20.1(1.8)    \\
             & 96    &    0.50091(48)    &    0.222(16)    &    20.1(1.9)    \\
             & 128   &    0.50025(51)    &    0.242(17)    &    21.4(2.2)    \\
    \end{tabular}
  \end{ruledtabular}
  \caption{Fits of the surface two-point function for $N=3$ to Eq.~(\ref{phi_ansatz}) as a function of the minimum lattice size $L_{\rm min}$, and of the maximum value of $(x/L)$ taken into account. For all fits we consider MC data for ${\bf x}\ge {\bf x}_{\rm min} = 4$.}
  \label{fits_phi_heisenberg_xmin4}
\end{table}

\begin{table}[t]
  \begin{ruledtabular}
    \begin{tabular}{llccc}
      $({\bf x}/L)_{\rm max}$ & $L_{\rm min}$ & ${\cal N}_\varphi$ & $C$ & $\chi^2/{\rm d.o.f.}$ \\
      \hline
            & 48    &    0.4818(21)    &    1.49(17)    &    1.47(42)    \\
      $1/8$ & 64    &    0.4818(21)    &    1.49(17)    &    1.51(43)    \\
            & 96    &    0.4821(22)    &    1.46(18)    &    1.61(46)    \\
            & 128   &    0.4816(23)    &    1.48(20)    &    1.64(53)    \\
      \hline
      $1/12$ & 96    &    0.4809(23)    &    1.56(19)    &    1.52(59)    \\
             & 128   &    0.4805(24)    &    1.58(21)    &    1.37(62)    \\
    \end{tabular}
  \end{ruledtabular}
  \caption{Same as Table \ref{fits_phi_heisenberg_xmin4} for ${\bf x}_{\rm min}=6$.}
  \label{fits_phi_heisenberg_xmin6}
\end{table}

In Tables \ref{fits_phi_heisenberg_xmin4} and \ref{fits_phi_heisenberg_xmin6} we report results of fits to Eq.~(\ref{phi_ansatz}), as a function of $L_{\rm min}$, $({\bf x}/L)_{\rm max}$, and for two values of the minimum separation ${\bf x}_{\rm min}$ considered.
Fits for ${\bf x}_{\rm min} = 4$ (Table \ref{fits_phi_heisenberg_xmin4}) exhibit a large value of $\chi^2/{\rm d.o.f.}$.
Increasing ${\bf x}_{\rm min}$ to ${\bf x}_{\rm min} = 6$ (Table \ref{fits_phi_heisenberg_xmin6}) significantly improves the value of $\chi^2/{\rm d.o.f.}$.
Although in the fits of Table \ref{fits_phi_heisenberg_xmin6}
the central value of $\chi^2/{\rm d.o.f.}$ remains somewhat large,
we notice that it is nevertheless compatible with $1$ within one estimated standard deviation.
The inclusion of a finite-size correction $\propto ({\bf x}/L)^3$ in the fit Ansatz of Eq.~(\ref{phi_ansatz}) does not significantly change the $\chi^2/{\rm d.o.f.}$: indeed, Fig.~\ref{plots_surface}(b) suggests a size dependence which is smaller than the statistical error bars.
From Table \ref{fits_phi_heisenberg_xmin6} we obtain the estimates given in Eq.~(\ref{Nphi_heisenberg}).

\subsubsection{Magnetization profile}
\label{app:fits:3:profile}
The rescaled order-parameter profile $\langle \sigma(z)\rangle$ shown in Fig.~\ref{plots_surface}(d) exhibits scaling corrections and finite-size effects, similar to the $N=2$ case.

\begin{table*}[t]
  \begin{ruledtabular}
    \begin{tabular}{llcccc}
      $(z/L)_{\rm max}$ & $L_{\rm min}$ & $M_\sigma$ & $B_\sigma$ & $z_0$ & $\chi^2/{\rm d.o.f.}$ \\
      \hline
      &   32    &  0.706517(95)   &  1.1503(42)   &  1.03951(97)   &  9.3(1.2)   \\
      &   48    &  0.706468(95)   &  1.1640(55)   &  1.03866(97)   &  7.7(1.2)   \\
$1/4$ &   64    &  0.706470(94)   &  1.1604(70)   &  1.03862(96)   &  7.9(1.2)   \\
      &   96    &  0.706473(92)   &  1.1559(91)   &  1.03866(91)   &  8.6(1.3)   \\
      &  128    &  0.706490(96)   &  1.163(13)   &  1.03883(97)   &  8.0(1.5)   \\
      \hline
      &   32    &  0.706625(93)   &  1.065(13)   &  1.04086(91)   &  9.8(1.9)   \\
      &   48    &  0.706605(96)   &  1.083(17)   &  1.04057(95)   &  9.7(1.9)   \\
$1/8$ &   64    &  0.70663(10)   &  1.046(23)   &  1.0410(10)   &  9.3(2.1)   \\
      &   96    &  0.70673(10)   &  0.927(34)   &  1.0424(10)   &  7.0(2.1)   \\
      &  128    &  0.706734(99)   &  0.903(42)   &  1.0425(10)   &  8.2(2.6)   \\
    \end{tabular}
  \end{ruledtabular}
  \caption{Fits of the order-parameter profile for $N=3$ to Eq.~(\ref{sigma_ansatz}) as a function of the minimum lattice size $L_{\rm min}$, and of the maximum value of $(z/L)$ taken into account. For all fits we consider MC data for $z\ge z_{\rm min} = 4$.
  The quoted error bars are the sum of the statistical uncertainty originating from the fit, and the dependence of the results on varying $\Delta_\phi = 0.518920(25)$ \cite{Hasenbusch-20} within one error bar.}
  \label{fits_sigma_heisenberg_xmin4}
\end{table*}

\begin{table*}[t]
  \begin{ruledtabular}
    \begin{tabular}{llcccc}
      $(z/L)_{\rm max}$ & $L_{\rm min}$ & $M_\sigma$ & $B_\sigma$ & $z_0$ & $\chi^2/{\rm d.o.f.}$ \\
      \hline
      &   32    &  0.70626(12)   &  1.1659(41)   &  1.0328(18)   &  5.11(95)   \\
      &   48    &  0.70617(13)   &  1.1826(55)   &  1.0307(19)   &  3.15(91)   \\
$1/4$ &   64    &  0.70614(13)   &  1.1870(72)   &  1.0300(19)   &  3.09(95)   \\
      &   96    &  0.70612(12)   &  1.1914(90)   &  1.0293(18)   &  3.2(1.1)   \\
      &  128    &  0.70612(12)   &  1.206(13)   &  1.0289(18)   &  1.41(51)   \\
      \hline
      &   48    &  0.70630(12)   &  1.101(18)   &  1.0326(17)   &  1.45(94)   \\
$1/8$ &   64    &  0.70631(13)   &  1.093(24)   &  1.0328(19)   &  1.45(98)   \\
      &   96    &  0.70637(15)   &  1.050(42)   &  1.0341(22)   &  1.2(1.0)   \\
      &  128    &  0.70630(15)   &  1.107(55)   &  1.0325(24)   &  1.09(92)   \\
    \end{tabular}
  \end{ruledtabular}
  \caption{Same as Table \ref{fits_sigma_heisenberg_xmin4} for $z_{\rm min} = 6$.}
  \label{fits_sigma_heisenberg_xmin6}
\end{table*}

\begin{table*}[t]
  \begin{ruledtabular}
    \begin{tabular}{llcccc}
      $(z/L)_{\rm max}$ & $L_{\rm min}$ & $M_\sigma$ & $B_\sigma$ & $z_0$ & $\chi^2/{\rm d.o.f.}$ \\
      \hline
      &   32    &  0.70617(15)   &  1.1782(47)   &  1.0303(30)   &  3.75(94)   \\
      &   48    &  0.70610(16)   &  1.1872(57)   &  1.0286(31)   &  3.00(98)   \\
$1/4$ &   64    &  0.70606(17)   &  1.1926(79)   &  1.0273(34)   &  2.9(1.1)   \\
      &   96    &  0.70600(16)   &  1.2013(97)   &  1.0253(32)   &  2.9(1.2)   \\
      &  128    &  0.70593(16)   &  1.225(13)   &  1.0222(31)   &  0.50(25)   \\
      \hline
      &   64    &  0.70622(17)   &  1.091(28)   &  1.0299(30)   &  1.00(92)   \\
$1/8$ &   96    &  0.70626(19)   &  1.072(45)   &  1.0309(37)   &  0.95(97)   \\
      &  128    &  0.70608(23)   &  1.171(71)   &  1.0263(46)   &  0.17(28)   \\
    \end{tabular}
  \end{ruledtabular}
  \caption{Same as Table \ref{fits_sigma_heisenberg_xmin4} for $z_{\rm min} = 8$.}
  \label{fits_sigma_heisenberg_xmin8}
\end{table*}

In Tables \ref{fits_sigma_heisenberg_xmin4}-\ref{fits_sigma_heisenberg_xmin8} we report fit results, as a function of a minimum value $z_{\rm min}$ of $z$, the maximum value $(z/L)_{\rm max}$ of $z/L$ and the minimum lattice size $L_{\rm min}$ taken into account.
As in the case $N=2$, we additionally vary the exponent $\Delta_\phi = 0.518920(25)$ \cite{Hasenbusch-20} within one error bar, and add the resulting variation to the statistical error bars arising from the fits.

Fits for $z_{\rm min}=4$ (Table \ref{fits_sigma_heisenberg_xmin4}) exhibit a large $\chi^2/{\rm d.o.f.}$.
Its value is significantly reduced when we increase $z_{\rm min}$ to $z_{\rm min}=6$ (Table \ref{fits_sigma_heisenberg_xmin6}) and $z_{\rm min}=8$ (Table \ref{fits_sigma_heisenberg_xmin8}).
Nevertheless, we observe also in these cases a large $\chi^2/{\rm d.o.f.}$ for $(z/L)_{\rm max}=1/4$, suggesting that, within the precision of MC data, the order-parameter profile is not well approximated by Eq.~(\ref{sigma_ansatz}), when $(z/L)_{\rm max} \sim 1/4$.
Restricting the fit to $(z/L)_{\rm max}=1/8$, we obtain a good $\chi^2/{\rm d.o.f.}$ for $z_{\rm min}=8$, while for $z_{\rm min}=6$ the value of $\chi^2/{\rm d.o.f.}$ is nevertheless compatible with $1$, within one standard deviation.
Irrespective of the above considerations on the quality of the fit, we notice that the fitted value of $M_\sigma$ in Tables \ref{fits_sigma_heisenberg_xmin6} and \ref{fits_sigma_heisenberg_xmin8} is extremely stable.
Judging conservatively the variation in the results of Tables \ref{fits_sigma_heisenberg_xmin6} and \ref{fits_sigma_heisenberg_xmin8} for $(z/L)_{\rm max}=1/8$, we can obtain the estimate given in Eq.~(\ref{sigma_heisenberg}).

\subsubsection{Surface-bulk correlations}
\label{app:fits:3:surface-bulk}

\begin{table*}
  \begin{ruledtabular}
    \begin{tabular}{llcccc}
      $(z/L)_{\rm max}$ & $L_{\rm min}$ & $M_\varphi$ & $B_\varphi$ & $C$ & $\chi^2/{\rm d.o.f.}$ \\
      \hline
      &   32    &  0.46290(37)   &&  3.318(50)   &  18.1(1.4)   \\
      &   48    &  0.46299(37)   &&  3.314(50)   &  15.9(1.3)   \\
$1/4$ &   64    &  0.46315(37)   &&  3.305(50)   &  15.0(1.3)   \\
      &   96    &  0.46325(37)   &&  3.299(50)   &  15.9(1.5)   \\
      &  128    &  0.46334(38)   &&  3.294(51)   &  16.3(1.7)   \\
      \hline
      &   32    &  0.46329(37)   &  -0.474(42)   &  3.296(49)   &  15.4(1.3)   \\
      &   48    &  0.46318(38)   &  -0.284(74)   &  3.302(49)   &  15.6(1.3)   \\
$1/4$ &   64    &  0.46307(38)   &  0.17(11)   &  3.310(49)   &  15.1(1.3)   \\
      &   96    &  0.46298(39)   &  0.82(19)   &  3.315(50)   &  15.3(1.4)   \\
      &  128    &  0.46304(40)   &  1.34(28)   &  3.312(50)   &  15.4(1.5)   \\
      \hline
      &   32    &  0.46301(37)   &&  3.311(49)   &  33.2(2.8)   \\
      &   48    &  0.46300(37)   &&  3.313(49)   &  33.0(2.8)   \\
$1/8$ &   64    &  0.46308(37)   &&  3.310(49)   &  31.9(2.9)   \\
      &   96    &  0.46318(37)   &&  3.304(50)   &  33.5(3.2)   \\
      &  128    &  0.46329(38)   &&  3.297(51)   &  34.3(3.5)   \\
      \hline
      &   32    &  0.46319(37)   &  -0.546(76)   &  3.304(49)   &  32.4(2.8)   \\
      &   48    &  0.46316(38)   &  -0.47(11)   &  3.305(49)   &  33.0(2.9)   \\
$1/8$ &   64    &  0.46290(38)   &  0.59(18)   &  3.319(49)   &  32.1(3.0)   \\
      &   96    &  0.46237(40)   &  3.51(36)   &  3.351(49)   &  28.5(3.0)   \\
      &  128    &  0.46225(42)   &  6.16(58)   &  3.359(50)   &  26.1(3.0)   \\
    \end{tabular}
  \end{ruledtabular}
  \caption{Fits of the surface-bulk correlations for $N=3$ to Eq.~(\ref{phiperp_ansatz2}) as a function of the minimum lattice size $L_{\rm min}$, and of the maximum value of $(z/L)$ taken into account.
    For all fits we use $z_0=1.031(4)$ (Eq.~(\ref{sigma_heisenberg})) and consider MC data for $z\ge z_{\rm min} = 4$.
    The quoted error bars are the sum of the uncertainty stemming from the fit, and the spread of the results on varying $z_0$ within one error bar quoted in Eq.~(\ref{sigma_heisenberg}).
    An absent $B_\varphi$ indicates that we fixed it to $B_\varphi=0$, see main text.}
  \label{fits_phiperp_heisenberg_zmin4}
\end{table*}

\begin{table*}
  \begin{ruledtabular}
    \begin{tabular}{llcccc}
      $(z/L)_{\rm max}$ & $L_{\rm min}$ & $M_\varphi$ & $B_\varphi$ & $C$ & $\chi^2/{\rm d.o.f.}$ \\
      \hline
      &   32    &  0.46599(39)   &&  2.901(74)   &  5.73(67)   \\
      &   48    &  0.46602(39)   &&  2.906(74)   &  3.97(56)   \\
$1/4$ &   64    &  0.46622(39)   &&  2.891(73)   &  2.15(43)   \\
      &   96    &  0.46644(39)   &&  2.869(74)   &  1.67(38)   \\
      &  128    &  0.46656(42)   &&  2.858(77)   &  1.43(41)   \\
      \hline
      &   32    &  0.46692(39)   &  -0.638(41)   &  2.813(73)   &  1.30(27)   \\
      &   48    &  0.46704(40)   &  -0.728(68)   &  2.800(73)   &  1.26(27)   \\
$1/4$ &   64    &  0.46693(42)   &  -0.62(11)   &  2.814(74)   &  1.20(28)   \\
      &   96    &  0.46693(45)   &  -0.62(18)   &  2.814(76)   &  1.30(31)   \\
      &  128    &  0.46686(47)   &  -0.54(27)   &  2.824(78)   &  1.28(38)   \\
      \hline
      &   48    &  0.46664(41)   &&  2.834(74)   &  2.55(59)   \\
$1/8$ &   64    &  0.46658(41)   &&  2.847(74)   &  1.88(53)   \\
      &   96    &  0.46660(41)   &&  2.848(75)   &  1.71(52)   \\
      &  128    &  0.46664(43)   &&  2.848(77)   &  1.51(56)   \\
      \hline
      &   48    &  0.46694(42)   &  -0.70(13)   &  2.813(74)   &  1.65(48)   \\
$1/8$ &   64    &  0.46684(44)   &  -0.53(20)   &  2.824(76)   &  1.63(48)   \\
      &   96    &  0.46675(51)   &  -0.30(40)   &  2.833(81)   &  1.72(52)   \\
      &  128    &  0.46666(55)   &  -0.06(63)   &  2.846(83)   &  1.57(58)   \\
    \end{tabular}
  \end{ruledtabular}
  \caption{Same as Table \ref{fits_phiperp_heisenberg_zmin4} for $z_{\rm min} = 6$.}
  \label{fits_phiperp_heisenberg_zmin6}
\end{table*}

\begin{table*}
  \begin{ruledtabular}
    \begin{tabular}{llcccc}
      $(z/L)_{\rm max}$ & $L_{\rm min}$ & $M_\varphi$ & $B_\varphi$ & $C$ & $\chi^2/{\rm d.o.f.}$ \\
      \hline
      &   32    &  0.46642(44)   &  2.80(12)   &  4.15(59)   \\
      &   48    &  0.46630(44)   &  2.85(12)   &  3.38(53)   \\
$1/4$ &   64    &  0.46640(44)   &  2.86(12)   &  2.05(44)   \\
      &   96    &  0.46673(45)   &  2.81(12)   &  1.51(39)   \\
      &  128    &  0.46707(51)   &  2.73(13)   &  1.26(39)   \\
      \hline
      &   32    &  0.46763(44)   &  -0.660(54)   &  2.65(12)   &  0.97(23)   \\
      &   48    &  0.46782(47)   &  -0.739(72)   &  2.62(12)   &  0.90(22)   \\
$1/4$ &   64    &  0.46774(50)   &  -0.71(11)   &  2.64(12)   &  0.92(23)   \\
      &   96    &  0.46793(57)   &  -0.82(19)   &  2.60(13)   &  0.93(24)   \\
      &  128    &  0.46802(61)   &  -0.89(28)   &  2.56(14)   &  0.89(29)   \\
      \hline
      &   64    &  0.46739(49)   &  2.67(12)   &  0.93(36)   \\
$1/8$ &   96    &  0.46735(50)   &  2.68(12)   &  0.91(35)   \\
      &  128    &  0.46741(53)   &  2.66(13)   &  0.84(38)   \\
      \hline
      &   64    &  0.46761(53)   &  -0.39(27)   &  2.64(12)   &  0.86(32)   \\
$1/8$ &   96    &  0.46762(65)   &  -0.40(45)   &  2.64(14)   &  0.89(33)   \\
      &  128    &  0.46797(77)   &  -0.94(77)   &  2.57(16)   &  0.75(32)   \\
    \end{tabular}
  \end{ruledtabular}
  \caption{Same as Table \ref{fits_phiperp_heisenberg_zmin4} for $z_{\rm min} = 8$.}
  \label{fits_phiperp_heisenberg_zmin8}
\end{table*}

Similar to the $N=2$ case (Fig.~\ref{plots_surface}(e)), in rescaled surface-bulk correlation function $\<\vec{\varphi}(0)\cdot \vec{\varphi}(0,z)\>$ shown in Fig.~\ref{plots_surface}(f) we also observe significant scaling corrections, and comparatively smaller finite-size corrections.
A quantitative analysis of the correlations $\<\vec{\varphi}(0)\cdot \vec{\varphi}(0,z)\>$ presents the same challenges as for the $N=2$ case.
 In particular, fits to Eq.~(\ref{phiperp_ansatz}) provide unstable results.
As in the $N=2$ case, we fit the MC data to Eq.~(\ref{phiperp_ansatz2}), using the value of $z_0$ determined in Eq.~(\ref{sigma_heisenberg}) from the order-parameter profile.
In Tables \ref{fits_phiperp_heisenberg_zmin4}-\ref{fits_phiperp_heisenberg_zmin8} we present the fit results.
While fits for $z_{\rm min}=4$ (Table \ref{fits_phiperp_heisenberg_zmin4}) give a large $\chi^2/{\rm d.o.f.}$, restricting the data to $z_{\rm min}=6$ (Table \ref{fits_phiperp_heisenberg_zmin6}) and $z_{\rm min}=8$ (Table \ref{fits_phiperp_heisenberg_zmin8}) gives stable and consistent results especially for the amplitude $M_\varphi$, while the amplitude of the finite-size correction $B_\varphi$ can be determined only with a limited precision.
Judging conservatively the variation of the results in Tables \ref{fits_phiperp_heisenberg_zmin6} and \ref{fits_phiperp_heisenberg_zmin8} we arrive to the estimates of Eq.~(\ref{Mphi_heisenberg}).

\end{document}